\documentclass[12pt,a4paper,twoside]{report}

\usepackage{lscape}
\usepackage[spanish]{babel}
\usepackage{amssymb,amsmath}
\usepackage{graphicx,epsfig,color}
\usepackage{amsfonts,latexsym}
\usepackage{longtable}

\def\nt{{\noindent }}

\newtheorem{theorem}{Theorem}[section]

\newtheorem{teorema}[theorem]{Teorema}
\newtheorem{proposicion}[theorem]{Proposici\'{o}n}
\newtheorem{lema}[theorem]{Lema}
\newtheorem{corolario}[theorem]{Corolario}
\newtheorem{observacion}[theorem]{Observaci\'{o}n}
\newtheorem{definicion}[theorem]{Definici\'{o}n}

\newtheorem{supuesto}[theorem]{Supuesto}

\begin{document}
\renewcommand{\tablename}{Tabla}
\renewcommand{\listtablename}{\'Indice de tablas}

\pagestyle{empty}
\vspace{10 cm}
\begin{center}
{\bf  \huge TESTS DE BONDAD DE AJUSTE\\[.6 cm]
	 PARA LA DISTRIBUCI\'ON\\[1 cm]
	  POISSON BIVARIANTE}\\[1 cm]

\Large Francisco Novoa Mu\~noz\\[2.1 cm]

\vspace{.8cm}

Concepci\'on, Agosto de 2018
\end{center}

\thispagestyle{empty}

\pagenumbering{roman}

\newpage

\hfill

\newpage

\thispagestyle{empty}
\setcounter{tocdepth}{3}
\setcounter{secnumdepth}{3}

\pagestyle{plain}
\pagenumbering{roman}
\tableofcontents
\listoftables



\chapter*{Introducci\'on}\label{Introduccion}
\addcontentsline{toc}{chapter}{Introducci\'on}
\pagenumbering{arabic}
Los datos de conteo pueden aparecer bajo diferentes circunstancias. En un marco univariante, la distribuci\'on Poisson es la distribuci\'on que con mayor frecuencia ha sido empleada para modelar tales datos (ver por ejemplo, Haight (1967, pp. 100--107) \cite{Hai67}, Johnson y Kotz (1969, pp. 88--90) \cite{JoKo69}, Sahai y Khurshid (1993) \cite{SaKh93}).

En la pr\'actica, los datos de conteo bivariantes surgen en varias disciplinas diferentes y la distribuci\'on Poisson bivariante (DPB), siendo una generalizaci\'on de la distribuci\'on Poisson, juega un rol importante al momento de modelarlos, siempre que dichos datos presenten una correlaci\'on no negativa.

Esta distribuci\'on ha sido usada para modelar datos que aparecen en un amplio rango de campos incluyendo medicina, para mediciones en el pretratamiento y en el postratamiento en los mismos pacientes, o el n\'umero de accidentes por trabajador en una factor\'ia durante dos intervalos dados de tiempo (ver por ejemplo, Hamdan (1972) \cite{Ham72}); en marketing, para el n\'umero de compras de diferentes productos; en epidemiolog\'ia, para el n\'umero de incidentes de distintos tipos de muertes en una serie de distritos; en deportes, para el n\'umero de goles marcados por cada uno de los dos equipos oponentes en un partido de balompi\'e (ver por ejemplo, Maher (1982) \cite{Mah82}, Karlis y Ntzoufras (2000) \cite{KaNt00}, (2003a) \cite{KaNt03a}, (2003b) \cite{KaNt03b}, (2005) \cite{KaNt05}, Rue y Salvesen (2000) \cite{RuSa00}); en biolog\'ia, para el n\'umero de semillas y plantas que crecen en una parcela (ver por ejemplo, Lakshminarayana, Pandit y Rao (1999) \cite{LaPaRa99}); en econometr\'ia, para el n\'umero de cambios de trabajo voluntarios e involuntarios (ver por ejemplo, Jung y Winkelmann (1993) \cite{JuWi93}); en datos de turismo (ver por ejemplo, Berkhout y Plug (2004) \cite{BePl04}); en seguros de coches (ver por ejemplo, Berm\'udez (2009) \cite{Ber09}); en sanidad (ver por ejemplo, Karlis y Ntzoufras (2005) \cite{KaNt05}, Karlis y Tsiamyrtzis (2008) \cite{KaTs08}); en la industria textil, para el n\'umero de dos tipos de defectos en muestras de fibras de textil (ver por ejemplo, Ho y Singer (2001) \cite{HoSi01}), entre muchos otros.

En el caso multivariante, la distribuci\'on Poisson ha sido usada para modelar redes sociales multi-relacionales (ver por ejemplo, Dai, Chua y Lim (2012) \cite{DaChLi12}).

Contrastar la bondad de ajuste de las observaciones dadas con un modelo probabil\'istico es un aspecto crucial del an\'alisis de datos. Para el caso univariante, se han construido muchos tests de bondad de ajuste con la finalidad de comprobar si los datos provienen de una distribuci\'on Poisson (para una revisi\'on detallada, ver G\"urtler y Henze, (2000) \cite{GuHe00}). En comparaci\'on, la literatura sobre tests de bondad de ajuste para la DPB es m\'as bien escasa. Hasta donde conocemos, podemos mencionar el test propuesto por Crockett (1979) \cite{Cro79}, el test desarrollado por Loukas y Kemp (1986) \cite{LoKe86}, que se basa en una extensi\'on del \'indice de dispersi\'on univariante, y el test sugerido por Rayner y Best (1995) \cite{RaBe95}, que consiste en una modificaci\'on del test dado por Loukas y Kemp (1986) \cite{LoKe86}. La principal desventaja de estos tests de bondad de ajuste es que no son consistentes contra cada alternativa fija.

El objetivo de esta texto es proponer y estudiar tests de bondad de ajuste para la DPB, que sean consistentes. Dado que la funci\'on generatriz de probabilidad (fgp) caracteriza la distribuci\'on de un vector aleatorio y se puede estimar consistentemente por la funci\'on generatriz de probabilidad emp\'irica (fgpe), los tests que proponemos son funciones de la fgpe. El primer test estad\'istico compara la fgpe de los datos con un estimador de la fgp de la DPB. Luego, mostramos que la fgp de la DPB es la \'unica fgp que satisface cierto sistema de ecuaciones diferenciales parciales, lo cual nos lleva a proponer dos tests estad\'isticos basados en el an\'alogo emp\'irico de dicho sistema, uno de ellos de tipo Cram\'er-von Mises y el otro se basa en los coeficientes de los polinomios de la versi\'on emp\'irica. Los tests que proponemos pueden ser vistos como extensiones al caso bivariante de algunos tests de bondad de ajuste dise\~nados para el caso univariante.

Con el fin de decidir cu\'ando rechazar la hip\'otesis nula, debemos conocer la distribuci\'on nula del test estad\'istico o, al menos, una aproximaci\'on de la misma. Puesto que, para los tests propuestos, no es posible obtener las distribuciones nulas para tama\~nos de muestra finito, las aproximamos por las asint\'oticas, las cuales resultaron depender de cantidades desconocidas, por lo tanto no son \'utiles como estimaciones de la distribuci\'on nula. As\'i, para aproximar la distribuci\'on nula, proponemos un estimador bootstrap param\'etrico.

En cuanto a la potencia, obtuvimos que los tests que proponemos son consistentes contra alternativas fijas. Adem\'as, analizamos el comportamiento asint\'otico de los tests estad\'isticos bajo alternativas contiguas y encontramos que son capaces de detectar alternativas que convergen a la nula a raz\'on de $n^{-1/2}$.

Todas las propiedades estudiadas de los tests introducidos en este texto son asint\'oticas, es decir, describen el comportamiento de los tests para muestras de tama\~no grande. Con la finalidad de evaluar el comportamiento de los tests propuestos para muestras de tama\~no finito, realizamos un estudio de simulaci\'on. En todos los casos considerados, el m\'etodo bootstrap proporciona una buena aproximaci\'on a la distribuci\'on nula. En cuanto a la potencia, a diferencia de los tests estad\'isticos dise\~nados por los investigadores  Crockett (1979) \cite{Cro79}, Loukas y Kemp (1986) \cite{LoKe86}, y Rayner y Best (1995) \cite{RaBe95}, los tests que proponemos fueron capaces de detectar todas las alternativas seleccionadas.

Con el prop\'osito de mantener la notaci\'on tan simple como sea posible, desarrollamos el an\'alisis te\'orico para el caso bivariante, pero los m\'etodos se pueden extender de manera natural al caso multivariante.

El presente texto se organiza de la siguiente manera. En el Cap\'itulo \ref{Resultados previos y definiciones} presentamos algunos resultados preliminares que nos servir\'an en los cap\'itulos siguientes, tambi\'en damos la definici\'on de la DPB con algunas de sus propiedades y adem\'as, mostramos contrastes de bondad de ajuste para la distribuci\'on Poisson tanto para datos univariantes como para datos bivariantes.

El Cap\'itulo \ref{Estadisticos-tipoCramer-von-Mises} contiene los dos primeros tests estad\'isticos que proponemos, que son de tipo Cram\'er-von Mises. Aqu\'i, tambi\'en mostramos la distribuci\'on asint\'otica nula de los tests estad\'isticos y proporcionamos estimadores bootstraps consistentes. En la parte final, estudiamos la potencia de los tests propuestos frente a alternativas fijas y locales.

En el Cap\'itulo \ref{Estadistico-Wn} presentamos el tercer estad\'istico que proponemos y describimos sus caracter\'isticas. Estudiamos su distribuci\'on asint\'otica nula y aproximamos su distribuci\'on nula por medio de un estimador bootstrap consistente. Tambi\'en, analizamos su potencia frente a alternativas fijas y contiguas.

El Cap\'itulo \ref{Resultados-numericos} est\'a dedicado a mostrar los resultados de un estudio de simulaci\'on y la aplicaci\'on de los tests propuestos a dos conjuntos de datos reales. Dicho estudio de simulaci\'on fue realizado con el objetivo de evaluar el comportamiento de los tests que proponemos y comparar la potencia tanto entre ellos como con otros tests que encontramos en la literatura estad\'istica.

En el Cap\'itulo \ref{Expresiones-matematicas} entregamos las expresiones matem\'aticas de los tests que hemos desarrollado y damos algunos detalles t\'ecnicos que son muy \'utiles al momento de implementar algoritmos o subrutinas en alg\'un lenguaje de programaci\'on.

El Cap\'itulo \ref{Extensiones} muestra c\'omo los tests propuestos se pueden extender al caso multivariante general.


\chapter{Resultados previos y definiciones} \label{Resultados previos y definiciones}
\section{Notaci\'on}\label{Notacion}
\begin{itemize}
  \item Todos los vectores a utilizar en este texto ser\'an vectores filas y $v^\top$ es el traspuesto del vector fila $v$.
  \item Para cualquier vector $v,\, v_k$ denota su $k-$\'esima coordenada.
  \item $\mathbb{N}_0=\{0,1,2,3,\ldots\}$.
  \item $\Theta=\{\theta= \,(\theta_1,\theta_2,\theta_3)\in \mathbb{R}^3\ :\ \theta_1>\theta_3, \ \theta_2>\theta_3,\ \theta_3> 0\}$.
  \item $\langle\, \cdot \, ,\cdot\,\rangle_{_V}$ denota el producto escalar en el espacio vectorial $V$.
  \item $\|\cdot\|_{_V}$ denota la norma en el espacio vectorial $V$ y $\|\cdot\|$ denota la norma Euclidea de $\mathbb{R}^d,\, d\in \mathbb{N}$.
  \item $I\{A\}$ denota la funci\'on indicadora del conjunto $A$.
  \item $P_{\vartheta}$ denota la ley de probabilidades de una DPB con par\'ametro $\vartheta$.
  \item $E_{\vartheta}$ denota la esperanza con respecto a la funci\'on de probabilidad $P_{\vartheta}$.
  \item $\mathop{\longrightarrow} \limits^{L}\ $ denota convergencia en ley (o en distribuci\'on).
\item $\mathop{\longrightarrow} \limits^{P}\ $ denota convergencia en probabilidad.
  \item $\mathop{\longrightarrow} \limits^{c.s.}\ $ denota convergencia casi segura (c.s.).
  \item Si $\{C_n\}$ es una sucesi\'on de variables aleatorias (v.a.) y $\, \epsilon\in \mathbb{R}$, entonces
      \begin{itemize}
        \item \vskip -.1 cm $C_n=O_{_P}(n^{-\epsilon})\,$ significa que $\,n^{\epsilon}\,C_n\,$ est\'a acotada en probabilidad, cuando $\,n\rightarrow\infty$.
        \item $C_n=o_{_P}(n^{-\epsilon})\,$ significa que $\,n^{\epsilon}\,C_n\ \mathop{\longrightarrow} \limits^{\!P} \ 0$, cuando $\,n\rightarrow\infty$.
        \item $C_n=o(n^{-\epsilon})\,$ significa que $\,n^{\epsilon}\,C_n\ \mathop{\longrightarrow} \limits^{\!c.s.} \ 0$, cuando $\,n\rightarrow\infty$.
      \end{itemize}
\end{itemize}

\section{Resultados preliminares}
Para demostrar la Proposici\'on \ref{Conv-FuncGenProbBiv}, dada a continuaci\'on, primero presentaremos un lema, que es una extensi\'on del Lema 8.2.6 en Athreya y Lahiri (2006, p. 242) \cite{AtLa06}. Previamente, consideremos la siguiente notaci\'on: para un conjunto arbitrario $S$, $\partial S$ e $int S$ denotan los conjuntos de puntos frontera y puntos interiores de $S$, respectivamente.

\begin{lema} \label{polya-d}
Sean $\{f_n\}_{n\geq 1}$ y $f$ una colecci\'on de funciones reales y no decrecientes definidas sobre $Q=[b_1,c_1]\times [b_2,c_2]\times\cdots\times [b_d,c_d] \subseteq \mathbb{R}^d$, con $-\infty<b_j\leq c_j<\infty$, $j=1,2,\ldots,d$. Sea $D=D_1\cup D_2 \cup D_3$, donde $D_1$ es el conjunto de v\'ertices de $Q$, $D_2$ es un conjunto denso en $\partial Q$ y $D_3$ es un conjunto denso en $int Q$. Si $f$ es continua en $Q$ y
\[\left|f_{n}(x)-f(x)\right|\to 0, \quad  \forall x \in D,\]
entonces \[\displaystyle \sup_{x\in Q}|f_n(x)-f(x)| \to 0.\]
\end{lema}
\nt {\bf Demostraci\'on} \hspace{2pt} Sea $\varepsilon>0$ arbitrario pero fijo. Como $f$ es una funci\'on continua y $Q$ es un conjunto compacto, entonces $f$ es uniformemente continua en $Q$ y por tanto
\begin{equation}\label{f(x)-f(y)<e}
\exists\delta=\delta(\varepsilon)>0\ \, \text{tal que}\ \, \forall x,y\in Q:\, \|x-y\|<\delta\, \Longrightarrow \, |f(x)-f(y)|<\varepsilon.
\end{equation}
Sean
\[H_i=[u_{1i},v_{1i}]\times [u_{2i},v_{2i}]\times\cdots\times [u_{di},v_{di}],\ \ i=1,2,\ldots,M,\]
con $\|u_i-v_i\|< \delta$, $u_i=(u_{1i},u_{2i},\ldots,u_{di})\in D$ y $v_i=(v_{1i},v_{2i},\ldots,v_{di})\in D$, $1\leq i \leq M$, de modo que
\[Q\subseteq \bigcup_{i=1}^M H_i.\]
Consideremos
\begin{equation}\label{Delta-n}
\Delta_n=\max_{x\in V} \left |f_n(x)-f(x)\right |,
\end{equation}
donde $V=\{u_i,v_i: i=1,2,\ldots,M\}$.
Por la convergencia puntual de $f_n(\cdot)$ a $f(\cdot)$ en $D$, obtenemos
\begin{equation}\label{Delta-n-a-cero}
\Delta_n \to 0.
\end{equation}

Por otro lado, para cada $x\in Q$, existe $i\in\{1,2,\ldots,M\}$, tal que $x\in H_i$, con lo cual, $u_i\leq x\leq v_i$.
La monoticidad de las funciones $f_n$ y $f$, junto con (\ref{f(x)-f(y)<e}) y (\ref{Delta-n}), implican
\[f_n(x)-f(x)\leq f_n(v_i)-f(u_i)=f_n(v_i)-f(v_i)+f(v_i)-f(u_i)\leq \Delta_n+\varepsilon.\]
An\'alogamente
\[f_n(x)-f(x)\geq -\Delta_n-\varepsilon.\]

De (\ref{Delta-n-a-cero}) y puesto que $\varepsilon >0$ es arbitrario se logra
\[\sup_{x\in Q}|f_n(x)-f(x)|  \to 0.\ \square\] 

Feuerverger (1989) \cite{Fue89} demostr\'o que la funci\'on generatriz de momentos emp\'irica converg\'ia c.s. a la funci\'on generatriz de momentos (fgm), en conjuntos compactos, para v.a.

Aunque existe una relaci\'on uno a uno entre la fgm y la fgp, hay problemas en el origen, por esta raz\'on, probaremos a continuaci\'on, que la fgpe converge c.s. a la fgp, para el caso multivariante en general, en conjuntos de la forma $Q=[b_1,c_1]\times [b_2,c_2]\times\cdots\times [b_d,c_d]$, donde $0\leq b_j\leq c_j<\infty,\ 1\leq j\leq d$, $d\in\mathbb{N}$.

Tambi\'en probaremos, que las derivadas de la fgpe convergen c.s. a las derivadas de la fgp, en conjuntos $Q$ como los definidos en el p\'arrafo anterior. Sea $d\in\mathbb{N}$. Para cualquier funci\'on $h:S\subseteq\mathbb{R}^d \to \mathbb{R}$, denotaremos
\[D^{k_1k_2\cdots k_d}h(u)=\frac{\partial^k }{\partial u_1^{k_1} \partial u_2^{k_2}\cdots \partial u_d^{k_d}}\ h(u),\]
 $\forall  k_1,k_2,\ldots,k_d\in \mathbb{N}_0$ tal que $k=k_1+k_2+\cdots+k_d$.

Sean $\boldsymbol{X}_1=(X_{11},\ldots,X_{d1}), \boldsymbol{X}_2=(X_{12},\ldots,X_{d2}),\ldots, \boldsymbol{X}_n=(X_{1n},\ldots,X_{dn})$ vectores aleatorios independientes e id\'enticamente distribuidos (iid) definidos sobre el espacio de probabilidad $(\Omega,\mathcal{A},P)$ y que toman valores en $\mathbb{N}_0^d$. En lo que sigue, sea
\[g_n(u)=\frac{1}{n}\sum_{i=1}^n u_1^{X_{1i}}u_2^{X_{2i}}\cdots u_d^{X_{di}},\quad u\in W,\]
la fgpe de $\boldsymbol{X}_1, \boldsymbol{X}_2,\ldots,\boldsymbol{X}_n$, para alg\'un conjunto apropiado $W \subseteq \mathbb{R}^d$. De manera m\'as formal,
\[g_n(u,\omega)=\frac{1}{n}\sum_{i=1}^n u_1^{X_{1i}(\omega)}u_2^{X_{2i}(\omega)}\cdots u_d^{X_{di}(\omega)},\quad u\in W,\quad \omega\in \Omega,\]
pero la dependencia de $\omega$ es usualmente suprimida.

\begin{proposicion}\label{Conv-FuncGenProbBiv}
Sean $\boldsymbol{X}_1,\ldots,\boldsymbol{X}_n$ vectores aleatorios iid de $\boldsymbol{X}=(X_1,\ldots,X_d)\in \mathbb{N}_0^d$. Sea $g(u)=E\!\left(u_1^{X_1}u_2^{X_2}\cdots u_d^{X_d}\right)$ la fgp de $\boldsymbol{X}$, definida sobre $W \subseteq \mathbb{R}^d$. Adem\'as, sean $0\leq b_j\leq c_j<\infty,\ 1\leq j\leq d$, tal que $Q=[b_1,c_1]\times [b_2,c_2]\times\cdots\times [b_d,c_d] \subseteq W$, entonces
\begin{equation}\label{gn-converge-g}
\sup_{u\in Q} |g_n(u)-g(u)|\ \mathop{\longrightarrow}\limits^{c.s.} \ 0.
\end{equation}
Si $D^{k_1k_2\cdots k_d}g(u)$ existe en $Q$, entonces
\begin{equation}\label{Dgn-converge-Dg}
\sup_{u\in Q} \left|D^{k_1k_2\cdots k_d} g_n(u)-D^{k_1k_2\cdots k_d} g(u)\right|\ \mathop{\longrightarrow}\limits^{c.s.} \ 0.
\end{equation}
\end{proposicion}

\nt {\bf Demostraci\'on} \hspace{2pt}  Sea $D$ un conjunto denso numerable en $Q$, de acuerdo al Lema \ref{polya-d}. Por la ley fuerte de los grandes n\'umeros, existe un conjunto $A_D \in \mathcal{A}$ tal que $P(A_D)=1$ y para todo $\omega \in A_D$: \[g_n(u,\omega)\ \mathop{\longrightarrow}\limits^{c.s.} \ g(u), \forall u \in D.\]

Puesto que $X_i\geq 0\,$ y $\,u_i\geq 0$, $1\leq i \leq d$, tenemos que $g_n(u,\omega)$ y $g(u)$ son funciones no decrecientes.

Adem\'as, $g$ es una funci\'on continua definida sobre $Q$.

Ahora, (\ref{gn-converge-g}) se sigue del Lema \ref{polya-d}.

La demostraci\'on de (\ref{Dgn-converge-Dg}) sigue pasos similares a los dados para demostrar (\ref{gn-converge-g}). $\square$\\

Para probar algunos de nuestros resultados aplicaremos los lemas que establecemos a continuaci\'on, los que presentamos aqu\'i para facilitar la lectura de nuestras demostraciones.
\begin{lema}[\textnormal{M\'etodo Delta}]\label{Metodo-Delta}
(Lehmann y Romano(2005, p. 436) \cite{LehRom05})\
Supongamos que $X$ y $X_1, X_2, \ldots\,$ son vectores aleatorios en $\mathbb{R}^k$. Supongamos que $\ \displaystyle\tau_n(X_n-\mu)\ \mathop{\longrightarrow}^L \ X$ donde $\mu$ es un vector constante y $\tau_n$ es una sucesi\'on de constantes tales que $\ \tau_n\to \infty$.
\begin{enumerate}
  \item [$(a)$] Supongamos que $g$ es una funci\'on desde $\mathbb{R}^k$ a $\mathbb{R}$ que es diferenciable en $\mu$ con gradiente (vector de primeras derivadas parciales) de dimensi\'on $1\times k$ en $\mu$ igual a $\dot{g}(\mu)\neq \mathbf{0}$. Entonces, \[\tau_n\left[\,g(X_n)-g(\mu)\right] \ \mathop{\longrightarrow}^L \ \dot{g}(\mu)X^{\top}.\] En particular, si $X$ es normal multivariante en $\mathbb{R}^k$ con vector de media $\mathbf{0}$ y matriz de covarianzas $\Sigma$, entonces \[\tau_n\left[\,g(X_n)-g(\mu)\right] \ \mathop{\longrightarrow}^L\ N\left(\mathbf{0},\dot{g}(\mu)\,\Sigma\ \dot{g}(\mu)^{\top}\right).\]
  \item [$(b)$]\vskip .2 cm M\'as generalmente, supongamos que $g=(g_1,g_2,\ldots,g_q)^{\top}$ es una funci\'on desde $\mathbb{R}^k$ a $\mathbb{R}^q$, donde $g_i$ es una funci\'on desde $\mathbb{R}^k$ a $\mathbb{R}$ que es diferenciable en $\mu$. Sea $D$ una matriz no nula de orden $q\times k$, donde el elemento $(i,j)$ es $\displaystyle\frac{\partial}{\partial y_j}g_i(y_1,y_2,\ldots,y_k)$ evaluada en $\mu$. Entonces \begin{equation}\notag \tau_n\left[\,g(X_n)-g(\mu)\right]=\tau_n\left[\,g_1(X_n)-g_1(\mu), \ldots,g_q(X_n)-g_q(\mu)\right]^{\top} \ \mathop{\longrightarrow}^L \ DX^{\top}. \end{equation} En particular, si $X$ es normal multivariante en $\mathbb{R}^k$ con vector de media $\mathbf{0}$ y matriz de covarianzas $\Sigma$, entonces \begin{equation}\notag\tau_n\left[\,g(X_n)-g(\mu)\right] \ \mathop{\longrightarrow}^L\ N\left(\mathbf{0},D\Sigma D^{\top}\right). \end{equation}
\end{enumerate}
\end{lema}

Para los lemas siguientes, con $H$ denotaremos un espacio de Hilbert de dimensi\'on infinita, separable y real.

\begin{lema}\label{TCL-Hilbert} (Teorema central del l\'imite en espacios de Hilbert, van der Vaart y Wellner (1996, pp. 50--51)\cite{vanWel96}) Si $X_1, X_2, X_3, \ldots$ son elementos aleatorios iid medibles \mbox{Borel} en un espacio de Hilbert $H$ con media cero (es decir, $E\left(\langle X_1, h\rangle_{_H}\right)=0$ para cada $h$), y $E\left(\|X_1\|^2_{_H}\right)< \infty$, entonces la sucesi\'on $\displaystyle \frac{1}{\sqrt{n}}\sum_{i=1}^n X_i$ converge en distribuci\'on a la variable Gaussiana $G$. La distribuci\'on de $G$ est\'a determinada por la distribuci\'on de sus marginales $\langle G, h\rangle_{_H}$, que se distribuyen seg\'un una ley $N\left(0,E\left(\langle X, h\rangle^2_{_H}\right)\right)$ para cada $h\in H$.
\end{lema}

\begin{lema} \label{kundu} (Teorema 1.1 en Kundu et al. (2000) \cite{KuMaMu00}) Sea $\{e_k: \, k \geq 0\}$ una base ortonormal de $H$. Para cada $n\geq 1$, sea $Y_{n1}, Y_{n2}, \ldots, Y_{nn}$ un arreglo triangular de elementos aleatorios independientes $H$-valuados con medias cero y momentos segundos finitos, sea $Y_n=\displaystyle \sum_{j=1}^n Y_{nj}$. Sea $C_n$ el operador de covarianza de $Y_n$. Supongamos que se cumplen las siguientes condiciones:
\begin{itemize} \itemsep=0pt
\item[(i)] $\displaystyle \lim_{n\to \infty} \langle C_ne_k,e_l\rangle_{_{\!H}}=a_{kl}$, existe para todo $ k,l \geq 0$.
\item[(ii)] \vskip .1 cm $\displaystyle \lim_{n\to \infty}\sum_{k=0}^{\infty}\langle C_ne_k,e_k\rangle_{_{\!H}}=\sum_{k=0}^{\infty} a_{kk}<\infty  $.
\item[(iii)] \vskip .1 cm $\displaystyle \lim_{n\to \infty} L_n(\varepsilon, e_k)=0$ para cada $\varepsilon>0$ y cada $k \geq 0$, donde, para $b \in H$, \[L_n(\varepsilon, b)=\sum_{j=1}^n E\left(\langle Y_{nj}, b\rangle_{_{\!H}}^2 \,I{\left\{|\langle Y_{nj}, b\rangle_{_{\!H}}|>\varepsilon\right\}}\right).\]
\end{itemize}
Entonces \[Y_n \stackrel{L}{\longrightarrow} N(0,C),\] en $H$, donde el operador de covarianza $C$ es caracterizado por
$\langle C f,e_l\rangle_{_{\!H}}=\displaystyle \sum_{k=0}^{\infty}\langle f,e_k \rangle_{_{\!H}} a_{kl}$, para cada $l \geq 0$.
\end{lema}

\section{Definici\'on de la distribuci\'on Poisson bivariante}
Se han dado varias definiciones  para la DPB (ver por ejemplo, Kocherlakota y Kocherlakota (1992, pp. 87--90) \cite{KoKo92}, para una revisi\'on detallada). En este texto, consideraremos la siguiente, que es la que ha recibido m\'as atenci\'on en la literatura estad\'istica (ver por ejemplo, Holgate (1964) \cite{Hol64}; Johnson, Kotz y Balakrishnan (1997) \cite{JoKoBa97}).

\begin{definicion}\label{Dist_Pois_Biv}
(Johnson, Kotz y Balakrishnan (1997, pp. 124--125) \cite{JoKoBa97})  \hspace{2pt}  Sean
\[X_1=Y_1+Y_{3}\, \quad \textnormal{ y } \quad X_2=Y_2+Y_{3}\, ,\]
donde $\,Y_1,\,Y_2\,$ e $\,Y_{3}\,$ son v.a. Poisson, mutuamente independientes con medias dadas por $\theta'_1=\theta_1-\theta_{3}>0$, $\theta'_2=\theta_2-\theta_{3}>0\ $ y $\ \theta_{3}\geq0$, respectivamente.

A la distribuci\'on conjunta del vector $(X_1,  X_2)$ se le denomina {\bf distribuci\'on Poisson bivariante} (DPB) con par\'ametro $\theta=(\theta_1, \theta_2, \theta_{3})$, lo cual denotaremos mediante $(X_1, X_2)\sim PB(\theta_1, \theta_2, \theta_{3})$ o simplemente $(X_1, X_2)\sim PB(\theta)$.

La funci\'on de probabilidad conjunta de $X_1$ y $X_2$ est\'a dada por
\[P_{\theta}(X_1=x_1,X_2=x_2)=\exp(\theta_{3}-\theta_{1}-\theta_2) \sum_{i=0}^{\min\{x_{1},x_2\}} \frac{(\theta_{1}-\theta_{3})^{x_{1}-i}\,(\theta_2-\theta_{3})^{x_2-i}\, \theta_{3}^i}{(x_1-i)!\ (x_2-i)!\ i!},\]
donde $x_1, x_2\in \mathbb{N}_0$.
\end{definicion}

\begin{observacion}
Si $\theta_3=0$, entonces $X_1=Y_1$ y $X_2=Y_2$, y por tanto $X_1$ y $X_2$ son independientes, pues $Y_1$ e $Y_2$ son v.a. Poisson mutuamente independientes. En los Cap\'itulos \ref{Estadisticos-tipoCramer-von-Mises} y \ref{Estadistico-Wn} supondremos que $\theta_3>0$. El caso $\theta_3=0$ ser\'a descrito en el Cap\'itulo \ref{Extensiones}.
\end{observacion}

Tal como sucede en el caso univariante, una de las formas como se obtuvo la funci\'on de probabilidad conjunta de la DPB fue como el l\'imite de la distribuci\'on binomial bivariante.

En primer lugar daremos la definici\'on de la distribuci\'on binomial bivariante y luego presentaremos el resultado que aproxima la distribuci\'on binomial bivariante a la DPB, cuya demostraci\'on puede verse en Hamdan y Al-Bayyati (1969) \cite{HaAl69}.
\begin{definicion}\label{Dist-Binom-Biv}(Johnson, Kotz y Balakrishnan (1997, p. 125) \cite{JoKoBa97})  \hspace{2pt}
Supongamos que cada individuo de una poblaci\'on es clasificado ya sea como $A$ o $A^{c}$ y simult\'aneamente como $B$ o $B^{c}$, con probabilidades dadas por
\begin{center}
\begin{tabular}[t]{c|cc|c}
&$B$&$B^c$&\\
\hline
$A$&$p_{11}$&$p_{10}$&$p_1$\\
$A^c$&$p_{01}$&$p_{00}$&$q_1$\\
\hline &$p_2$&$q_2$&$1$
\end{tabular}
\end{center}

Consideremos una muestra aleatoria de tama\~no $n$, seleccionada con reemplazo de la poblaci\'on anterior. Sean las v.a.:

$X_1=$ n\'umero de individuos que son clasificados como $A$,

$X_2=$ n\'umero de individuos que son clasificados como $B$.

Estas v.a. tienen distribuci\'on binomial bivariante conjunta, con funci\'on de pro\-babilidad
\[P(X_1=x_1, X_2=x_2) = \sum_{k=0}^{\min\{x_1,
x_2\!\}}\frac{n! \ p_{11}^k \, p_{10}^{x_1-k}\, p_{01}^{x_2-k}\,
p_{00}^{n+k-x_1-x_2}}{k!\ (x_1-k)!\ (x_2-k)!\ (n+k-x_1-x_2)!}\,,\]
lo que se representar\'a por $(X_1,X_2)\sim BB(n; p_{10}, p_{01}, p_{11})$.
\end{definicion}

\begin{teorema}\label{Aprox-Binom-Biv}
(Hamdan y Al-Bayyati (1969) \cite{HaAl69}) \newline
Sea $(X_1,X_2)\sim BB(n; p_{10}, p_{01}, p_{11})$. Suponer que $\,p_{10}, p_{01}, p_{11}\rightarrow 0$, cuando $\,n \rightarrow \infty,\,$ de modo que $\,n p_{10}=\theta_1-\theta_3, \, n p_{01}=\theta_2-\theta_3\,$ y $\,n p_{11}=\theta_3$. Entonces
\begin{align}
\lim_{n \rightarrow \infty}P(X_1=x_1 &, X_2=x_2)\notag\\
&\!=
\exp(\theta_3-\theta_1-\theta_2)\sum_{k=0}^{\min\{x_1,
x_2\}}\frac{(\theta_1-\theta_3)^{x_1-k} (\theta_2-\theta_3)^{x_2-k}\,
\theta_3^k}{(x_1-k)!\, (x_2-k)!\, k!}, \  x_1, x_2\in \mathbb{N}_0.\notag
\end{align}
\end{teorema}

\section[Algunas caracter\'isticas y propiedades de la DPB]{Algunas caracter\'isticas y propiedades de la DPB}
A continuaci\'on presentamos ciertos resultados de la DPB que nos ser\'an de utilidad en el desarrollo de este texto. Quiz\'as el de mayor importancia es el dado a continuaci\'on, pues es la base de nuestro trabajo.

Para ello consideremos el vector aleatorio $(X_1,X_2)\sim PB(\theta)$, como el establecido en la Definici\'on \ref{Dist_Pois_Biv}.

\subsection{Funci\'on generatriz de probabilidad}
Para deducir la fgp de la DPB, recordemos que la fgp de una v.a. $X$ que se distribuye Poisson (univariante) con par\'ametro $\lambda > 0$, se define y calcula mediante
\[E_{\lambda}\!\left(t^X\right) = \sum_{x=0}^\infty t^x \frac{\lambda^x\exp(-\lambda)}{x!}
= \exp(-\lambda)\sum_{x=0}^\infty \frac{(\lambda t)^x}{x!}=
\exp\{\lambda(t-1)\},\ \forall\,
t\in \mathbb{R}.\]

Ahora, como las v.a. $Y_1, Y_2$ e $Y_3$ se distribuyen seg\'un una ley de Poisson, entonces sus fgp se pueden expresar por
\[E_{\theta'_1}\!\left(t^{Y_1}\right)=\exp\{\theta'_1(t-1)\},\ \, E_{\theta'_2}\!\left(t^{Y_2}\right)=\exp\{\theta'_2(t-1)\},\ \, E_{\theta_3}\!\left(t^{Y_3}\right)=\exp\{\theta_3(t-1)\},\, \forall\,
t\!\in\! \mathbb{R}.\]

Adem\'as, como las v.a. $Y_1, Y_2$ e $Y_3$ son mutuamente independientes, entonces, la fgp conjunta, $g(u;\theta)$, de la DPB se obtiene mediante
\begin{align}
g(u;\theta)&=E_{\theta}\!\left(u_1^{X_1}\,u_2^{X_2} \right),\notag\\[.3 cm]
&=E_{\theta}\!\left(u_1^{Y_1+Y_3}\,u_2^{Y_2+Y_3}\right)= E_{\theta'_1}\!\left(u_1^{Y_1}\right)E_{\theta'_2}\!\left(u_2^{Y_2}\right) E_{\theta_3}\!\left\{(u_1u_2)^{Y_3}\right\}\notag\\[.3 cm]
&=\exp\{\theta'_1(u_1-1)+\theta'_2(u_2-1)+ \theta_3(u_1u_2-1)\}, \notag\\[.3 cm]
&=\exp\{(\theta_1-\theta_3)(u_1-1)+(\theta_2-\theta_3)(u_2-1)+ \theta_3(u_1u_2-1)\}, \notag\\[.3 cm]
&=\exp\{\theta_1(u_1-1)+\theta_2(u_2-1)+\theta_3(u_1-1)(u_2-1) \},\label{fgp3-DPB}
\end{align}
$\forall\,u=(u_1,u_2)\in \mathbb{R}^{2}$, $\forall\,\theta\in\Theta$.

\subsection{Funci\'on generatriz de momentos}
Recordemos que la fgm de una v.a. $X$ que se distribuye seg\'un una ley de Poisson (univariante) con par\'ametro $\lambda > 0$, se define y calcula mediante
\[M_X(t) = E_{\lambda}\!\left\{\exp(tX)\right\}=\exp(-\lambda)\sum_{x=0}^\infty
\frac{\left(\lambda e^t\right)^x}{x!}= \exp\!\left\{\lambda\left(e^t-1\right)\right\},\ \forall\, t\in \mathbb{R}.\]

Puesto que las v.a. $Y_1, Y_2$ e $Y_3$ se distribuyen seg\'un una ley de Poisson, entonces sus fgm se pueden escribir como
\[M_{Y_1}(t)=\exp\!\left\{\theta'_1\!\left(e^t\!-\!1\right)\right\}\!, \, M_{Y_2}(t)=\exp\!\left\{\theta'_2\!\left(e^t\!-\!1\right)\right\}\!, \, M_{Y_3}(t)=\exp\!\left\{\theta_3\!\left(e^t\!-\!1\right)\right\}\!, \forall\,t\!\in\! \mathbb{R}.\]

Adem\'as, como las v.a. $Y_1, Y_2$ e $Y_3$ son mutuamente independientes, entonces, la fgm conjunta, se obtiene como
\begin{align}
M_{(X_1, X_2)}(u_1, u_2) &= E_{\theta}\!\left\{\exp(u_1X_1+u_2X_2)\right\}\notag\\[.3 cm]
&= E_{\theta}\!\left[\exp\!\left\{u_1Y_1+u_2Y_2+(u_1+u_2)Y_3\right\}\right]\notag\\[.3 cm]
&= E_{\theta'_1}\!\!\left(e^{u_1 Y_1}\right)E_{\theta'_2}\!\!\left(e^{u_2 Y_2}\right)E_{\theta_3}\!\!\left\{e^{(u_1+u_2)Y_3}\right\}\notag\\[.3 cm]
&=\exp\!\left\{\theta'_1(e^{u_1}-1)+\theta'_2(e^{u_2}-1)+\theta_3
(e^{u_1+\,u_2}-1)\right\},\notag\\[.3 cm]
&=\exp\!\left\{(\theta_1-\theta_3)\!\left(e^{u_1}-1\right)+(\theta_2- \theta_3)\!\left(e^{u_2}-1\right)+\theta_3\!\left(e^{u_1+u_2}-1\right)\right\}, \notag\\[.3 cm]
&=\exp\!\left\{\theta_1\!\left(e^{u_1}-1\right)+\theta_2\!\left(e^{u_2}-1\right)+ \theta_3\!\left(e^{u_1}-1\right)\!\left(e^{u_2}-1\right)\right\},\notag
\end{align}
$\forall\,u=(u_1,u_2)\in \mathbb{R}^{2}$, $\forall\,\theta\in\Theta$.

\subsection{Momentos}\label{Momentos-DPB}
Recordemos que si una v.a. $X$ se distribuye seg\'un una ley de Poisson (univariante) con par\'ametro $\lambda > 0$, entonces su $k-$\'esimo momento en torno al cero es
\[\mu_k^{'}=E_{\lambda}\!\left(X^k\right)= \sum_{i=0}^k \lambda^i\,S(k,i),\]
donde $S(a,b)$ es llamado {\bf n\'umero Stirling de segundo tipo} y satisface las relaciones
\begin{align}
S(n,0)&=0,\  S(0,0)=S(n,1)=S(n,n)=1, \ \text{para}\ n\in \mathbb{N},\notag\\[.2 cm]
S(n,j)=S&(n-1,j-1)+jS(n-1,j), \ \text{para}\
j=1,2,\ldots,n-1.\notag
\end{align}

En particular, para $k=1,2,3,$ obtenemos
\[E_{\lambda}\!\left(X\right)=\lambda,\quad E_{\lambda}\!\left(X^2\right)= \lambda+\lambda^2,\quad E_{\lambda}\!\left(X^3\right)=\lambda+3\lambda^2+ \lambda^3.\]

El correspondiente momento central de $X$ es
\[\mu_k=E_{\lambda}\!\left\{(X-\lambda)^k\right\} = \sum_{i=0}^k
\binom{k}{i}(-\lambda)^{k-i}\, \mu_i^{'}.\]

Para el modelo Poisson bivariante, el $r=(r_1,r_2)-$\'esimo momento en torno al origen es
\begin{align}
\mu_{r_1,r_2}^{'}& =\mu_r^{'}(X_1,X_2)= E_{\theta}\!\left(X_1^{r_1}X_2^{r_2}\right)= E_{\theta}\!\left\{(Y_1+Y_3)^{r_1}(Y_2+Y_3)^{r_2}\right\}\notag\\[.2 cm]
&=E_{\theta}\!\left\{\sum_{i_1=0}^{r_1} \binom{r_1}{i_1} Y_1^{i_1}\
Y_3^{r_1-i_1}\,\sum_{i_2=0}^{r_2} \binom{r_2}{i_2} Y_2^{i_2}\ Y_3^{r_2-i_2}\right\}\notag\\[.2 cm]
&=\sum_{i_1=0}^{r_1}\sum_{i_2=0}^{r_2}\binom{r_1}{i_1}\binom{r_2}{i_2}
E_{\theta'_1}\!\left(Y_1^{i_1}\right)\, E_{\theta'_2}\!\left(Y_2^{i_2}\right)\, E_{\theta_3}\!\left(Y_3^{r_1+r_2-i_1-i_2}\right)\notag\\[.2 cm]
&=\sum_{i_1=0}^{r_1}\sum_{i_2=0}^{r_2}\binom{r_1}{i_1}\binom{r_2}{i_2}
\sum_{j_1=0}^{i_1}(\theta_1-\theta_3)^{j_1}S(i_1,j_1)
\sum_{j_2=0}^{i_2}(\theta_2-\theta_3)^{j_2} S(i_2,j_2)\notag\\
&\hspace{76 mm} \times\sum_{j_3=0}^{r_1+r_2-i_1-i_2}\theta_3^{j_3}
S(r_1+r_2-i_1-i_2,j_3).\notag
\end{align}

En particular,
\begin{align}
E_{\theta}(X_k) &= \theta_{k},\ \ \text{para} \ k=1,2,\notag\\[.15 cm]
E_{\theta}(X^2_k) &= \theta_{k}+\theta_{k}^2,\ \ \text{para} \ k=1,2,\notag\\[.15 cm]
E_{\theta}(X_1\,X_2) &= \theta_{1}\,\theta_{2}+\theta_3,\notag\\[.15 cm]
E_{\theta}(X_1^2\,X_2) &= \theta_{1}\,\theta_{2}+\theta_{1}^2\,\theta_{2}+2\,\theta_{1}\,\theta_3+ \theta_3,\notag\\[.15 cm]
E_{\theta}(X_1\,X_2^2) &= \theta_{1}\,\theta_{2}+\theta_{1}\,\theta_{2}^2+2\,\theta_{2}\,\theta_3+ \theta_3,\notag\\[.15 cm]
E_{\theta}(X_1^2\,X_2^2) &= \theta_{1}\,\theta_{2}+\theta_{1}\,\theta_{2}^2+\theta_{1}^2\,\theta_{2}+ \theta_{1}^2\,\theta_{2}^2+4\,\theta_{1}\,\theta_{2}\,\theta_3+2\,\theta_{1} \,\theta_3+2\,\theta_{2}\,\theta_3+\theta_3+2\,\theta_3^2.\notag
\end{align}
\begin{observacion} De los resultados dados en las ecuaciones anteriores se sigue que la DPB tiene momentos de todos los \'ordenes.
\end{observacion}

Escribiremos el correspondiente $r=(r_1,r_2)-$\'esimo momento central para el modelo Poisson bivariante mediante
\begin{align}
\mu_{r_1,r_2} &= \mu_r(X_1,X_2)=E_{\theta}\! \left[\{X_1-E_{\theta}(X_1)\}^{r_1}\{X_2-E_{\theta}(X_2)\}^{r_2}\right] \notag\\[.15 cm] &=E_{\theta}\!\left[\{X_1-\theta_1\}^{r_1}\{X_2-\theta_2\}^{r_2}\right]\notag\\[.15 cm]
&= \sum_{i=0}^{r_1}\sum_{j=0}^{r_2}\binom{r_1}{i}\binom{r_2}{j} (-\theta_1)^{r_1-i}(-\theta_2)^{r_2-j}E_{\theta}\!\left(X_1^iX_2^j\right) \notag\\[.15 cm]
&=  \sum_{i=0}^{r_1}\sum_{j=0}^{r_2}\binom{r_1}{i}\binom{r_2}{j} (-\theta_1)^{r_1-i}(-\theta_2)^{r_2-j} \mu_{i,j}^{'}.\notag
\end{align}

Por otra parte, como las v.a. $X_1$ y $X_2$ se distribuyen seg\'un una ley de Poisson (univariante), entonces sus varianzas son
\[\text{var}(X_k) = \theta_k,\ k=1, 2.\]

Adem\'as, como $Y_1$, $Y_2$ e $Y_3$ son variables independientes, la covarianza entre $X_1$ y $X_{2}$ es
\begin{align}
\text{cov}\left(X_1,X_2\right) &= \text{cov}\left(Y_1+Y_3,
Y_2+Y_3\right)\notag\\[.2 cm]
&= \text{cov}\left(Y_1, Y_2\right)+\text{cov}\left(Y_1, Y_3\right)+\text{cov}\left(Y_3, Y_2\right)+\text{cov}\left(Y_3, Y_3\right)\notag\\[.2 cm]
&=\text{var}\left(Y_3\right)=\theta_3\,.\notag
\end{align}

Por lo tanto, el coeficiente de correlaci\'on entre $X_1$ y
$X_2$ es
\[\rho=\text{corr}\left(X_1,X_2\right)=
\frac{\text{cov}\left(X_1,X_2\right)}{\sqrt{\text{var}(X_1)\ \text{var}(X_2)}}=\frac{\theta_3}{\sqrt{\theta_1\,
\theta_2}}\,.\]

Este coeficiente de correlaci\'on no puede exceder de
$\theta_3\left(\min\!\left\{\theta_1,\theta_2\right\}\right)^{-1}$, porque
\[\min\left\{\theta_1,\theta_2\right\}\leq \theta_k,\ k=1,2, \text{ luego  } \theta_1\,\theta_2\geq \left(\min\left\{\theta_1,\theta_2\right\}\right)^2.\]

O bien, como lo se\~nal\'o Holgate (1964) \cite{Hol64},
\[0<\rho\leq \min\!\left(\sqrt{\frac{\theta_1}{\theta_2}},
\sqrt{\frac{\theta_2}{\theta_1}}\,\right),\]
pues, de la Definici\'on \ref{Dist_Pois_Biv}
\[\sqrt{\frac{\theta_1}{\theta_2}}=\sqrt{\frac{\theta_1}{\theta_2} \frac{\theta_1}{\theta_1}}=\frac{\theta_1}
{\sqrt{\theta_1 \theta_2}}=\frac{\theta'_1+\theta_3}
{\sqrt{\theta_1 \theta_2}}\geq \rho,\]
de igual forma se consigue que $\rho\leq \sqrt{\frac{\theta_2}{\theta_1}}$.

\subsection{F\'ormulas de recursi\'on}
Para calcular num\'ericamente los valores de
$P(r, s;\theta)=P_{\theta}(X_1=r, X_2=s)$, son \'utiles las
relaciones de recurrencia presentadas por Johnson, Kotz y Balakrishnan (1997, p. 125) \cite{JoKoBa97} que se enuncian como sigue
\begin{proposicion} (Johnson et al. (1997, p. 125) \cite{JoKoBa97}) Si $(X_1, X_2)\sim PB(\theta_{1}, \theta_{2}, \theta_3)$, entonces
\begin{equation}\label{form-rec-DPB}
\begin{array}{l}
rP(r, s;\theta)= (\theta_1-\theta_{3}) P(r-1, s;\theta)+\theta_{3}\, P(r-1, s-1;\theta),\\[.3 cm]
sP(r, s;\theta) = (\theta_2-\theta_{3}) \,P(r, s-1;\theta)+\theta_{3}\, P(r-1, s-1;\theta),\\[.3 cm]
\text{Si} \ r<0\ \ \ \text{o} \ \ s<0,\ \text{entonces considerar} \ P(r, s;\theta)=0.
\end{array}
\end{equation}
\end{proposicion}

Adem\'as de las relaciones dadas en la Proposici\'on anterior, tambi\'en nos ser\'an de gran utilidad las relaciones de recurrencia dadas en Johnson, Kotz y Balakrishnan (1997, p. 127) \cite{JoKoBa97}, que se enuncian como sigue.
\begin{proposicion} Sea $(X_1, X_2)\sim PB(\theta_{1}, \theta_{2}, \theta_3)$, entonces
\begin{align}
\frac{\partial}{\partial\theta_1}P(r, s;\theta) &= P(r-1, s;\theta)- P(r, s;\theta),\notag \\[.1 cm]
\frac{\partial} {\partial\theta_2}P(r, s;\theta) &=P(r, s-1;\theta)-P(r, s;\theta)\label{form-rec-derDPB} \\[.1 cm]
\frac{\partial}{\partial\theta_3}P(r, s;\theta) &= P(r-1, s-1;\theta)-P(r-1, s;\theta)-P(r, s-1;\theta)+P(r, s;\theta).\notag
\end{align}
\end{proposicion}

\subsection{Estimaci\'on puntual}\label{Estimacion-puntual}
Los estimadores m\'as usados com\'unmente son los obtenidos por los siguientes m\'etodos:
 \begin{itemize}
   \item M\'etodo de m\'axima verosimilitud (Kocherlakota y Kocherlakota (1992, pp. 43-45) \cite{KoKo92}).

{\bf Propiedades} Seg\'un Kocherlakota y Kocherlakota (1992, p. 45) \cite{KoKo92}, el estimador de m\'axima verosimilitud, $\hat{\theta}_{MV}$, es consistente, asint\'oticamente normal y asint\'oticamente eficiente para $\theta$.
\begin{itemize}
  \item El estimador m\'aximo veros\'imil de $\theta$ debe satisfacer
\[\hat{\theta}_k=\bar{X}_{k}=\frac{1}{n}\sum_{i=1}^n X_{ki}\, ,\ \ k=1,2\quad \text{y} \quad \bar{R}=1,\]
donde
\[\bar{R}=\frac{1}{n}\mathop{\sum}\limits_{i=1}^n\frac{P(X_{1i}-1,
X_{2i}-1;\theta)}{P(X_{1i}, X_{2i};\theta)}\,.\]

\begin{observacion}
El estimador m\'aximo veros\'imil del par\'ametro $\theta_3$ no tiene una forma expl\'icita y por lo tanto debe ser calculado por m\'etodos num\'ericos utilizando la ecuaci\'on $\,\bar{R}=1$. En nuestro caso, para calcular el estimador $\hat{\theta}_3$ empleamos el m\'etodo iterativo de Newton-Raphson est\'andar.
\end{observacion}

  La matriz de varianzas y covarianzas asint\'otica de los estimadores m\'aximo veros\'{\i}miles, $\Sigma_{MV}$, est\'a dada por
\[\Sigma_{MV}=\left(\begin{array}{ccc}
\theta_1 & \theta_3 & \theta_3 \\[.2 cm]
\theta_3 & \theta_{2} & \theta_3 \\[.2 cm]
\theta_3 & \theta_3 & \displaystyle\frac{\theta^2_3\left(\theta_1+\theta_2- 2\theta_3\right)\left(Q-1\right)-\theta_3^2+(\theta_1-2\theta_3) (\theta_2-2\theta_3)}{(\theta_{1}\theta_{2}-\theta_3^2)(Q-1)- \theta_{1}-\theta_2+2\theta_3}
\end{array}\right),\]
donde
\begin{equation}\label{Q}
Q=\sum_{r=1}^\infty\,\sum_{s=1}^\infty\frac{P^2(r-1,s-1;\theta)} {P(r,s;\theta)} \,.
\end{equation}

  \item Por las propiedades de los estimadores m\'aximo veros\'imiles, se tiene \[\sqrt{n}\left(\hat{\theta}_{MV}-\theta\right)\ \mathop{\longrightarrow}^L \ N_3(\mathbf{0},\Sigma_{MV}).\]
\end{itemize}

   \item \vskip .1 cm M\'etodo de los momentos (Kocherlakota y Kocherlakota (1992, pp. 34-35) \cite{KoKo92}).
\begin{itemize}
  \item El estimador de $\theta$ por el m\'etodo de los momentos, $\hat{\theta}_{MM}$, es consistente y est\'a dado por
\[\hat{\theta}_k=\bar{X}_k, \, k=1,2 \ \ \ \text{y}\ \ \ \hat{\theta}_3=s_{11}=\frac{1}{n}\sum_{i=1}^n X_{1i}X_{2i}-\bar{X}_1\bar{X}_2.\]
Por otro lado, la matriz de varianzas y covarianzas asint\'otica de los estimadores por el m\'etodo de los momentos, $\Sigma_{MM}$, est\'a dada por
\[\Sigma_{MM} =\left(\begin{array}{ccc}
\theta_1 & \theta_3 & \theta_3 \\[.2 cm]
\theta_3 & \theta_{2} & \theta_3 \\[.2 cm]
\theta_3 & \theta_3 & \theta_{1}\theta_{2}+\theta_3+\theta_3^2
\end{array}\right).\]

  \item Por ser $\hat{\theta}_{MM}$ funci\'on de momentos muestrales, aplicando el M\'etodo Delta (Teorema \ref{Metodo-Delta}) se tiene que
      \[\sqrt{n}\left(\hat{\theta}_{MM}-\theta\right)\ \mathop{\longrightarrow}^L \ N_3(\mathbf{0},\Sigma_{MM}).\]
\end{itemize}
\item M\'etodo del doble cero (Kocherlakota y Kocherlakota (1992, pp. 42-43) \cite{KoKo92}).
\begin{itemize}
  \item El estimador de $\theta$ por el m\'etodo del doble cero, $\hat{\theta}_{DC}$, es consistente y est\'a dado por
\[\hat{\theta}_k=\bar{X}_k, \ k=1,2 \ \ \ \text{y}\ \ \ \hat{\theta}_3=\bar{X}_1+\bar{X}_2+\log \hat{\phi},\]
donde $\phi=\exp(\theta_3-\theta_1-\theta_2)$ y $\hat{\phi}$ es la proporci\'on observada de los datos $(0,0)$ en la muestra.

La matriz de varianzas y covarianzas asint\'otica de los estimadores por el m\'etodo del doble cero, $\Sigma_{DC}$, est\'a dada por
\[\Sigma_{DC}=
\left(\begin{array}{ccc}
\theta_1 & \theta_3 & -\theta_1\,\phi \\
\theta_3 & \theta_2 & -\theta_2\,\phi \\
-\theta_1\,\phi & -\theta_2\,\phi & \phi\left(1-\phi\right) \end{array}
\right).\]

  \item Por ser $\hat{\theta}_{DC}$ funci\'on de momentos muestrales, aplicando el M\'etodo Delta (Teorema \ref{Metodo-Delta}) se tiene que
      \[\sqrt{n}\left(\hat{\theta}_{DC}-\theta\right)\ \mathop{\longrightarrow}^L \ N_3(\mathbf{0},\Sigma_{DC}).\]
\end{itemize}
   \item M\'etodo de los puntos pares (Kocherlakota y Kocherlakota (1992, pp. 40-41) \cite{KoKo92}).
\begin{itemize}
  \item El estimador de $\theta$ por el m\'etodo de los puntos pares, $\hat{\theta}_{PP}$, es consistente y est\'a dado por
\[\hat{\theta}_1=\bar{X}_1, \ \ \hat{\theta}_2=\bar{X}_2 \ \ \ \text{y} \ \ \ \hat{\theta}_3=\frac{1}{2}\,(\bar{X}_1+\bar{X}_2)+\frac{1}{4}\log \! \left(\frac{2\hat{A}}{n}-1\right), \ \text{si}\ \hat{A}>\frac{n}{2}\,,\]
donde $2A=n\left[1+\exp\{-2(\theta_1+\theta_2-2\theta_3)\}\right]$ y $\hat{A}$ es la suma de las frecuencias de ocurrencia de los pares $(X_{1i}, X_{2i})$, $i=1,2,\ldots,n$ para los cuales el valor de ambas variables tiene la misma paridad, es decir, en ambas variables el valor es par o en ambas variables el valor es impar.

La matriz de varianzas y covarianzas asint\'otica de los estimadores por el m\'etodo de los puntos pares, $\Sigma_{PP}$, est\'a dada por
\[\Sigma_{PP}=
\left(\begin{array}{ccc}
\theta_1 & \theta_3 & \theta_3 \\[.2 cm]
\theta_3 & \theta_2 & \theta_3 \\[.2 cm]
\theta_3 & \theta_3 &  \frac{1}{4}(6\theta_3-\theta_1-\theta_2)+\frac{1}{16}\!\left[\exp\{4(\theta_1+ \theta_2-2\theta_3)\}-1\right] \\
  \end{array}
\right).\]

  \item Por ser $\hat{\theta}_{PP}$ funci\'on de momentos muestrales, aplicando el M\'etodo Delta (Teorema \ref{Metodo-Delta}) se tiene que
      \[\sqrt{n}\left(\hat{\theta}_{PP}-\theta\right)\ \mathop{\longrightarrow}^L \ N_3(\mathbf{0},\Sigma_{PP}).\]
\end{itemize}
   \item M\'etodo de los puntos pares condicionados (Papageorgiou y Loukas (1988) \cite{PaLo88}).
\begin{itemize}
  \item El estimador de $\theta$ por el m\'etodo de los puntos pares condicionados, $\hat{\theta}_{PC}$, es consistente y est\'a dado por
\[\hat{\theta}_1=\bar{X}_1, \ \ \hat{\theta}_2=\bar{X}_2 \ \ \ \text{y} \ \ \ \hat{\theta}_3=\bar{X}_2+\frac{1}{2}\log \! \left(\frac{2\hat{A}}{\hat{A}+\hat{B}}-1\right), \ \text{si}\ \hat{A}>\hat{B}\,,\]
donde $2A=B\left[1+\exp\{2(\theta_3-\theta_2)\}\right]$, $A$ y $B$ son las sumas de las frecuencias observadas en los puntos $(0,2s)$ y $(0,2s+1)$, $s\in \mathbb{N}_0$, respectivamente.

La matriz de varianzas y covarianzas asint\'otica de los estimadores por el m\'etodo de los puntos pares condicionados, $\Sigma_{PC}$, est\'a dada por
\[\Sigma_{PC}=\frac{1}{4}
\left(  \begin{array}{ccc}
    C & D & E \\
    D & F & F-G+H \\
    E & F-G+H & F+2(H-G)+J \\
  \end{array}
\right),\]
donde
\begin{equation}\label{Notacion-PPC}
\begin{array}{ll}
\alpha=\exp(-\theta_1), & \beta=\exp\{2(\theta_3-\theta_2)\},\\[.3 cm]
C=\alpha(\beta+1)\{2+\alpha(\beta+1)\}, & D=\alpha^2\left(\beta^2-1\right),\\[.3 cm] E=\displaystyle\frac{D(\theta_1\!+\!\alpha^2\beta)\!+\!2\theta_3\alpha^2 (\beta\!+\!1)\!-\!\alpha^2\beta H}{\alpha^2 \beta}, & F=\alpha(1-\beta)\{2\!+\!\alpha(\beta-1)\}, \\[.3 cm] G=\beta^{-1}\theta_1(\beta-1)^2, & H=(\beta+1)\{\theta_3(\beta^{-1}+1)-2\theta_2\}, \\[.3 cm] J=\displaystyle\frac{\theta_1(\beta\!-\!1)^2\!+\!2\theta_3(\beta^2\!-\!1) \!+\!\theta_2(\beta\!+\!1)^2}{\alpha^2 \beta^2} .
\end{array}
\end{equation}

  \item Por ser $\hat{\theta}_{PC}$ funci\'on de momentos muestrales, aplicando el M\'etodo Delta (Teorema \ref{Metodo-Delta}) se tiene que
      \[\sqrt{n}\left(\hat{\theta}_{PC}-\theta\right)\ \mathop{\longrightarrow}^L \ N_3(\mathbf{0},\Sigma_{PC}).\]
\end{itemize}
 \end{itemize}

{\bf Eficiencia relativa asint\'otica}

La eficiencia relativa asint\'otica es la raz\'on entre la varianza generalizada (determinante de la matriz de varianzas y covarianzas) de los estimadores m\'aximo veros\'imiles y la del estimador bajo consideraci\'on. Las expresiones de las varianzas generalizadas de los estimadores discutidos anteriormente son:
\begin{align}
|\Sigma_{MV}|&=\displaystyle\frac{(\theta_1-\theta_3)^2(\theta_{2}-\theta_3)^2} {(\theta_1\theta_2-\theta_3^2)(Q-1)-\theta_1-\theta_2+2\theta_3}, \notag\\[.25 cm]
  |\Sigma_{MM}|&=\theta_1^2\theta_2^2+\theta_1\theta_2\theta_3- (\theta_{1}+ \theta_2)\theta_3^2+\theta_3^3-\theta_3^4, \notag\\[.25 cm]
  |\Sigma_{DC}|&=\left(\theta_1\theta_2-\theta_3^2\right) \{\exp(\theta_1+\theta_2-\theta_3)-1\}-\theta_{1}\theta_{2}(\theta_{1}+ \theta_{2}-2\,\theta_3), \notag\\[.25 cm]
  |\Sigma_{PP}|&=\frac{1}{16}\Bigl[\left(\theta_1\theta_2- \!\theta_3^2\right)\left[\exp\{4(\theta_1+\theta_2-2\theta_3)\}- 1\right] \Bigr.\notag\\ &\Bigl.\hspace{15mm}+4\left\{2\theta_3 (\theta_{1}-\theta_{3})(\theta_{2}-\theta_3)- \theta_{1}(\theta_{2}-\theta_3)^2-\theta_{2}(\theta_{1}- \!\theta_3)^2\right\}\Bigr], \notag\\[.25 cm]
  |\Sigma_{PC}|&=\frac{1}{64}\left[F\{CJ\!+\!E(2D\!-\!E)\}-(G\!-\!H) \{2D(E\!-\!D)\!+\!C(G\!-\!H)\}-D^2(F\!+\!J)\right],\notag
\end{align}
donde $Q$ est\'a definido en (\ref{Q}), y $C, D, E, F, G, H$ y $J$ est\'an definidas en (\ref{Notacion-PPC}).\vskip .3 cm

Un an\'alisis m\'as detallado sobre la eficiencia relativa asint\'otica del estimador de $\theta$ se puede encontrar en los trabajos desarrollados por los investigadores que han tratado el tema, ver por ejemplo, Holgate (1964) \cite{Hol64}, Loukas et al. (1986) \cite{LoKePa86}, Papageorgiou y Loukas (1988) \cite{PaLo88}, Paul y Ho (1989) \cite{PaHo89}, y Kocherlakota y Kocherlakota (1992) \cite{KoKo92}. Cabe se\~nalar que la eficiencia relativa asint\'otica para el estimador de $\theta$ por el m\'etodo de los puntos pares condicionados est\'a solamente tratado en Papageorgiou y Loukas (1988) \cite{PaLo88}.

\section{Bondad de Ajuste}
Un aspecto crucial en cualquier an\'alisis de datos es contrastar la bondad de ajuste de las observaciones con el modelo probabil\'istico supuesto, es decir, se contrasta si los datos provienen de la poblaci\'on que se supone.

Dadas las observaciones $X_1, X_2, \ldots, X_n$ iid, con distribuci\'on $F$, el objetivo es contrastar la hip\'otesis nula $H_0: ``F=F_0"$. La hip\'otesis alternativa ser\'a $H_1: ``F\neq F_0"$, $F_0$ puede ser una distribuci\'on totalmente especificada, o bien, especificada salvo por un n\'umero finito de par\'ametros.

\subsection{Tests de bondad de ajuste en dimensi\'on uno}\label{Test-bondad-ajuste-dim-1}
En este apartado nuestro inter\'es es contrastar
\begin{equation}\label{H0-DPUnivariante}
\begin{array}{l}
H_0 : X\sim P(\vartheta), \ \text{para alg\'un}\ \vartheta>0,\\[.2 cm]
 H_1 : X \ \text{no se distribuye}\ P(\vartheta), \ \forall\,\vartheta>0.
\end{array}
\end{equation}

Sean $X_1, X_2,\ldots, X_n$ observaciones independientes de una v.a. $X$ que toma valores enteros no negativos. Sean
\[F(k\,;\vartheta)=\exp(-\vartheta)\sum_{j=0}^k\frac{\vartheta^j}{j!}, \ k\in \mathbb{N}_0,\]
la funci\'on de distribuci\'on de la distribuci\'on Poisson $P(\vartheta)$ y $F$ la funci\'on de distribuci\'on desconocida de $X$. Adem\'as, $\hat{\vartheta}_n=\hat{\vartheta}_n(X_1, X_2,\ldots, X_n)=\bar{X}_n$ denota el estimador m\'aximo veros\'imil de $\vartheta$.

Como es bien sabido (ver por ejemplo, G\"urtler y Henze (2000) \cite{GuHe00}), el estad\'istico $\chi^2$, que es una herramienta cl\'asica para contrastar la bondad de ajuste, presenta dos inconvenientes al tratar de comprobar la hip\'otesis que los datos se distribuyen Poisson: $(a)$ selecci\'on de celdas y $(b)$ estimaciones del par\'ametro $\vartheta$ para datos agrupados.\vskip .2 cm

G\"urtler y Henze (2000) \cite{GuHe00} presentan una variedad de procedimientos para contrastar (\ref{H0-DPUnivariante}). La distribuci\'on nula de los estad\'isticos considerados dependen del verdadero y desconocido valor del par\'ametro. Ni siquiera su distribuci\'on asint\'otica es libre, es decir, tambi\'en depende del par\'ametro. Por ello, estos autores emplean un bootstrap param\'etrico para estimar los valores cr\'iticos o los $p-$valores.

En las siguientes subsecciones estudiaremos algunos de los contrastes presentados por G\"urtler y Henze (2000) \cite{GuHe00}.

\subsubsection[\'Indice de dispersi\'on de Fisher. Estad\'istico $\hat{U}^{\,2}_{n2}$]{\'Indice de dispersi\'on de Fisher. Estad\'istico $\boldsymbol{\hat{U}^{\,2}_{n2}}$}

\nt {\bf \'Indice de dispersi\'on de Fisher}

Puesto que la media y la varianza de la distribuci\'on Poisson son iguales, entonces el cociente entre sus estimadores deber\'ia estar cercano a 1. Espec\'ificamente,
\begin{equation}\label{ecuac2.27-1}
\frac{\hat{\sigma}^2}{\hat{\vartheta}}=\frac{1}{n} \frac{\sum_{j=1}^n(X_j-\bar{X}_n)^2}{\bar{X}_n}\approx 1.
\end{equation}

Un procedimiento para contrastar el test estad\'istico dado en (\ref{H0-DPUnivariante}) es rechazar $H_0$ para valores peque\~nos o grandes del \'indice de dispersi\'on de Fisher, dado por
\[D_n=\sum_{j=1}^n\frac{(X_j-\bar{X}_n)^2}{\bar{X}_n}.\]

\nt {\bf Estad\'istico $\boldsymbol{\hat{U}^2_{n2}}$}

De (\ref{ecuac2.27-1}) y del \'indice de dispersi\'on de Fisher, resulta
$$D_n\approx n\quad \Longleftrightarrow\quad (D_n-n)^2\approx 0.$$
Por lo tanto, para contrastar las hip\'otesis en (\ref{H0-DPUnivariante}), se rechaza $H_0$ para valores ``grandes" \, del estad\'istico
\[\hat{U}^{\,2}_{n2}=\left\{\frac{1}{\sqrt{2n}}(D_n-n)\right\}^2.\]

Como
\[\hat{U}_{n2}^{\,2}=\frac{1}{2n}{(D_n-n)^2}\ \mathop{\longrightarrow}\limits_{H_0}^{L}\ \chi^2_{1},\]
(ve\'ase por ejemplo Rayner, Thas y Best (2009, p. 156) \cite{RaThBe09}).

As\'i, se rechaza $\,H_0: X\sim P(\vartheta)\ $ si $\ P_{H_0}\!\left(\hat{U}_{n2}^{\,2}>c\right)=\alpha,\ $ donde $\ c=\chi^2_{1,1-\alpha}$. Por lo tanto, se rechaza $\,H_0: X\sim P(\vartheta)\ $ si
$$\frac{1}{2n}\left(\sum_{j=1}^n\frac{(X_j- \bar{X}_n)^2}{\bar{X}_n}-n \right)^{\!\!2}\,>\,\chi^2_{1,1-\alpha}\,.$$

\begin{observacion}
El test basado en el estad\'istico $\hat{U}_{n2}^{\,2}$ no es un test consistente pues est\'a construido en base a momentos, espec\'ificamente se centra en la propiedad que la media y la varianza para la distribuci\'on Poisson son iguales.
\end{observacion}

\subsubsection{Contrastes basados en la funci\'on generatriz de probabilidad emp\'irica}\label{Contrastes-basados-en-fgpe}

\nt {\bf Estad\'istico} $\boldsymbol{R_{n,a}}$

\vskip .1 cm

Puesto que la distribuci\'on de $X$ est\'a caracterizada por su fgp, obtenemos que si $X$ se distribuye seg\'un una ley de Poisson, entonces su fgp est\'a dada por
$$g(t;\vartheta)=E_\vartheta\!\left(t^X\right)=\exp\{\vartheta(t-1)\},$$
la cual puede ser estimada por su fgpe
\[g_n(t)=\frac{1}{n}\sum_{j=1}^n t^{X_j} \ \ \text{de}\ \ X_1, X_2,\ldots, X_n.\]

\vskip .1 cm

Por tanto, parece natural basar un contraste de $H_0$ sobre
\begin{equation}\label{ecuac2.29}
G_n(t)=\sqrt{n}\left(g_n(t)-g(t\,;\hat{\vartheta}_n)\right),\ 0\leq t\leq 1.
\end{equation}

\vskip .2 cm

N\'otese que, bajo $H_0$, $G_n(t)$ es la diferencia de dos estimadores consistentes de $g(t;\vartheta)$. Por lo que un test razonable para contrastes deber\'ia rechazar $H_0$ para valores ``grandes" \,de $R_{n,a}$ definido por
\[R_{n,a}=\int_0^1 G^2_n(t) t^a dt,\]
donde $\,a\geq 0\,$ es una constante. Elegir un valor grande de $\,a\,$ significa colocar m\'as peso cerca del punto extremo $\,t=1$.

\begin{observacion}
Para $a=0$, se obtiene el estad\'istico sugerido por Rueda et al. (1991) \cite{RuPeOR91} y el caso general, cuando $a\geq 0$, fue propuesto por Baringhaus et al. (2000) \cite{BaGuHe00}.
\end{observacion}

\vskip .4 cm

\nt {\bf Estad\'istico} $\boldsymbol{T_{n,a}}$

\vskip .1 cm

Es otro estad\'istico para contrastar $H_0$, el cual est\'a basado en el hecho que $\,g(t;\vartheta)\,$ es la \'unica fgp que satisface la ecuaci\'on diferencial
\[\vartheta\, g(t)-g'(t)=0,\]
donde $g(t)$ denota la fgp de la variable aleatoria $X$.

\vskip .2 cm

Si $H_0$ es cierta, entonces
\[\bar{X}_n\, g_n(t)-g'_n(t)\approx 0,\ \forall t.\]

Por lo que un test razonable para contrastar la hip\'otesis nula deber\'ia rechazar $H_0$ para valores ``grandes" \, de $\,T_{n,a}\,$ definido por
\[T_{n,a}=n\int_0^1\Bigl\{\bar{X}_n\, g_n(t)-g'_n(t)\Bigr\}^{\!2} t^a\, dt,\]
donde $\,a\geq 0\,$ es una constante. Elegir un valor grande de $\,a\,$ significa colocar m\'as peso cerca del punto extremo $\,t=1$.

\vskip .3 cm

La distribuci\'on de los estad\'isticos $\,R_{n,a}\,$ y $\,T_{n,a}\,$ es desconocida para tama\~nos de muestra finito. Un modo cl\'asico de aproximar la distribuci\'on nula es mediante la distribuci\'on asint\'otica nula.

\vskip .3 cm

La convergencia d\'ebil de los estad\'isticos $\,R_{n,a}\,$ y $\,T_{n,a}\,$ est\'a dada en el resultado siguiente, cuya prueba sigue pasos similares a los dados en la demostraci\'on del Teorema 2.1 en Baringhaus y Henze (1992) \cite{BaHe92}.

\begin{teorema}\label{TeoConvDebilRnaTna}
(G\"urtler y Henze (2000) \cite{GuHe00})\ Sea $\{X_{n,1},X_{n,2},\ldots,X_{n,n}\},\ n\geq 1$, un arreglo triangular de v.a. iid en cada fila, tales que $X_{n,1}\sim P(\vartheta_n)$, donde $0<\vartheta=\mathop{\lim}\limits_{n\rightarrow \infty} \vartheta_n$ existe. Sean $\,G_n(t)$ definido como en (\ref{ecuac2.29}) y
$$\tilde{G}_n(t)=\sqrt{n}\left(\bar{X}_n\,g_n(t)-g'_n(t)\right),\, 0\leq t\leq 1,$$
donde
$$g_n(t)=\frac{1}{n}\sum_{j=1}^n t^{X_{n,j}},\quad g(t\,;\hat{\vartheta}_n)=\exp\!\left\{\hat{\vartheta}_n(t-1)\right\}$$
$$\hat{\vartheta}_n=\hat{\vartheta}_n(X_{n,1},X_{n,2},\ldots,X_{n,n})= \bar{X}_n=\frac{1}{n}\sum_{j=1}^n X_{n,j}.$$
Entonces se cumple lo siguiente:
\begin{itemize}
  \item [$(i)$] $\displaystyle R_{n,a}\mathop{\longrightarrow}\limits^{L} \int_0^1 Z^2(t)\,t^a dt,\ a\geq0$, donde $Z(\cdot)$ es un proceso Gaussiano centrado con n\'ucleo de covarianza
      $$K(u,v)=\exp\{\vartheta(uv-1)\}-\{1+\vartheta(u-1)(v-1)\}\exp\{\vartheta(u+v-2)\}, \ 0\leq u,v\leq 1.$$
 \item [$(ii)$] $\displaystyle T_{n,a}\mathop{\longrightarrow}\limits^{L} \int_0^1 \tilde{Z}^2(t)\,t^a dt,\ a\geq0$, donde $\tilde{Z}(\cdot)$ es un proceso Gaussiano centrado con n\'ucleo de covarianza
      $$\tilde{K}(u,v)=\vartheta\left\{1+\vartheta(1-u)(1-v)\right\} \,\exp\{\vartheta(uv-1)\} -\vartheta \exp\{\vartheta(u+v-2)\},$$
       donde $\,0\leq u,v\leq 1$.
\end{itemize}
\end{teorema}

Adem\'as, G\"urtler y Henze (2000) \cite{GuHe00} demuestran que los test basados en los estad\'isticos $R_{n,a}\,$ y $\, T_{n,a}$ son consistentes frente a alternativas fijas.\\

Un modo alternativo de aproximar la distribuci\'on nula de los estad\'isticos antes citados es usar un estimador bootstrap param\'etrico, que se describe como sigue:\vskip .25 cm

Denotemos por $W_n$ a uno cualquiera de los estad\'isticos anteriores ($R_{n,a}\,$ o $\, T_{n,a}$).  Sea $\,H_{n,\vartheta}(t)=P_{\vartheta}(W_n\leq t)\,$ la funci\'on de distribuci\'on de la distribuci\'on nula de $W_n$ cuando $\vartheta$ es el verdadero valor del par\'ametro. El m\'etodo bootstrap estima el cuantil de $\,H_{n,\vartheta}(t)$ mediante el $(1-\alpha)$-cuantil de $H_{n,\,\hat{\vartheta}_n}$, que se aproxima mediante simulaci\'on, empleando los siguientes pasos:
\begin{itemize}
\item [$\bullet$] Generar $B$ seudo-muestras aleatorias de tama\~no $n$ con distribuci\'on $P(\hat{\vartheta}_n)$, es decir, generar $X^*_{j1},X^*_{j2}, \ldots, X^*_{jn}, j\!=\!1,2,\ldots, B$ iid de acuerdo a $P(\hat{\vartheta}_n)$.

\item [$\bullet$] \vskip .2 cm Calcular $\,W^*_{j,n}=W_n(X^*_{j1}, X^*_{j2},\ldots, X^*_{jn})$, para $j=1,2 ,\ldots, B$.

\item [$\bullet$] \vskip .2 cm Sea $\,H^*_{n,B}(t)=\displaystyle\frac{1}{B}\sum_{j=1}^B I{\{W^*_{j,n}\leq t\}}\,$ para la funci\'on de distribuci\'on emp\'irica de $W^*_{1,n}$, $W^*_{2,n},\ldots, W^*_{B,n}\,$ y $\,W^*_{1:B}\leq W^*_{2:B}\leq\ldots\leq W^*_{B:B}\,$ para sus estad\'isticos de orden.

\vskip .2 cm

Finalmente, el valor cr\'itico, $c^*_{n,B}$, est\'a dado por
\[c^*_{n,B}=
\left\{\begin{array}{ll}
W^*_{B(1-\alpha):B}\,, & \text{si}\ B(1-\alpha)\ \text{es un entero,} \\ [.3 cm]
W^*_{[B(1-\alpha)]+1:B}\,, & \text{en otro caso,}
\end{array}
\right.\]
donde $[y]$ es la parte entera de $y$.
\end{itemize}

\nt {\bf Estad\'istico} $\boldsymbol{V_n}$

Motivados por el hecho que $\,\frac{\partial^2}{\partial t^2}\log g(t;\vartheta)\equiv 0$, Nakamura y P\'erez-Abreu (1993) \cite{NaPe93} propusieron la suma de los cuadrados de los coeficientes del polinomio $\,g^2_n(t)\,\frac{\partial^2}{\partial t^2}\log g_n(t)$ como un estad\'istico para contrastar (\ref{H0-DPUnivariante}).

Como 
\begin{equation}\label{Der2-log(gn)}
\frac{\partial^2}{\partial t^2}\log g_n(t)=\frac{g_n(t) \displaystyle\frac{\partial^2}{\partial t^2} g_n(t)-\left\{\frac{\partial}{\partial t} g_n(t)\right\}^2}{g^2_n(t)},
\end{equation}
llamando $N_n(t)$ al numerador y puesto que $g_n(t)=\displaystyle\sum_{i=1}^n t^{X_i}$, obtenemos
\begin{align}
N_n(t)&=g_n(t) \frac{\partial^2}{\partial t^2} g_n(t)-\left\{\frac{\partial}{\partial t} g_n(t)\right\}^2\notag\\
&=\frac{1}{n^2}\sum_{i,j=1}^n X_i(X_i-X_j-1)t^{X_i+X_j-2}.\notag
\end{align}

Debido a que $g^2_n(t)>0, \forall t$, entonces de (\ref{Der2-log(gn)})
\[\frac{\partial^2}{\partial t^2}\log g_n(t)\equiv 0\ \Longleftrightarrow \ N_n(t)=0, \ \forall t.\]

Haciendo $X_{(n)}=\max\{X_1, X_2,\ldots, X_n\}$ y notando que $N_n(t)$ es un polinomio aleatorio en $t$ de grado $d_n=2X_{(n)}-2$, obtenemos \[N_n(t)=\sum_{k=0}^{d_n} a_k t^k,\] donde
\[a_k=\frac{1}{n^2}\sum_{i,j=1}^n X_i(X_i-X_j-1)I{\{X_i+X_j-2=k\}},\ 0\leq k \leq d_n.\]

Por lo tanto
\[N_n(t)=0, \ \forall t\ \Longleftrightarrow \ a_k=0,\ 0\leq k \leq d_n.\]

As\'i, un estad\'istico para contrastar $H_0$ es considerar la suma de los cuadrados de los coeficientes polinomiales, esto es,
\begin{align}
V_n&=\sum_{k=0}^{d_n} a_k^2\notag\\
&=\frac{1}{n^4}\sum_{i,j,k,l=1}^n X_i(X_i-X_j-1)X_k(X_k-X_l-1)I{\{X_i+X_j=X_k+X_l\}},\notag
\end{align}
que expresa a $V_n$ como un $V$-estad\'istico con un n\'ucleo de orden 4.

La distribuci\'on nula y la distribuci\'on asint\'otica nula de $V_n$ son ambas desconocidas. No obstante, Nakamura y P\'erez-Abreu (1993) \cite{NaPe93} observaron num\'ericamente que la distribuci\'on nula de
\[V_n^*=\frac{nV_n}{(\bar{X}_n)^{1\text{.}45}}\]
es aproximadamente independiente de $\vartheta$.

\subsection{Tests de bondad de ajuste en dimensi\'on dos}\label{Test-bondad-ajuste-dim-2}
El objetivo de este apartado es contrastar
\begin{equation}\label{H0-DPBivariante}
\begin{array}{l}
H_0 : (X_1,X_2)\sim PB(\theta_1, \theta_2,\theta_3),\ \text{para alg\'un}\ \theta_1, \theta_2,\theta_3>0,\\[.2 cm]
H_1 : (X_1,X_2)\ \,\text{no se distribuye}\ PB(\theta_1, \theta_2,\theta_3), \ \forall\, \theta_1, \theta_2,\theta_3>0.
\end{array}
\end{equation}

Hasta donde conocemos, existen tres tests estad\'isticos para contrastar (\ref{H0-DPBivariante}). A continuaci\'on los describiremos brevemente y expondremos lo esencial de cada uno de ellos.

\subsubsection[Test $T$ de Crockett]{Test $\boldsymbol{T}$ de Crockett}
El estad\'istico propuesto por Crockett (1979) \cite{Cro79}, $T$, est\'a basado en una forma cuadr\'atica en $Z_{X_1}=S^2_{X_1}-\bar{X}_1$ y $Z_{X_2}=S^2_{X_2}-\bar{X}_2$, donde $\bar{X}_1, \bar{X}_2, S^2_{X_1}$ y $S^2_{X_2}$ son las medias y varianzas muestrales, respectivamente. El objetivo es encontrar la matriz de varianzas y covarianzas de $Z_{X_1}$ y $Z_{X_2}$.

 Se tiene que var$(Z_{X_1})=$ var$(S^2_{X_1})+$ var$(\bar{{X_1}})-2$ cov$(S^2_{X_1},\bar{X}_1)$. De Stuart y Ord (1987, Volumen I, p. 338) \cite{StOr87}, obtenemos que
\[\text{var}(S_{X_1}^2)=\frac{\mu_4-\mu_2^2}{n},\]
donde $\mu_k$ es el $k$-\'esimo momento central definido en la Subsecci\'on \ref{Momentos-DPB}.

Como $\mu_1=0$, de Stuart y Ord (1987, Volumen I, ejemplo 10.2, p. 323) \cite{StOr87}
\[\text{cov}(S_{X_1}^2,\bar{X}_1)=\frac{\mu_3}{n}.\]
As\'i,
\[\text{var}(Z_{X_1})=\frac{\theta_1+3\theta_1^2-\theta_1^2}{n}+\frac{\theta_1}{n}- 2\frac{\theta_1}{n}=2\frac{\theta_1^2}{n}\]
y similarmente,
\[\text{var}(Z_{X_2})=2\frac{\theta_2^2}{n}.\]

Puesto que,
\[\text{cov}(Z_{X_1}, Z_{X_2}) = \text{cov}(S^2_{X_1},S^2_{X_2})+\text{cov}(\bar{X}_1,\bar{X}_2)- \text{cov}(S^2_{X_1},\bar{X}_2)-\text{cov}(\bar{X}_1,S^2_{X_2}),\]
de la primera ecuaci\'on de (2.1.19) en Kocherlakota y Kocherlakota (1992, p. 40) \cite{KoKo92}
\[\text{cov}(S^2_{X_1},S^2_{X_2})=\text{cov}(m_{2,0},m_{0,2})=\frac{\mu_{2,2}- \mu_{2,0}\mu_{0,2}}{n},\]
donde $\mu_{r,s}$ es el $(r,s)$-\'esimo momento central, descrito en la Subsecci\'on \ref{Momentos-DPB}, adem\'as
\[m_{r,s}=\frac{1}{n}\sum_{i=1}^n (X_{1i}-\bar{X}_1)^r (X_{2i}-\bar{X}_2)^s.\]

De Kocherlakota y Kocherlakota (1992, p. 43) \cite{KoKo92} \[\text{cov}(\bar{X}_1,\bar{X}_2)=\frac{1}{n}\,\text{cov}(X_1,X_2)= \frac{\theta_3}{n}\] y de (2.1.17) en Kocherlakota y Kocherlakota (1992, p. 40) \cite{KoKo92}
\[\text{cov}(S^2_{X_1},\bar{X}_2)=\text{cov}(m_{2,0},m_{0,1})=\frac{\mu_{2,1}-\mu_{2,0}\mu_{0,1}+2\mu_{1,1} \mu_{1,0}\mu_{0,0}-\mu_{2,1}\mu_{0,0}-2\mu_{1,0}\mu_{1,1}}{n}.\]
Haciendo los c\'alculos respectivos obtenemos
\[\text{cov}(Z_{X_1}, Z_{X_2})=2\frac{\theta_3^2}{n}.\]

As\'i, si $V$ denota la matriz de varianzas y covarianzas de $Z=(Z_{X_1}, Z_{X_2})$, entonces los resultados de esta secci\'on muestran que
\[V=\frac{2}{n}\left(
    \begin{array}{cc}
      \theta_1^2 & \theta_3^2 \\[.1 cm]
      \theta_3^2 & \theta_2^2
    \end{array}
  \right).
\]

Usando los estimadores $\bar{X}_1, \bar{X}_2$ y $m_{1,1}=S_{X_1X_2}$ (covarianza muestral) para $\theta_1, \theta_2$ y $\theta_3$, respectivamente, Crockett (1979) \cite{Cro79} demuestra que, bajo $H_0$,
\[T=Z\hat{V}^{-1}Z^{\top}\ \mathop{\longrightarrow}\limits^{L}\ Y\sim \chi^2_{2},\]
con lo cual se consigue el estad\'istico propuesto por dicho autor, dado por
\[ T= \frac{n}{2}\frac{\bar{X}_2^{2} \left(S^2_{X_1}-\bar{X}_1\right)^2- 2S^2_{X_1X_2}\left(S^2_{X_1}-\bar{X}_1\right) \left(S^2_{X_2}-\bar{X}_2\right)+ \bar{X}_1^{2} \left(S^2_{X_2}-\bar{X}_2\right)^2} {\bar{X}_1^{2} \bar{X}_2^{2}-S^4_{X_1X_2}}.\]

As\'i, se rechaza $H_0$ si
\[T \geq \chi^2_{2,\alpha},\]
donde $\chi^2_{2, \alpha}$, para $0<\alpha<1$, denota el percentil $\alpha$ superior de la distribuci\'on $\chi^2$ con 2 grados de libertad.

\subsubsection[Test $I_B$ de Loukas y Kemp]{Test $\boldsymbol{I_B}$ de Loukas y Kemp}
Loukas y Kemp (1986) \cite{LoKe86} desarrollaron un test para la DPB basado en lo que llamaron \'indice de dispersi\'on bivariante y denotaron por $I_B$, pues es una extensi\'on del \'indice de dispersi\'on univariante.

Con el objetivo de obtener $I_B$, consideremos ahora una muestra aleatoria simple $(X_{11}, X_{21})$, $(X_{12}, X_{22}), \ldots, (X_{1n}, X_{2n})$, tal que $(X_{1k}, X_{2k})\sim PB(\theta_1,\theta_2,\theta_3)$, para $\, k=1, 2, \ldots, n$, donde $\theta_1>\theta_3,\,\theta_2>\theta_3\,$ y $\,\theta_3>0$.

Si $\theta_1,\theta_2$ y $\theta_3$ son conocidos, entonces el \'indice de dispersi\'on bivariante toma la forma
$$I_B=\frac{1}{1-\rho^2}\sum_{i=1}^n\left(W_{1i}^2-2\rho W_{1i}W_{2i}+W_{2i}^2\right),$$
donde $\,W_{ji}=\displaystyle\frac{X_{ji}-\theta_j}{\sqrt{\theta_j}}, j=1,2, \ i=1,2,\ldots,n\ $ y $\ \rho=\displaystyle\frac{\theta_3}{\sqrt{\theta_1\theta_2}}$.\vskip .3 cm

Bajo las condiciones  del Teorema 1 en Loukas y Kemp (1986) \cite{LoKe86}, dichos autores muestran que $I_B$ se distribuye aproximadamente como una variable $\chi^2$ con $2n$ grados de libertad.\vskip .25 cm

En la situaci\'on pr\'actica usual, cuando $\, \theta_1,\, \theta_2\, $ y $\, \theta_3\,$ son desconocidos, entonces Loukas y Kemp (1986) \cite{LoKe86} definen el \'indice de dispersi\'on bivariante como
\begin{align}
I_B&=\frac{\mathop{\sum}\limits_{i=1}^n\left\{\displaystyle\frac{(X_{1i}- \bar{X}_1)^2}{\bar{X}_1}-\frac{2S_{12}(X_{1i}- \bar{X}_1)(X_{2i}-\bar{X}_2)}{\bar{X}_1 \bar{X}_2}+\frac{(X_{2i}-\bar{X}_2)^2}{\bar{X}_2} \right\}}{1-\displaystyle\frac{S_{12}^2}{\bar{X}_1 \bar{X}_2}}\notag\\[.2 cm]
&=\frac{n(\bar{X}_2S_1^2-2S_{12}^2+\bar{X}_1S_2^2)} {\bar{X}_1\bar{X}_2-S_{12}^2},\notag
\end{align}
donde $\bar{X}_1$ y $\bar{X}_2$ son las medias muestrales, $S^2_1$ y $S^2_2$ son la varianzas muestrales y $S_{12}$ es la covariaza muestral.

Como en este proceso se debe estimar el par\'ametro $\theta$, entonces se pierden tres grados de libertad y por tanto, este nuevo $I_B$ se distribuye aproximadamente como una $\chi^2$ con $2n-3$ grados de libertad, como lo mencionan Loukas y Kemp (1986) \cite{LoKe86}.\vskip .2 cm

As\'i, se rechaza $H_0$ si
\[I_B \geq \chi^2_{2n-3,\alpha},\]
donde $\chi^2_{2n-3, \alpha}$, para $0<\alpha<1$, denota el percentil $\alpha$ superior de la distribuci\'on $\chi^2$ con $2n-3$ grados de libertad.

\begin{observacion} Si $\theta_1,\theta_2$ y $\theta_3$ son desconocidos, entonces Loukas y Kemp (1986) \cite{LoKe86} aproximan $\theta_1$ y $\theta_2$ por sus estimadores m\'aximo veros\'imiles, esto es, $\hat{\theta}_1=\bar{X}_1$ y $\,\hat{\theta}_2=\bar{X}_2$, y el estimador de $\theta_3$ que, como debe calcularse por m\'etodos num\'ericos, lo aproximan por la covarianza muestral, es decir, $\hat{\theta}_3=S_{12}$.
\end{observacion}

\subsubsection[Test $NI_B$ de Rayner y Best]{Test $\boldsymbol{NI_B}$ de Rayner y Best}
Rayner y Best (1995) \cite{RaBe95} expresaron el estad\'istico de Loukas y Kemp (1986) \cite{LoKe86}, $I_B$, como sigue
\begin{equation}\label{ecuac3.63}
I_B=\frac{n}{1-\hat{\rho}^{\,2}}\!\left(\frac{S_{X_1}^2}{\bar{X}_1}-2\, \frac{S_{X_1X_2}^2}{\bar{X}_1\bar{X}_2}+\frac{S_{X_2}^2} {\bar{X}_2}\right),
\end{equation}
donde $\displaystyle\hat{\rho}=\frac{S_{X_1X_2}}{\sqrt{\bar{X}_1\bar{X}_2}}$ es un estimador de $\rho$, $\bar{X}_1$ y $\bar{X}_2$ son las medias muestrales, $S^2_{X_1}$ y $S^2_{X_2}$ son la varianzas muestrales, y $S_{X_1X_2}$ es la covariaza muestral.\vskip .2 cm

De (\ref{ecuac3.63}) se observa que
\[\text{si}\ \ \hat{\rho}^2=\frac{S_{X_1X_2}^2}{\bar{X}_1\bar{X}_2} >\frac{1}{2} \left(\frac{S_{X_1}^2}{\bar{X}_1}+\frac{S_{X_2}^2} {\bar{X}_2}\right),\ \text{entonces} \ \, I_B<0.\]
Por lo tanto, cuando $\rho$ crece hay una probabilidad creciente que $I_B$ sea negativo y su distribuci\'on no sea bien aproximada por una $\chi^2$. Para remediar esta situaci\'on, proponen utilizar un nuevo estad\'istico, lo llaman $NI_B$ y lo definen mediante la expresi\'on
$$NI_B=\frac{n}{1-r^2}\left(\frac{S_{X_1}^2}{\bar{X}_1}-2\,r^2 \sqrt{\frac{S_{X_1}^2S_{X_2}^2}{\bar{X}_1\bar{X}_2}}+\frac{S_{X_2}^2} {\bar{X}_2}\right)\!,$$
donde, el estimador de $\rho$ es el coeficiente de correlaci\'on muestral, dado por
\[r=\frac{\mathop{\sum}\limits_{i=1}^n(X_{1i}-\bar{X}_{1})(X_{2i}-\bar{X}_{2})} {\sqrt{\mathop{\sum}\limits_{i=1}^n(X_{1i}-\bar{X}_{1})^2\ \mathop{\sum}\limits_{i=1}^n(X_{2i}-\bar{X}_{2})^2}}.\]

Al igual que $I_B$, bajo $H_0$, $NI_B\cong \chi_{2n-3}^2$. Por lo tanto, se rechaza $H_0$ si
\[NI_B \geq \chi^2_{2n-3,\alpha},\]
donde $\chi^2_{2n-3, \alpha}$, para $0<\alpha<1$, denota el percentil $\alpha$ superior de la distribuci\'on $\chi^2$ con $2n-3$ grados de libertad.

\begin{observacion} Los tests estad\'isticos $T, I_B$ y $NI_B$ no son consistentes, pues est\'an construidos en base a los momentos, espec\'ificamente se basan en que los dos primeros momentos poblacionales son iguales.
\end{observacion}

\begin{observacion} Para efectos de programar los estad\'isticos $T, I_B$ y $NI_B$, Rayner y Best (1995) \cite{RaBe95} hacen notar que Crockett (1979) \cite{Cro79} usa $(n-1)$ como divisor para calcular $S^2_{X_1}$ y $S^2_{X_2}$, y divisor $n$ en $S_{X_1X_2}$, en cambio Loukas y Kemp (1986) \cite{LoKe86} usan el divisor $n$ para calcular $S^2_{X_1}, S^2_{X_2}$ y $S_{X_1X_2}$.
\end{observacion}





\chapter{Estad\'isticos tipo Cram\'er-von Mises}\label{Estadisticos-tipoCramer-von-Mises}

\section{Tests estad\'isticos} \label{Tests-estadisticos}
Sean $\boldsymbol{X}_1=(X_ {11}, X_ {21}), \boldsymbol{X}_2=(X_ {12}, X_ {22}), \ldots, \boldsymbol{X}_n=(X_ {1n}, X_ {2n})$ vectores aleatorios iid de $\boldsymbol{X}=(X_1,X_2)\in \mathbb{N}_0^2$. Bas\'andonos en la muestra aleatoria $\boldsymbol{X}_1,\boldsymbol{X}_2,\ldots,\boldsymbol{X}_n$, nuestro objetivo es contrastar la hip\'otesis
\[H_0 : (X_1,X_2)\sim PB(\theta_1,\theta_2,\theta_3), \ \text{para alg\'un}\ (\theta_1,\theta_2,\theta_3)\in\Theta,\]
contra la alternativa
\[H_1 : (X_1,X_2)\nsim PB(\theta_1,\theta_2,\theta_3), \ \forall(\theta_1,\theta_2,\theta_3)\in\Theta.\]
Con este prop\'osito, aprovecharemos algunas de las propiedades de la fgp que nos permitir\'an proponer los siguientes dos tests estad\'isticos.
\begin{enumerate}
  \item De acuerdo a la Proposici\'on \ref{Conv-FuncGenProbBiv}, un estimador consistente de la fgp es la fgpe. Si $H_0$ es verdadera y $\hat{{\theta}}_n$ es un estimador consistente de ${\theta}$, entonces $g(u;\hat{{\theta}}_n)$ estima consistentemente la fgp de los datos.

Como la distribuci\'on de $\boldsymbol{X}$ es determinada de forma \'unica por su fgp, $g(u)$, $u\in [0,1]^2$, un test razonable para contrastar $H_0$ deber\'ia rechazar la hip\'otesis nula para valores ``grandes'' de $R_{n,w}(\hat{{\theta}}_n)$ definido por
      \begin{equation}\label{Estad-Rnw} R_{n,w}(\hat{{\theta}}_n)=\int_0^1\int_0^1 G^2_n(u;\hat{{\theta}}_n)w(u)du,
      \end{equation}
      donde \[G_n(u;\hat{\theta}_n)= \sqrt{n}\left\{g_n(u)-g(u; \hat{{\theta}}_n)\right\},\] $\hat{\theta}_n=\hat{{\theta}}_n(\boldsymbol{X}_1, \boldsymbol{X}_2,\ldots,\boldsymbol{X}_n)=(\hat{\theta}_{1n}, \hat{\theta}_{2n},\hat{\theta}_{3n})$ es un estimador consistente de $\theta$ y $w(u)$ es una funci\'on medible de peso tal que $w(u)\geq 0, \ \forall u\in [0,1]^2$, y
      \begin{equation} \label{int-funcion-peso}
      \int_0^1\int_0^1 w(u)du < \infty.
      \end{equation}
      Este \'ultimo supuesto sobre $w$ asegura que la integral doble en (\ref{Estad-Rnw}) es finita para cada $n$ fijo.

$R_{n,w}(\hat{{\theta}}_n)$ es una extensi\'on bivariante de los estad\'isticos propuestos por Rueda et al. (1991) \cite{RuPeOR91} y Baringhaus et al. (2000) \cite{BaGuHe00} para contrastar la bondad de ajuste a la distribuci\'on Poisson univariante, tal como se vi\'o en la Subsecci\'on \ref{Contrastes-basados-en-fgpe}.

  \item Puesto que la fgp de la distribuci\'on Poisson univariante con par\'ametro $\lambda$ es la \'unica fgp que satisface $g'(t)=\lambda g(t)$, Baringhaus y Henze (1992) \cite{BaHe92} propusieron un test estad\'istico que se basa en el an\'alogo emp\'irico de esta ecuaci\'on, presentado en la Subsecci\'on \ref{Contrastes-basados-en-fgpe}.

Con el fin de extender este test al caso bivariante, primero tenemos que encontrar una ecuaci\'on o un sistema de ecuaciones cuya \'unica soluci\'on sea la fgp de la DPB. La siguiente proposici\'on establece dicho sistema. \begin{proposicion}\label{Soluc-sistema-de-dos-EDP} Sea $G_2=\{g:[0,1]^2\to \mathbb{R}$, tal que $g$  es una fgp y $\frac{\partial}{\partial u_i}g (u_1,u_2)$, $i=1,2$, existen $\forall (u_1,u_2) \in [0,1]^2\}$. Sea $g(u_1,u_2;\theta)$ como en (\ref{fgp3-DPB}). Entonces $g(u_1,u_2;\theta)$ es la \'unica fgp en $G_2$ que satisface el siguiente sistema \begin{equation}\label{EDPs-fgp} D_i(u;\theta)=0, \quad i=1,2, \quad \forall u \in [0,1]^2,\end{equation} donde
\[\begin{array}{ll} D_1(u;\theta)&=\displaystyle\frac{\partial}{\partial u_1}g(u_1,u_2)-\{\theta_1+\theta_3(u_2-1)\}g(u_1,u_2),\\[.35 cm] D_2(u;\theta) &=\displaystyle\frac{\partial }{\partial u_2}g(u_1,u_2)-\{\theta_2+\theta_3(u_1-1)\}g(u_1,u_2).\end{array}\] \end{proposicion}
{\bf Demostraci\'on} \hspace{2pt} Sea $(X_1,X_2)$ un vector aleatorio y sea $g(u_1,u_2)=E\left(u_1^{X_1} u_2^{X_2}\right)$ su fgp. Entonces, de la primera expresi\'on en (\ref{EDPs-fgp}) \begin{equation}\label{EDP1-Modif-fgp} \frac{\partial}{\partial u_1}\log g(u_1,u_2)=\theta_1+\theta_3(u_2-1).\end{equation}
Integrando (\ref{EDP1-Modif-fgp}) sobre $u_1$, obtenemos \[\begin{array}{rcl}
       g(u_1,u_2) & = & \exp\{\phi_1(u_2)+\theta_1 u_1+\theta_3(u_2-1) u_1\},\\[.25 cm]
       & = & \exp\{\phi(u_2)+\theta_1(u_1-1)+\theta_3(u_1-1)(u_2-1)\}, \end{array}\]
donde $\phi(u_2)=\phi_1(u_2)+\theta_1+\theta_3(u_2-1)$.\vskip .2 cm

Procediendo similarmente, de la segunda ecuaci\'on en (\ref{EDPs-fgp}) obtenemos
\[g(u_1,u_2)=\exp\{\varphi(u_1)+\theta_2(u_2-1)+\theta_3(u_1-1)(u_2-1)\}.\] As\'i, necesariamente $\varphi(u_1)=\theta_1(u_1-1)$ y $\phi(u_2)=\theta_2(u_2-1)$, en otras palabras, la fgp de la DPB es la \'unica soluci\'on de (\ref{EDPs-fgp}). $\square$\vskip .3 cm

Por la Proposici\'on \ref{Conv-FuncGenProbBiv}, $g(u)$ y sus derivadas pueden ser estimadas consistentemente por la fgpe y las derivadas de la fgpe, respectivamente. As\'i, si $H_0$ fuera cierta, entonces las funciones
\begin{equation}\label{EDPs-empiricas}
\begin{array}{l}
D_{1n}(u;\hat{\theta}_n )=\displaystyle\frac{\partial }{\partial u_1}g_n(u_1,u_2)-\left\{\hat{\theta}_{1n}+\hat{\theta}_{3n}(u_2-1)\right\} g_n(u_1,u_2), \\[.35 cm]
      D_{2n}(u;\hat{\theta}_{n})=\displaystyle\frac{\partial }{\partial u_2}g_n(u_1,u_2)-\left\{\hat{\theta}_{2n}+\hat{\theta}_{3n}(u_1-1)\right\} g_n(u_1,u_2),
     \end{array}\end{equation}
     deber\'ian estar cerca de 0, donde $\hat{\theta}_n$ es un estimador consistente de $\theta$. As\'i, para contrastar $H_0$ consideramos el siguiente test estad\'istico
\begin{equation}\label{Estad-Snw}
S_{n,w}(\hat{\theta}_n)=n\int_0^1\int_0^1 \left\{D^2_{1n}(u;\hat{\theta}_n )+D^2_{2n}(u;\hat{\theta}_{n} )\right\}w(u)du,
\end{equation}
donde la funci\'on de peso $w(u)$ es como la definida en el \'item anterior.

     Notar que la fgp de $\boldsymbol{X}$ est\'a en $G_2$ s\'i y s\'olo si $E(X_i)$ existe, $i=1,2$.
\end{enumerate}

En los dos casos anteriores, un test razonable para contrastar $H_0$ deber\'ia rechazar la hip\'otesis nula para valores grandes de cada test estad\'istico. Ahora, para determinar qu\'e son los valores grandes en cada caso, debemos calcular la distribuci\'on nula de cada test estad\'istico o al menos una aproximaci\'on de cada una de ellas.

Puesto que las distribuciones nulas son desconocidas, trataremos de aproximarlas. Un modo cl\'asico de estimar la distribuci\'on nula es mediante la distribuci\'on asint\'otica nula. En la siguiente secci\'on estudiaremos esta situaci\'on.

\section{Aproximaci\'on de la distribuci\'on nula} \label{Aproximacion-distribucion-nula}
\subsection{Distribuci\'on asint\'otica nula}\label{Distribucion-asintotica-nula}
Como un primer intento de aproximar la distribuci\'on nula de $R_{n,w}(\hat{\theta}_n)$ y de $S_{n,w}(\hat{\theta}_n)$ obtendremos sus distribuciones asint\'oticas nulas. Con este prop\'osito, supondremos que el estimador $\hat{\theta}_n$ satisface la siguiente condici\'on de regularidad.
\begin{supuesto}\label{hat(theta)-theta}
Bajo $H_0$, si $\theta= (\theta_{1},\theta_{2},\theta_{3})\!\in\!\Theta$ denota el verdadero valor del par\'ametro, entonces
\[\sqrt{n}\left(\hat{\theta}_n-\theta\right) =\frac{1}{\sqrt{n}}\sum_{i=1}^n \boldsymbol{\ell}\left(\boldsymbol{X}_{i}; \theta\right)+\boldsymbol{o}_{_P}(1),\]
donde $\boldsymbol{\ell}:\mathbb{N}^2_0\times \Theta \longrightarrow \mathbb{R}^3$ es tal que \[E_{\theta}\bigl\{\boldsymbol{\ell} (\!\boldsymbol{X}_{\!{1}}; \theta)\bigr\}=\mathbf{0}\in \mathbb{R}^3\] y \[J(\theta)=E_{\theta}\!\left\{ \boldsymbol{\ell}\!\left(\boldsymbol{X}_{\!{1}}; \theta\right)^\top\boldsymbol{\ell} \!\left(\boldsymbol{X}_{\!{1}}; \theta\right)\right\}< \infty.\]
Aqu\'i, $\boldsymbol{o}_{_P}(1)$ es un vector que consta de 3 elementos $o_{_P}(1)$.
\end{supuesto}

\begin{observacion}\label{metodos-aprox}
El Supuesto \ref{hat(theta)-theta} no es restrictivo pues lo cumplen los estimadores m\'as usados com\'unmente, como son los citados en la Subsecci\'on \ref{Estimacion-puntual}.
\end{observacion}

Para obtener la distribuci\'on asint\'otica nula de $R_{n,w}(\hat{\theta}_n)$ y de $S_{n,w}(\hat{\theta}_n)$, la plataforma de trabajo que usaremos ser\'a el espacio de Hilbert separable $\mathcal{H}=L^2\left([0,1]^2,w\right)$ definido por
\[\mathcal{H}=\left\{\varphi:[0,1]^2\rightarrow \mathbb{R}, \text{ medible, tal que } \| \varphi\|_{_\mathcal{H}}^2=\int_0^1\int_0^1 \varphi(u)^2 w(u)du<\infty\right\},\]
con producto escalar
$ \langle \phi, \psi\rangle_{_\mathcal{H}}=\int_0^1\int_0^1 \phi(u)\psi(u) w(u)du$.

En dicho espacio de Hilbert $\mathcal{H}$, tenemos que
\begin{itemize}
  \item 
  $R_{n,w}(\hat{\theta}_n)= \|G_n(\hat{{\theta}}_n)\|_{_\mathcal{H}}^{2},\ \, \text{con}\ \, G_n(u;\hat{{\theta}}_n)=\displaystyle\frac{1}{\sqrt{n}}\sum_{i=1}^n G(\boldsymbol{X}_i;\hat{\theta}_n;u),\ \, u\in [0,1]^2$,

  donde
  \[G(\boldsymbol{X}_i;\hat{\theta}_n;u)= u_1^{X_{1i}}\,u_2^{X_{2i}}-g(u;\hat{\theta}_n),\, i=1,2,\ldots,n.\]
  \item 
  $S_{n,w}(\hat{\theta}_n)= \|Z_{1n}\|_{_\mathcal{H}}^{2}+ \|Z_{2n}\|_{_\mathcal{H}}^{2},\ \ \text{con}\ \ Z_{kn}(u)=\displaystyle\frac{1}{\sqrt{n}}\sum_{i=1}^n V_k(\boldsymbol{X}_i;\hat{\theta}_n;u),\ \,k=1,2$,$\ u\in [0,1]^2$, donde
  \[V_1(\boldsymbol{X}_i;\hat{\theta}_n;u)= X_{1i}\,I{\{X_{1i}\geq 1\}}\,u_1^{X_{1i}-1}\,u_2^{X_{2i}}-\left\{\hat{\theta}_{1n}+ \hat{\theta}_{3n}(u_2-1)\right\}u_1^{X_{1i}}\,u_2^{X_{2i}},\]
  \[V_2(\boldsymbol{X}_i;\hat{\theta}_n;u)= X_{2i}\,I{\{X_{2i}\geq 1\}}\,u_1^{X_{1i}}\,u_2^{X_{2i}-1}-\left\{\hat{\theta}_{2n}+ \hat{\theta}_{3n}(u_1-1)\right\}u_1^{X_{1i}}\,u_2^{X_{2i}},\]
  $i=1,2,\ldots,n$.
\end{itemize}

El siguiente resultado proporciona la distribuci\'on asint\'otica nula de  $R_{n,w}(\hat{\theta}_n)$ y $S_{n,w}(\hat{\theta}_n)$.

\begin{teorema}\label{ConvDebil-Rnw-Snw}
Sean $\boldsymbol{X}_{1},\boldsymbol{X}_{2},\ldots, \boldsymbol{X}_{n}$ vectores aleatorios iid de $\boldsymbol{X}=(X_{1},X_{2})\sim PB(\theta)$. Supongamos que se cumple el Supuesto \ref{hat(theta)-theta}. Entonces
\begin{itemize}
   \item [(a)] $R_{n,w}(\hat{\theta}_n)=\|R_n\|_{_\mathcal{H}}^2+r_n,$\\ donde $P_{{\theta}}(|r_n|>\varepsilon)\to 0$, $\forall \varepsilon>0$, $R_n(u)=\displaystyle\frac{1}{\sqrt{n}}\sum_{i=1}^n R^0(\boldsymbol{X}_i;{\theta};{u})$, con
      $$R^0(\boldsymbol{X}_i;{\theta}; {u})=u_1^{X_{1i}} u_2^{X_{2i}}-g({u};{\theta})  \left\{1+\boldsymbol{\ell}\left(\boldsymbol{X}_{i}; {\theta}\right)(u_1\!-1,u_2\!-1,(u_1\!-1)(u_2\!-1))^\top\right\},$$

  \item [(b)] \begin{equation} \label{Snw-Sn} S_{n,w}(\hat{\theta}_n)= \|S_{1n}\|_{_\mathcal{H}}^2+\|S_{2n}\|_{_\mathcal{H}}^2+s_n,\end{equation}
 donde $P_{{\theta}}(|s_n|>\varepsilon)\to 0$, $\forall \varepsilon>0$, $S_{kn}(u)=\displaystyle\frac{1}{\sqrt{n}}\sum_{i=1}^n S^0_k(\boldsymbol{X}_i;\theta; u),\ k=1,2,\ $ con
      \begin{align}
      S^0_1(\boldsymbol{X}_i;\theta; u)& = X_{1i}\,I{\{X_{1i}\geq 1\}}\,u_1^{X_{1i}-1}\,u_2^{X_{2i}}-\left\{\theta_{1}+\theta_{3}(u_2-1) \right\} u_1^{X_{1i}}\,u_2^{X_{2i}}\notag\\[.1 cm] &\hspace{8mm}-g(u;\theta) \boldsymbol{\ell}\left(\boldsymbol{X}_i; \theta\right)(1,0,u_2-1)^{\top},\notag\\[.25 cm]
      S^0_2(\boldsymbol{X}_i;\theta; u)& = X_{2i}\,I{\{X_{2i}\geq 1\}}\,u_1^{X_{1i}}\,u_2^{X_{2i}-1}-\left\{\theta_{2}+\theta_{3}(u_1-1) \right\} u_1^{X_{1i}}\,u_2^{X_{2i}}\notag\\[.1 cm]       &\hspace{8mm}-g(u;\theta) \boldsymbol{\ell}\left(\boldsymbol{X}_i; \theta\right)(0,1,u_1-1)^{\top}.\notag
      \end{align}
  \end{itemize}
Adem\'as,
\[R_{n,w}(\hat{\theta}_n)\mathop{\longrightarrow}\limits^{L} \sum_{j\geq 1}\lambda_j^R\chi^2_{1j}.\]
\[S_{n,w}(\hat{\theta}_n)\mathop{\longrightarrow}\limits^{L} \sum_{j\geq 1}\lambda_j^S\chi^2_{1j}.\]
donde $\chi^2_{11},\chi^2_{12},\ldots$ son v.a. independientes $\chi^2$ con 1 grado de libertad y los conjuntos $\{\lambda_j^R\}_{j\geq 1}$ y $\{\lambda_j^S\}_{j\geq 1}$ son los autovalores no nulos de los operadores $C_R(\theta)$ y $C_S(\theta)$, respectivamente, definidos sobre el espacio de funciones

$\left\{\tau:\mathbb{N}_0^2\to \mathbb{R},  \text{tal que} \ E_{\theta}\!\left\{\tau^2(\boldsymbol{X})\right\}<\infty,\forall \theta\in\Theta\right\}$, como sigue
\begin{equation} \label{Operador-C}
C_R(\theta) \tau({x})= E_{\theta}\{h_R({x},\boldsymbol{Y};\theta) \tau(\boldsymbol{Y})\} \quad \text{y}\quad C_S(\theta) \tau({x})= E_{\theta}\{h_S({x},\boldsymbol{Y};\theta) \tau(\boldsymbol{Y})\}
\end{equation}
donde
\[h_R({x},{y};\theta)=\int_0^1\int_0^1  R^0({x}; \theta; u) R^0({y}; \theta; u)w(u)du,\]
\begin{equation} \label{h-S} h_S({x},{y};\theta)=\int_0^1\int_0^1  \sum_{k=1}^2 S_k^0({x}; \theta; u) S_k^0({y}; \theta; u)w(u)du.
\end{equation}
\end{teorema}

\nt {\bf Demostraci\'on} \hspace{2pt} \hspace{2pt} Solamente presentaremos la demostraci\'on de la parte (b), pues la demostraci\'on de la parte (a) sigue pasos similares.

Por definici\'on,
$S_{n,w}(\hat{\theta}_n)= \|Z_{1n}\|_{_\mathcal{H}}^{2} +\|Z_{2n}\|_{_\mathcal{H}}^{2}$, con
\[Z_{kn}(u)=\frac{1}{\sqrt{n}}\sum_{i=1}^n V_k(\boldsymbol{X}_i;\hat{\theta}_n;u), \ k=1,2.\]
Por desarrollo en serie de Taylor,
\begin{equation}\label{Z-kn-aprox}
Z_{kn}(u)=\frac{1}{\sqrt{n}}\sum_{i=1}^n V_k(\boldsymbol{X}_i;\theta; u)+\frac{1}{n}\sum_{i=1}^nQ^{(1)}_k(\boldsymbol{X}_i;\theta; u)\,\sqrt{n}(\hat{\theta}_n-\theta)^{\top}+q_{kn},
\end{equation}
$k=1,2$, donde
\[q_{kn}=\frac{1}{2}\sqrt{n}\bigl(\hat{\theta}_n-\theta\bigr) \frac{1}{\sqrt{n}}\frac{1}{n}\sum_{i=1}^n Q^{(2)}_k\bigl(\boldsymbol{X}_i;\widetilde{\theta};u\bigr) \,\sqrt{n}\bigl(\hat{\theta}_n-\theta\bigr)^{\top},\]
$\widetilde{\theta}=\alpha \hat{\theta}_n+(1-\alpha)\theta$, para alg\'un $0<\alpha<1\,$ y  $\,Q^{(2)}_k\bigl(x;\vartheta; u\bigr)$ es la matriz de orden $3\times 3$ que contiene las derivadas de segundo orden de $V_k\bigl(x;\vartheta; u\bigr)$ con respecto a $\vartheta$, para $k=1,2$, \[Q^{(1)}_k(x;\vartheta; u)=
\left( Q^{(1)}_{k1}(x;\vartheta; u), Q^{(1)}_{k2}(x;\vartheta; u), Q^{(1)}_{k3}(x;\vartheta; u)\right),\]
donde $\ Q^{(1)}_{kj}(x;\vartheta;u)=\frac{\partial}{\partial \vartheta_j} V_k(x;\vartheta; u)$, para $k=1,2$, $j=1,2,3$.\vskip .15 cm

Como las primeras derivadas son
\begin{equation}\label{Deriv-Vk}
Q^{(1)}_1(x;\vartheta;u)=-u_1^{x_1} u_2^{x_2}(1,0,u_2-1),\ \ \ Q^{(1)}_2(x;\vartheta;u)=-u_1^{x_1} u_2^{x_2}(0,1,u_1-1),
\end{equation}
entonces, $\left|Q^{(1)}_{kj}\!\left(\boldsymbol{X}_{\!1};\theta; u\right)\right|\!\leq\! 1$, para $k=1,2$, $j=1,2,3$, $\forall u \!\in\! [0,1]^2$.

As\'i, considerando (\ref{int-funcion-peso}), resulta
\begin{equation}\label{E-Qkj-acotada}
E_{\theta}\left\{\left\|Q^{(1)}_{kj}\left(\boldsymbol{X}_{1};\theta; u\right)\right\|_{_\mathcal{H}}^2\right\}<\infty,\ k=1,2,\ j=1,2,3.
\end{equation}
Como, para $k=1,2$, $j=1,2,3$, tenemos que
\[E_{\theta}\!\left[\left\{\frac{1}{n}\sum_{i=1}^n Q^{(1)}_{kj}(\boldsymbol{X}_i;\theta; u) \right\}^2\right]= \frac{1}{n} E_{\theta}\!\left[\left\{Q^{(1)}_{kj}(\boldsymbol{X}_1;\theta; u) \right\}^2\right]+\frac{n-1}{n}\!\left[E_{\theta}\! \left\{Q^{(1)}_{kj} (\boldsymbol{X}_1;\theta; u) \right\}\right]^2,\]
entonces
\begin{align}
E_{\theta}\!\left[\left\{\frac{1}{n}\sum_{i=1}^n Q^{(1)}_{kj} (\boldsymbol{X}_i;\theta; u)-E_{\theta}\! \left\{Q^{(1)}_{kj} (\boldsymbol{X}_1;\theta; u)\right\} \right\}^2\right]&\notag\\[.2 cm] &\hspace{-38mm}=\frac{1}{n}E_{\theta}\!\left[\left\{Q^{(1)}_{kj} (\boldsymbol{X}_1;\theta; u)\right\}^2\right]-\frac{1}{n}\!\left[E_{\theta}\! \left\{Q^{(1)}_{kj} (\boldsymbol{X}_1;\theta; u) \right\}\right]^2\notag\\[.2 cm]
&\hspace{-38mm}\leq\frac{1}{n}E_{\theta}\!\left[\left\{Q^{(1)}_{kj} (\boldsymbol{X}_1;\theta; u)\right\}^2\right].\notag
\end{align}
Usando la desigualdad de Markov y (\ref{E-Qkj-acotada}), se logra
\begin{align}
&P_{\theta}\!\left[\left\| \frac{1}{n}\sum_{i=1}^n Q^{(1)}_{kj}(\boldsymbol{X}_i;\theta; u)-E_{\theta}\!\left\{Q^{(1)}_{kj}(\boldsymbol{X}_1;\theta; u)\right\} \right\|_{_\mathcal{H}}>\varepsilon\right]\notag\\[.2 cm]
&\hspace{85mm}\leq \frac{1}{n\, \varepsilon^2}\,E_{\theta}\!\left[\left\|Q^{(1)}_{kj} (\boldsymbol{X}_1;\theta; u)\right\|_{_\mathcal{H}}^2\right]\to 0,\notag
\end{align}
para $k=1,2$, $j=1,2,3$.

As\'i, obtenemos que
\begin{equation}\label{E-Qk}
\frac{1}{n}\sum_{i=1}^n Q^{(1)}_{k}(\boldsymbol{X}_i;\theta; u)\mathop{\longrightarrow} \limits^{P}E_{\theta}\left\{Q^{(1)}_{k}(\boldsymbol{X}_1;\theta; u)\right\}=
\left\{
 \begin{array}{ll}
    -g(u;\theta)(1,0,u_2-1), & \hbox{si } k=1,\\[.2 cm]
    -g(u;\theta)(0,1,u_1-1), & \hbox{si } k=2,
  \end{array}
\right.
\end{equation}
en $\mathcal{H}$.

De (\ref{Deriv-Vk}) se sigue que $Q^{(2)}_k\!\bigl(x;\vartheta; u\bigr)= \boldsymbol{0}\,$ (la matriz nula de orden $3\times 3$), $k=1,2$ y as\'i $q_{kn}=0$.

Ahora, por el Supuesto \ref{hat(theta)-theta} y como $q_{kn}=0$, entonces, (\ref{Z-kn-aprox}) se puede escribir como
\begin{align}
Z_{kn}(u)&=\frac{1}{\sqrt{n}}\sum_{i=1}^n \left[V_k(\boldsymbol{X}_i;\theta; u)+E_{\theta}\left\{Q^{(1)}_{k}(\boldsymbol{X}_1;\theta; u)\right\}  \boldsymbol{\ell}\left(\boldsymbol{X}_i; \theta\right)^{\top}\right]\notag\\[.2 cm]
&\hspace{22mm}+ E_{\theta}\left\{Q^{(1)}_{k}(\boldsymbol{X}_1;\theta; u)\right\}\, \boldsymbol{o}_{_P}(1)^{\top}.\notag
\end{align}
As\'i, de esta \'ultima ecuaci\'on, y de (\ref{E-Qkj-acotada}) y (\ref{E-Qk})
\[
Z_{kn}(u)=S_{kn}(u)+s_{kn},
\]
donde $\|s_{kn}\|_{\mathcal{H}}=o_{_P}(1)$, $k=1,2$.

Por otra parte, observar que
\[
\|S_{1n}\|_{_\mathcal{H}}^{2}+\|S_{2n}\|_{_\mathcal{H}}^{2}=\frac{1}{n} \sum_{i=1}^n\sum_{j=1}^n h_S(\boldsymbol{X}_i,\boldsymbol{X}_j;\theta),\]
donde, $h_S(x,y;\theta)$ es como el definido en (\ref{h-S}) y satisface $\,h_S(x,y;\theta)= h_S(y,x;\theta)$,
$E_{\theta}\left\{h_S^2(\boldsymbol{X}_1,\boldsymbol{X}_2;\theta)\right\}< \infty\,$ y $\,E_{\theta} \left\{|h_S(\boldsymbol{X}_1,\boldsymbol{X}_1;\theta)|\right\} < \infty$.

Adem\'as, por la Proposici\'on \ref{Soluc-sistema-de-dos-EDP} y el Supuesto \ref{hat(theta)-theta}
\[E_{\theta}\!\left\{h_S(\boldsymbol{X}_1,\boldsymbol{X}_2;\theta)\right\}= \int_0^1\int_0^1 \sum_{k=1}^2 E_{\theta}\!\left\{S_k^0(\boldsymbol{X}_1; \theta; u)\right\} E_{\theta}\!\left\{S_k^0(\boldsymbol{X}_2; \theta; u)\right\}w(u)\,du=0.\]
Por \'ultimo, como $h_S$ es degenerado, $E_{\theta}\!\left\{h_S(\boldsymbol{X}_1,\boldsymbol{X}_2;\theta)/ \boldsymbol{X}_1\right\}=0$, entonces, por el Teorema 6.4.1.B en Serfling (1980) \cite{Ser80}
\begin{equation}\label{norma-Skn-converge}
\|S_{1n}\|_{_\mathcal{H}}^{2}+\|S_{2n}\|_{_\mathcal{H}}^{2}
\mathop{\longrightarrow} \limits^{\!L}\ \sum_{j\geq 1}\lambda_j^S\,\chi^2_{1j}\,,
\end{equation}
donde $\chi^2_{11},\chi^2_{12},\ldots$ y el conjunto $\{\lambda_j^S\}_{j\geq 1}$ son como los definidos en el Teorema.\vskip .15 cm

De (\ref{norma-Skn-converge}) se desprende que $\|S_{1n}\|_{_\mathcal{H}}^{2}+\|S_{2n}\|_{_\mathcal{H}}^{2}=O_P(1)$.

Ahora, como \[\|Z_{kn}\|_{_\mathcal{H}}^{\,2}=\|S_{kn}\|_{_\mathcal{H}}^{\,2}+ \|s_{kn}\|_{_\mathcal{H}}^{\,2}+2\langle S_{kn},s_{kn}\rangle_{_\mathcal{H}}, \ k=1,2,\]

\nt y puesto que $\ \langle S_{kn},s_{kn}\rangle_{_\mathcal{H}}\leq\|S_{kn}\|_{_\mathcal{H}}\, \|s_{kn}\|_{_\mathcal{H}}=o_{_P}(1)$,
entonces
\[S_{n,w}(\hat{\theta}_n)\mathop{\longrightarrow} \limits^{\!L}\ \sum_{j\geq 1}\lambda_j^S\,\chi^2_{1j}\,.\]
Lo cual concluye la demostraci\'on del resultado. $\square$\\

La distribuci\'on asint\'otica nula de $R_{n,w}(\hat{\theta}_n)$ y de $S_{n,w}(\hat{\theta}_n)$ depende del desconocido verdadero valor del par\'ametro $\theta$, por tanto, en la pr\'actica, no proporcionan una soluci\'on \'util al problema de estimar la distribuci\'on nula de los respectivos tests estad\'isticos. Esto podr\'ia solucionarse al reemplazar $\theta$ por $\hat{\theta}_n$.

Pero una dificultad mayor es determinar los conjuntos $\{\lambda_j^R\}_{j\geq 1}$ y $\{\lambda_j^S\}_{j\geq 1}$, puesto que, en la mayor\'ia de los casos, calcular los autovalores de un operador no es una tarea simple y en nuestro caso, debemos obtener tambi\'en las expresiones de $h_R(x,y;\theta)$ y $h_S(x,y;\theta)$, que no son f\'aciles de encontrar, pues dependen de la funci\'on $\boldsymbol{\ell}$, que por lo general no tiene una expresi\'on sencilla.

As\'i, en la siguiente subsecci\'on consideramos otra forma de aproximar la distribuci\'on nula de los tests estad\'isticos, el m\'etodo bootstrap param\'etrico.

\subsection{Aproximaci\'on bootstrap de la distribuci\'on nula}\label{Aproximacion-bootstrap}
Un modo alternativo de estimar la distribuci\'on nula es mediante el m\'etodo bootstrap param\'etrico.

Sean $\boldsymbol{X}_{1}, \boldsymbol{X}_{2},\ldots, \boldsymbol{X}_{n}$ vectores aleatorios iid que toman valores en $\mathbb{N}_0^2$. Supongamos que $\hat{\theta}_n= \hat{\theta}_n(\boldsymbol{X}_{\!1}, \boldsymbol{X}_{2},\ldots, \boldsymbol{X}_{n}) \in \Theta$.

Sean $\boldsymbol{X}^*_{1}, \boldsymbol{X}^*_{2},\ldots, \boldsymbol{X}^*_{n}$ vectores aleatorios iid de una poblaci\'on que se distribuye seg\'un la ley $PB(\hat{\theta}_{1n},\hat{\theta}_{2n}, \hat{\theta}_{3n})$, dado $\boldsymbol{X}_{1}, \boldsymbol{X}_{2},\ldots, \boldsymbol{X}_{n}$ y sea $R^*_{n,w}(\hat{\theta}_n^*)$ la versi\'on bootstrap de $R_{n,w}(\hat{\theta}_n)$ obtenida al reemplazar $\boldsymbol{X}_{\!1}, \boldsymbol{X}_{2},\ldots, \boldsymbol{X}_{n}$ y
$\hat{\theta}_n= \hat{\theta}_n(\boldsymbol{X}_{\!1}, \boldsymbol{X}_{2},\ldots, \boldsymbol{X}_{n})$ por $\boldsymbol{X}^*_{\!1}, \boldsymbol{X}^*_{2},\ldots, \boldsymbol{X}^*_{n}$ y $\hat{\theta}_n^*= \hat{\theta}_n(\boldsymbol{X}^*_{1}, \boldsymbol{X}^*_{2},\ldots, \boldsymbol{X}^*_{n})$, respectivamente, en la expresi\'on de $R_{n,w}(\hat{\theta}_n)$.

Adem\'as, sea $P_*$ la ley de probabilidad condicional bootstrap, dado $\boldsymbol{X}_{1}, \boldsymbol{X}_{2},\ldots, \boldsymbol{X}_{n}$.

En forma an\'aloga se describe $S^*_{n,w}(\hat{\theta}_n^*)$, la versi\'on bootstrap de $S_{n,w}(\hat{\theta}_n)$.

Para probar que el m\'etodo bootstrap se puede usar para aproximar consistentemente la distribuci\'on nula de $R_{n,w}(\hat{\theta}_n)$, o de $S_{n,w}(\hat{\theta}_n)$, supondremos las siguientes hip\'otesis que son un poco m\'as fuertes que el Supuesto \ref{hat(theta)-theta}.

\begin{supuesto}\label{E(l*l)-l-cont} Se cumple el Supuesto \ref{hat(theta)-theta} y las funciones  $\boldsymbol{\ell}$ y $J$ satisfacen
 \begin{enumerate}
 \item $\sup_{\vartheta\in\,\Theta_0} E_{\vartheta}\Bigl[\|\boldsymbol{\ell} (\boldsymbol{X}; \vartheta)\|^2 I{\left\{\|\boldsymbol{\ell}(\boldsymbol{X}; \vartheta)\|>\gamma \right\}}\Bigr]\rightarrow 0$, cuando $\gamma \to \infty$, donde $\Theta_0 \subseteq \Theta$ es una vecindad abierta de $\theta$.
\item $\boldsymbol{\ell}(\boldsymbol{X};\vartheta)$ y $J(\vartheta)$ son continuas como funciones de $\vartheta$ en $\vartheta=\theta$ y $J(\vartheta)$ es finita $\forall \vartheta \in \Theta_0$.
 \end{enumerate}
\end{supuesto}

Tal como se estableci\'o en la Observaci\'on \ref{metodos-aprox}, el Supuesto \ref{E(l*l)-l-cont} no es restrictivo pues lo cumplen estimadores com\'unmente usados.

Para mostrar que el m\'etodo bootstrap aproxima consistentemente a la distribuci\'on nula de $R_{n,w}(\hat{\theta}_n)$, o de $S_{n,w}(\hat{\theta}_n)$, nos ser\'a \'util el siguiente resultado, que lo volveremos a usar m\'as adelante.

\begin{lema}\label{P(hat(t))*l(hat(t))-P(t)*l(t)}
Supongamos que se verifica el Supuesto \ref{E(l*l)-l-cont} y que $\hat{\theta}_n
\mathop{\longrightarrow}\limits^{c.s.} \theta$, para alg\'un $\theta \in \Theta$. Entonces
\begin{itemize}
  \item [(a)] $\displaystyle \sum_{r,s \geq 0} r\left|\boldsymbol{\ell}_i(r,s;\hat{\theta}_n) P(r,s;\hat{\theta}_n)- \boldsymbol{\ell}_i(r,s;\theta) P(r,s;\theta)\right|=o(1),\ i=1,2,3,$
  \item [(b)] $\displaystyle \sum_{r,s \geq 0} s\left|\boldsymbol{\ell}_i(r,s;\hat{\theta}_n) P(r,s;\hat{\theta}_n)- \boldsymbol{\ell}_i(r,s;\theta) P(r,s;\theta)\right|=o(1),\ i=1,2,3,$
\end{itemize}
donde $P(r,s;\vartheta)=P_{\vartheta}(X_1=r, X_2=s),\, \forall \vartheta\in \Theta_0$.
\end{lema}
\nt {\bf Demostraci\'on} \hspace{2pt} Solamente presentaremos la demostraci\'on de la parte (b), pues la demostraci\'on de la parte (a) sigue pasos similares.

Como
\[\sum_{r,s \geq 0}=\sum_{r,s \geq 0;\, \|\boldsymbol{\ell}(r,s; \vartheta)\|>\gamma}+\sum_{r,s \geq 0;\, \|\boldsymbol{\ell}(r,s; \vartheta)\|\leq \gamma},\]
consideraremos por separado cada uno de estos sumandos.

\nt \underline{Caso $\|\boldsymbol{\ell}(r,s; \vartheta)\|>\gamma$} \hspace{2pt} Tengamos presente que $E_{\theta}\left(X^2_2\right)=\theta_2+\theta_2^2$.
Sea $\varepsilon>0$ arbitrario pero fijo. Por el Supuesto \ref{E(l*l)-l-cont} (1), $\exists \gamma=\gamma(\varepsilon)>0$ tal que
\begin{equation}\label{Sup-acotado}
\sup_{\|\boldsymbol{\ell}(r,s;\vartheta)\|>\gamma} \sum_{r,s} \|\boldsymbol{\ell}(r,s;\vartheta)\|^2 P(r,s;\vartheta)<\frac{\varepsilon^2}{4\left(\theta_2+\theta_2^2\right)}, \ \ \forall \vartheta \in \Theta_0
\end{equation}
y por tanto
\begin{equation}\label{Sup-suma-acotado}
\sup_{\|\boldsymbol{\ell}(r,s;\vartheta)\|>\gamma} \sum_{r,s} s\left|\boldsymbol{\ell}_i(r,s;\vartheta) P(r,s;\vartheta)- \boldsymbol{\ell}_i(r,s;\theta) P(r,s;\theta)\right|<\varepsilon,
\end{equation}
$i=1,2,3$, pues, por la desigualdad de Cauchy-Schwarz
\begin{align}
&\sum_{r,s} s\left|\boldsymbol{\ell}_i(r,s;\vartheta) P(r,s;\vartheta)- \boldsymbol{\ell}_i(r,s;\theta) P(r,s;\theta)\right|\notag\\
&\ \ \ \ \leq  \sum_{r,s} s \left|\boldsymbol{\ell}_i(r,s;\vartheta)\right| P(r,s;\vartheta) + \sum_{r,s} s \left|\boldsymbol{\ell}_i(r,s;\theta)\right| P(r,s;\theta)\notag\\
&\ \ \ \ \leq \!\left(\theta_2+\theta_2^2\right)^{\!\frac{1}{2}}\! \left\{\sum_{r,s} \|\boldsymbol{\ell}(r,s;\vartheta)\|^2 P(r,s;\vartheta)\right\}^{\!\!\frac{1}{2}}\! + \! \left(\theta_2+\theta_2^2\right)^{\!\frac{1}{2}} \!\left\{\sum_{r,s} \|\boldsymbol{\ell}(r,s;\theta)\|^2 P(r,s;\theta)\right\}^{\!\!\frac{1}{2}}\!.\notag
\end{align}
De lo anterior y considerando (\ref{Sup-acotado}), obtenemos (\ref{Sup-suma-acotado}).\vskip .1 cm

\nt \underline{Caso $\|\boldsymbol{\ell}(r,s; \vartheta)\|\leq\gamma$} \hspace{2pt} Para esta situaci\'on, primero mostraremos la validez de (b) cuando $r,s\geq M$ y luego para $r,s< M$, donde $M$ es una constante positiva que determinaremos m\'as adelante.

Como $\sum_{r,s \geq 0} s P(r,s;\vartheta)=E_{\vartheta}(X_2)=\vartheta_2$ y las $P(r,s;\vartheta)$ son continuas como funciones de $\vartheta$, entonces dado $\varepsilon_1> 0, \,\exists M=M(\varepsilon_1)>0$, de modo que
\[\sum_{r,s < M} s P(r,s;\vartheta)> \vartheta_2-\frac{\varepsilon_1}{2\gamma}, \ \forall \vartheta \in \Theta_1\subseteq\Theta_0,\]
siempre que $\|\vartheta-\theta\|<\delta_1$, para cierto $\delta_1 >0$.
Por tanto, si $\|\boldsymbol{\ell}(r,s;\vartheta)\| \leq \gamma$, entonces
\begin{equation}\label{Suma-acotada}
\sum_{r,s\geq M} s\left|\boldsymbol{\ell}_i(r,s;\vartheta) P(r,s;\vartheta)- \boldsymbol{\ell}_i(r,s;\theta) P(r,s;\theta)\right|<\varepsilon_1,\ \forall \vartheta\in \Theta_0 \cap\left\{\|\vartheta-\theta\|<\delta_1\right\},\ \,
\end{equation}
$ i=1,2,3$, pues
\[\vartheta_2=\sum_{r,s \geq 0} s P(r,s;\vartheta)=\sum_{r,s \geq M} s P(r,s;\vartheta)+\sum_{r,s<M} s P(r,s;\vartheta)>\sum_{r,s \geq M} s P(r,s;\vartheta)+\vartheta_2-\frac{\varepsilon_1}{2\gamma},\]
de donde
\[\sum_{r,s \geq M} s P(r,s;\vartheta)<\frac{\varepsilon_1}{2\gamma}.\]

Considerando esta \'ultima desigualdad y el hecho que $|\boldsymbol{\ell}_i(r,s;\vartheta)|\leq\|\boldsymbol{\ell}(r,s; \vartheta)\|\leq\gamma$, para $i=1,2,3$, entonces
\begin{align}
\sum_{r,s\geq M} s &\left|\boldsymbol{\ell}_i(r,s;\vartheta) P(r,s;\vartheta) - \boldsymbol{\ell}_i(r,s;\theta) P(r,s;\theta)\right|\notag \\
&\hspace{43mm}\leq\gamma\left\{ \sum_{r,s\geq M} s P(r,s;\vartheta) + \sum_{r,s\geq M} s P(r,s;\theta) \right\}< \varepsilon_1,\notag
\end{align}
lo cual prueba (\ref{Suma-acotada}).

Para $r,s< M$. Como $\boldsymbol{\ell}_i(r,s;\vartheta)P(r,s;\vartheta)$ son funciones continuas en $\vartheta$, para $ i=1,2,3$, se tendr\'a que $\,\forall \varepsilon_2> 0, \,\exists \delta_2=\delta_2(\varepsilon_2)>0$, tal que $\|\vartheta-\theta\|<\delta_2$ implica que
\[\sup_{r,s < M}\left| \boldsymbol{\ell}_i(r,s;\vartheta)P(r,s;\vartheta) -\boldsymbol{\ell}_i(r,s;\theta)P(r,s;\theta)\right|<\frac{\varepsilon_2}{M^3}\]
y por tanto,
\begin{align}
\sum_{r,s< M} s &\left|\boldsymbol{\ell}_i(r,s;\vartheta) P(r,s;\vartheta)- \boldsymbol{\ell}_i(r,s;\theta) P(r,s;\theta)\right|\notag\\
&\hspace{15mm}<M\sum_{r,s< M} \left|\boldsymbol{\ell}_i(r,s;\vartheta) P(r,s;\vartheta)- \boldsymbol{\ell}_i(r,s;\theta) P(r,s;\theta)\right|< \varepsilon_2,\notag
\end{align}
siempre que $\,\vartheta \in \left\{\|\vartheta-\theta\|<\delta_2\right\}\cap \Theta_0$, para cierto $\delta_2>0$. Lo cual concluye la demostraci\'on, pues $\varepsilon_1$ y $\varepsilon_2$ son arbitrarios. $\square$\vskip .3 cm

Tambi\'en nos ser\'a de utilidad el siguiente resultado.

\begin{proposicion}\label{Nucleos-Rnw-Snw}
Sean  $\boldsymbol{X}_{1},\boldsymbol{X}_{2},\ldots, \boldsymbol{X}_{n}$ vectores aleatorios iid de $\boldsymbol{X}=(X_{1},X_{2}) \in \mathbb{N}_0^2$. Supongamos que se verifica el Supuesto \ref{E(l*l)-l-cont} y que $\hat{\theta}_n \mathop{\longrightarrow}\limits^{c.s.} \theta$, para alg\'un $\theta \in \Theta$. Entonces
\begin{itemize}
  \item [(a)] $\displaystyle\sup_{u,v \in [0,1]^2} \left| K^R_n(u,v)-K^R(u,v)\right|\stackrel{c.s.}{\longrightarrow}0$, donde
      \begin{align}
      K^R_n(u,v)&=E_*\left\{R^0(\boldsymbol{X}_1^*;\hat{\theta}_n; u) R^0(\boldsymbol{X}_1^*;\hat{\theta}_n; v)\right\}, \notag\\[.2 cm]
      K^R(u,v)&=E_{\theta}\left\{R^0(\boldsymbol{X}_1;\theta; u) R^0(\boldsymbol{X}_1;\theta; v)\right\},\notag
      \end{align} con $R^0(\boldsymbol{X}_1;\theta; u)$ el definido en el Teorema \ref{ConvDebil-Rnw-Snw}(a).

  \item [(b)] \vskip .1 cm $\displaystyle\sup_{u,v \in [0,1]^2} \left| K^S_n(u,v)-K^S(u,v)\right|\stackrel{c.s.}{\longrightarrow}0$, donde
      \begin{align}
      K^S_n(u,v)&=\left(E_*\left\{S_i^0(\boldsymbol{X}_1^*;\hat{\theta}_n; u) S_j^0(\boldsymbol{X}_1^*;\hat{\theta}_n; v)\right\}\right), 1\leq i,j \leq 2. \notag\\[.2 cm]
      K^S(u,v)&=\Bigl(E_{\theta}\left\{S_i^0(\boldsymbol{X}_1;\theta; u) S_j^0(\boldsymbol{X}_1;\theta; v)\right\}\Bigr), 1\leq i,j \leq 2.\notag
      \end{align} con $S_i^0(\boldsymbol{X}_1;\theta; u)$ el definido en el Teorema \ref{ConvDebil-Rnw-Snw}(b), $\,i=1,2$.
\end{itemize}
\end{proposicion}
\nt {\bf Demostraci\'on} \hspace{2pt}
Solamente presentaremos la demostraci\'on de la parte (b), pues la demostraci\'on de la parte (a) sigue pasos similares.

El Supuesto \ref{E(l*l)-l-cont} y $\hat{\theta}_n \mathop{\longrightarrow}\limits^{c.s.} \theta$ implican que
\begin{equation}\label{E*S-EtS}
\sup_{u,v \in [0,1]^2} \left| E_*\left\{S_i^0(\boldsymbol{X}_1^*;\hat{\theta}_n; u) S_j^0(\boldsymbol{X}_1^*;\hat{\theta}_n; v)\right\}- E_{\theta}\left\{S_i^0(\boldsymbol{X}_1;\theta; u) S_j^0(\boldsymbol{X}_1;\theta; v)\right\}\right| \stackrel{c.s.}{\longrightarrow}0,
\end{equation}
$1\leq i,j \leq 2$.

Para probar este resultado, primero hacemos la resta de las esperanzas en (\ref{E*S-EtS}) y encontramos, por ejemplo, que
\begin{align}
&E_*\!\left\{S_1^0(\boldsymbol{X}_1^*;\hat{\theta}_n; u) S_2^0(\boldsymbol{X}_1^*;\hat{\theta}_n; v)\right\}- E_{\theta}\!\left\{S_1^0(\boldsymbol{X}_1;\theta; u) S_2^0(\boldsymbol{X}_1;\theta; v)\right\}\notag\\[.21 cm]
&=v_1u_2\left[\left\{\hat{\theta}_{3n}+\phi_1(r;\hat{\theta}_n) \phi_2(r;\hat{\theta}_n)\right\} g(r;\hat{\theta}_n)-\left\{\theta_3+ \phi_1(r;\theta) \phi_2(r;\theta)\right\}g(r;\theta)\right]\notag\\[.21 cm]
&\quad -v_1\left\{\phi_1(r;\hat{\theta}_n) \phi_2(v;\hat{\theta}_n) g(r;\hat{\theta}_n)- \phi_1(r;\theta)\phi_2(v;\theta) g(r;\theta)\right\}\notag\\[.21 cm]
&\quad -u_2\left\{\phi_1(u;\hat{\theta}_n) \phi_2(r;\hat{\theta}_n) g(r;\hat{\theta}_n)-\phi_1(u;\theta)\phi_2(r;\theta) g(r;\theta)\right\}\notag\\[.21 cm]
&\quad +\phi_1(u;\hat{\theta}_n) \phi_2(v;\hat{\theta}_n) g(r;\hat{\theta}_n)- \phi_1(u;\theta)\phi_2(v;\theta) g(r;\theta)\notag\\[.21 cm]
&\quad+A(u)^{\top}\!\left\{g(u;\hat{\theta}_n)g(v;\hat{\theta}_n)J(\hat{\theta}_n) -g(u;\theta)g(v;\theta)J(\theta)\right\}B(v)\notag\\[.21 cm]
&\quad+\!\left[g(v;\hat{\theta}_n)\phi_1(u;\hat{\theta}_n) E_*\!\!\left\{u_1^{X^*_{1}}u_2^{X^*_{2}} \! \boldsymbol{\ell}(\boldsymbol{X}^*_1;\hat{\theta}_n)\!\right\}\!-\!g(v;\theta) \phi_1(u;\theta)E_{\theta}\!\left\{u_1^{X_{1}}u_2^{X_{2}} \boldsymbol{\ell}(\boldsymbol{X}_{1};\theta)\right\}\right]B(v)\notag\\[.21 cm]
&\quad+\!\left[g(u;\hat{\theta}_n)\phi_2(v;\hat{\theta}_n) E_*\!\!\left\{v_1^{X^*_{1}}v_2^{X^*_{2}}\! \boldsymbol{\ell}(\boldsymbol{X}^*_1;\hat{\theta}_n)\!\right\}\!-\!g(u;\theta) \phi_2(v;\theta)E_{\theta}\{v_1^{X_{1}}v_2^{X_{2}} \boldsymbol{\ell}(\boldsymbol{X}_{1};\theta)\}\right]A(u),\notag\\[.21 cm]
&\quad-\left[g(v;\hat{\theta}_n)E_*\!\!\left\{\psi_1(u;\boldsymbol{X}^*_1; \hat{\theta}_n)\right\}-g(v;\theta)E_{\theta}\!\left\{\psi_1(u;\boldsymbol{X}_1; \theta)\right\}\right]B(v)\notag\\[.21 cm]
&\quad-\left[g(u;\hat{\theta}_n)E_*\!\!\left\{\psi_2(v;\boldsymbol{X}^*_1; \hat{\theta}_n)\!\right\}-g(u;\theta)E_{\theta}\! \left\{\psi_2(v;\boldsymbol{X}_{\!1}; \theta)\right\}\right]A(u)\notag
\end{align}
donde
\begin{align}
&r_1=u_1v_1,\ r_2=u_2v_2,\ r=(r_1,r_2),\ \ A(u)=(1,0,u_2-1)^{\top},\ B(v)=(0,1,v_1-1)^{\top},\notag\\[.22 cm]
&\phi_1(w;\vartheta)=\vartheta_1+\vartheta_3(w_2-1), \quad \phi_2(w;\vartheta)=\vartheta_2+\vartheta_3(w_1-1),\quad w=(w_1,w_2),\notag\\[.22 cm]
&\psi_1(u;\boldsymbol{X}_1;\vartheta)=X_1\,I{\{X_1\geq 1\}}\,u_1^{X_1-1}\,u_2^{X_2}\,\boldsymbol{\ell}(\boldsymbol{X}_1;\vartheta), \notag\\[.22 cm]
&\psi_2(u;\boldsymbol{X}_1;\vartheta)=X_2\,I{\{X_2\geq 1\}}\,u_1^{X_1}\,u_2^{X_2-1}\,\boldsymbol{\ell}(\boldsymbol{X}_1;\vartheta), \notag
\end{align}

Las otras diferencias de (\ref{E*S-EtS}) tienen expresiones an\'alogas a la diferencia de esperanzas anterior. Las situaciones que aparecen en tales esperanzas son de alguno de los siguientes casos:
\begin{itemize}
  \item Caso 1: sea $f(u,v;\vartheta)$ una funci\'on polinomial en las variables $u, v\in [0,1]^2$ y $\vartheta\in\Theta_1\subseteq\Theta_0$, donde $\Theta_1$ es un conjunto compacto que contiene a $\theta$. Luego, $f(u,v;\theta)g(u;\theta)$ es continua como funci\'on de $u, v$ y $\theta$, por lo tanto, es uniformemente continua en $[0,1]^2\times\Theta_1$ y como $\hat{\theta}_n \mathop{\longrightarrow}\limits^{c.s.} \theta$, entonces \[\sup_{u,v\in [0,1]^2} \left|f(u,v;\hat{\theta}_n)g(u;\hat{\theta}_n)- f(u,v;\theta)g(u;\theta)\right| =o(1).\]

  \item \vskip .2 cm Caso 2: como $g(u;\theta)g(v;\theta)J(\theta)$ es continua como funci\'on de $\theta$, entonces es uniformemente continua en $\Theta_1$, donde $\Theta_1$ es un conjunto compacto que contiene a $\theta$. Adem\'as, como $u,v\in [0,1]^2\,$ y $\,\hat{\theta}_n \mathop{\longrightarrow}\limits^{c.s.} \theta$, entonces
      \[\sup_{u,v\in [0,1]^2} \left|A(u)^{\top} \left[g(u;\hat{\theta}_n)g(v;\hat{\theta}_n)J(\hat{\theta}_n) -g(u;\theta)g(v;\theta)J(\theta)\right]B(v)\right|=o(1).\]

  \item Caso 3: sea $f(u,v;\vartheta)$ una funci\'on polinomial en las variables $u, v\in [0,1]^2$ y $\vartheta\in\Theta_1\subseteq\Theta_0$, donde $\Theta_1$ es un conjunto compacto que contiene a $\theta$. Por lo tanto $f$ es continua como funci\'on de $\vartheta$ y tambi\'en lo es $g(u;\vartheta)$, luego $f(u,v;\hat{\theta}_n)=f(u,v;\theta)+o(1)$ y $g(u;\hat{\theta}_n)=g(u;\theta)+o(1)$, con lo cual
      \begin{align}
      f(u,v;\hat{\theta}_n)&g(u;\hat{\theta}_n)E_*\! \left\{v_1^{X^*_{1}}v_2^{X^*_{2}} \boldsymbol{\ell}(\boldsymbol{X}^*_1;\hat{\theta}_n)\right\}- f(u,v;\theta)g(u;\theta)E_{\theta}\!\left\{v_1^{X_{1}}v_2^{X_{2}} \boldsymbol{\ell}(\boldsymbol{X}_1;\theta)\right\}\notag\\[.2 cm]
      &\hspace{4mm}=f(u,v;\theta)g(u;\theta)\left[E_*\!\left\{v_1^{X^*_{1}}v_2^{X^*_{2}} \boldsymbol{\ell}(\boldsymbol{X}^*_1;\hat{\theta}_n)\right\}- E_{\theta}\!\left\{v_1^{X_{1}}v_2^{X_{2}} \boldsymbol{\ell}(\boldsymbol{X}_1;\theta)\right\}\right]\notag\\[.2 cm]
      &\hspace{12mm}+o(1)\{f(u,v;\theta) +g(u;\theta)+1\}E_*\!\left\{v_1^{X^*_{1}}v_2^{X^*_{2}} \boldsymbol{\ell}(\boldsymbol{X}^*_1;\hat{\theta}_n)\right\}.\notag
      \end{align}

      Usando el Lema \ref{P(hat(t))*l(hat(t))-P(t)*l(t)} y el hecho que $\hat{\theta}_n\in \Theta_0$, resulta
      \begin{align}
      E_*\!\left\{v_1^{X^*_{1}}v_2^{X^*_{2}} \boldsymbol{\ell}_i(\boldsymbol{X}^*_1;\hat{\theta}_n)\right\}&- E_{\theta}\{v_1^{X_{1}}v_2^{X_{2}} \boldsymbol{\ell}_i(\boldsymbol{X}_1;\theta)\}\notag\\[.1 cm]
      &=\sum_{r,s \geq 0} v_1^r v_2^s\left\{\boldsymbol{\ell}_i(r,s;\hat{\theta}_n) P(r,s;\hat{\theta}_n)-\boldsymbol{\ell}_i(r,s;\theta) P(r,s;\theta)\right\}\notag\\[.1 cm]
      &\leq \sum_{r,s \geq 0}\left|\boldsymbol{\ell}_i(r,s;\hat{\theta}_n) P(r,s;\hat{\theta}_n)-\boldsymbol{\ell}_i(r,s;\theta) P(r,s;\theta)\right|=o(1),\notag
      \end{align}
      para $i=1,2,3$. Por lo tanto,
      \[\left|E_*\!\left\{v_1^{X^*_{1}}v_2^{X^*_{2}} \boldsymbol{\ell}(\boldsymbol{X}^*_1;\hat{\theta}_n)\right\}- E_{\theta}\!\left\{v_1^{X_{1}}v_2^{X_{2}} \boldsymbol{\ell}(\boldsymbol{X}_1;\theta)\right\}\right|=\mathbf{o}(1).\]
      Ahora, puesto que $J(\hat{\theta}_n)< \infty$, entonces
      \[\left\|E_*\!\left\{v_1^{X^*_{1}}v_2^{X^*_{2}} \boldsymbol{\ell}(\boldsymbol{X}^*_1;\hat{\theta}_n)\right\}\right\|\leq \!\left\{ E_*\!\left(\|\boldsymbol{\ell}(\boldsymbol{X}^*_1;\hat{\theta}_n)\|^2\right) \right\}^{1/2}< \infty.\]

      Adem\'as, como $|g(u;\theta)|\leq 1$ y $|f(u,v;\theta)|< \infty$, $\forall u, v\in [0,1]^2$, entonces
      \[\sup_{u,v\in [0,1]^2}\left|\left[\phi(u,v;\hat{\theta}_n) E_*\!\left\{\varphi(v;\boldsymbol{X}^*_{1};\hat{\theta}_n)\right\} -\phi(u,v;\theta)E_{\theta}\{\varphi(v;\boldsymbol{X}_{1};\theta)\} \right]A(u)\right|=o(1),\]
      donde $\,\phi(u,v;\vartheta)=f(u,v;\vartheta) g(u;\vartheta)\ $  y $\ \varphi(v;\boldsymbol{X}_{1};\vartheta)=v_1^{X_{1}}v_2^{X_{2}} \boldsymbol{\ell}(\boldsymbol{X}_{1};\vartheta)$.

  \item \vskip .15 cm Caso 4: como $g(u;\hat{\theta}_n)=g(u;\theta)+o(1)$, pues es continua como funci\'on de $\vartheta\in \Theta_0$, entonces
      \begin{align}
      &g(u;\hat{\theta}_n)\,E_*\!\left\{X^*_{2}\,I{\{X^*_{2}\geq 1\}}\,v_1^{X^*_{1}}\,v_2^{X^*_{2}-1}\boldsymbol{\ell}(\boldsymbol{X}^*_1; \hat{\theta}_n)\right\}\notag\\[.1 cm]
      &\hspace{40 mm}-g(u;\theta)E_{\theta}\left\{X_{2}\,I{\{X_{2}\geq 1\}}\,v_1^{X_{1}}\,v_2^{X_{2}-1}\boldsymbol{\ell}(\boldsymbol{X}_1; \theta)\right\}\notag\\[.2 cm]
      &\hspace{35 mm} =g(u;\theta)\biggl[E_*\!\left\{X^*_{2}\,I{\{X^*_{2}\geq 1\}}\,v_1^{X^*_{1}}\,v_2^{X^*_{2}-1}\boldsymbol{\ell}(\boldsymbol{X}^*_1; \hat{\theta}_n)\right\}\biggr.\notag\\
      &\hspace{55 mm}\biggl.-E_{\theta}\left\{X_{2}\,I{\{X_{2}\geq 1\}}\,v_1^{X_{1}}\,v_2^{X_{2}-1}\boldsymbol{\ell}(\boldsymbol{X}_1; \theta)\right\}\biggr]\notag\\[.2 cm]
      &\hspace{55 mm}+o(1)\,E_*\!\left\{X^*_{2}\,I{\{X^*_{2}\geq 1\}}\,v_1^{X^*_{1}}\,v_2^{X^*_{2}-1}\boldsymbol{\ell}(\boldsymbol{X}^*_1; \hat{\theta}_n)\right\}.\notag
      \end{align}

      Usando el Lema \ref{P(hat(t))*l(hat(t))-P(t)*l(t)} y el hecho que $\hat{\theta}_n\in \Theta_0$, obtenemos
      \begin{align}
      E_*\!\left\{X^*_{2}I{\{X^*_{2}\geq 1\}}v_1^{X^*_{1}}v_2^{X^*_{2}-1}\boldsymbol{\ell}_i(\boldsymbol{X}^*_1; \hat{\theta}_n)\right\}&-E_{\theta}\!\left\{X_{2}I{\{X_{2}\geq 1\}}v_1^{X_{1}}v_2^{X_{2}-1}\boldsymbol{\ell}_i(\boldsymbol{X}_1; \theta)\right\}\notag\\[.2 cm]
      &\hspace{-38mm}=\sum_{r\geq 0,s \geq 1}s \,v_1^r v_2^{s-1}\left\{\boldsymbol{\ell}_i(r,s;\hat{\theta}_n) P(r,s;\hat{\theta}_n)-\boldsymbol{\ell}_i(r,s;\theta) P(r,s;\theta)\right\}\notag\\[.2 cm]
      &\hspace{-38mm}\leq\sum_{r,s \geq 0} s \left|\boldsymbol{\ell}_i(r,s;\hat{\theta}_n) P(r,s;\hat{\theta}_n)-\boldsymbol{\ell}_i(r,s;\theta) P(r,s;\theta)\right|=o(1),\notag
      \end{align}
      para $i=1,2,3$. Por lo tanto,
      \[\left|E_*\!\left\{\psi_2(v;\boldsymbol{X}_1^*;\hat{\theta}_n)\right\} -E_{\theta}\!\left\{\psi_2(v;\boldsymbol{X}_1;\theta)\right\}\right| =\mathbf{o}(1),\]
      donde $\, \psi_2(v;\boldsymbol{X}_1;\vartheta)=X_2\,I{\{X_2\geq 1\}}\,v_1^{X_1}\,v_2^{X_2-1}\boldsymbol{\ell}(\boldsymbol{X}_1;\vartheta)$.

      Por otra parte,
      \[E_*\!\left\{\psi_2(v;\boldsymbol{X}_1^*;\hat{\theta}_n)\right\}\leq \left\{E_*\!\left(X^{*2}_{2}\right)\right\}^{1/2} \left\{E_*\!\left(\|\boldsymbol{\ell}(\boldsymbol{X}^*_1; \hat{\theta}_n)\|^2\right)\right\}^{1/2}< \infty,\]
      con lo cual,
      \[\sup_{u,v \in [0,1]^2}\left|\left[g(u;\hat{\theta}_n)E_*\! \left\{\psi_2(v;\boldsymbol{X}^*_1;\hat{\theta}_n)\!\right\}-g(u;\theta) E_{\theta}\!\left\{\psi_2(v;\boldsymbol{X}_1;\theta)\right\}\right]A(u) \right|=o(1).\]
\end{itemize}

La presentaci\'on de estos cuatro casos generales muestran que se verifica (\ref{E*S-EtS}) y con ello se consigue el resultado. $\square$

\begin{teorema}\label{Cons-boot-Rnw-Snw}
 Sean  $\boldsymbol{X}_{1},\boldsymbol{X}_{2},\ldots, \boldsymbol{X}_{n}$ vectores aleatorios iid de $\boldsymbol{X}=(X_{1},X_{2}) \in \mathbb{N}_0^2$. Supongamos que se verifica el Supuesto \ref{E(l*l)-l-cont} y que $\hat{\theta}_n\mathop{\longrightarrow}\limits^{c.s.} \theta$, para alg\'un $\theta \in \Theta$. Entonces
\begin{itemize}
  \item [(a)]$\displaystyle\sup_{x\in\mathbb{R}} \left|P_*\left\{R^*_{n,w}(\hat{\theta}^{*}_n) \leq x\right\}-P_{\theta}\left\{R_{n,w} (\hat{\theta}_n )\leq x\right\}\right|\ \mathop{\longrightarrow}\limits^{c.s.}\ 0.$
   \item [(b)]$\displaystyle\sup_{x\in\mathbb{R}} \left|P_*\left\{S^*_{n,w}(\hat{\theta}^{*}_n) \leq x\right\}-P_{\theta}\left\{S_{n,w} (\hat{\theta}_n )\leq x\right\}\right|\ \mathop{\longrightarrow}\limits^{c.s.}\ 0.$
\end{itemize}
\end{teorema}
 \nt {\bf Demostraci\'on} \hspace{2pt}Solamente presentaremos la demostraci\'on de la parte (b), pues la demostraci\'on de la parte (a) sigue pasos similares.

Por definici\'on, $S_{n,w}^*(\hat{\theta}_n^*)= \|Z_{1n}^*\|_{_\mathcal{H}}^{2} +\|Z_{2n}^*\|_{_\mathcal{H}}^{2}$, con
\[Z_{kn}^*(u)=\frac{1}{\sqrt{n}}\sum_{i=1}^n V_k(\boldsymbol{X}_i^*;\hat{\theta}_n^*; u),\, k=1,2.\]
Siguiendo pasos similares a los dados en la demostraci\'on del Teorema \ref{ConvDebil-Rnw-Snw}(b) se puede ver que
\[Z_{kn}^*(u)=S_{kn}^*(u)+s^*_{kn},\ k=1,2,\]
con $\|s_{kn}^*\|_{\mathcal{H}}=o_{_{P_*}}(1)$ c.s., $k=1,2$, donde $S_{kn}^*(u)$ es definido como $S_{kn}(u)$ que aparece en el Teorema \ref{ConvDebil-Rnw-Snw}, con
$\boldsymbol{X}_i$ y $\theta$ reemplazados por $\boldsymbol{X}^*_i$ y $\hat{\theta}_n$, respectivamente.

Por conveniencia anal\'itica, ahora consideraremos el siguiente espacio de Hilbert separable:
\begin{align}
\mathcal{H}_1&=\Biggl\{\varphi:[0,1]^2\rightarrow \mathbb{R}^2, \text{ donde } \varphi(u)=(\varphi_1(u), \varphi_2(u)) \text{ es una funci\'on medible tal que} \Biggr.\notag\\[.2 cm]
&\hspace{45 mm} \Biggl.\|\varphi\|_{_{\mathcal{H}_1}}^{2}\!=\!\!  \int_{[0,1]^2}\! \left\{\varphi_1^2(u)+ \varphi_2^2(u)\right\}w(u)du<\infty \Biggr\}.\notag
\end{align}
con producto escalar \[\langle(\phi_1, \phi_2), (\psi_1,\psi_2) \rangle_{_{\mathcal{H}_1}}=\int_{[0,1]^2} \!\! \left\{\phi_1(u)\psi_1(u)+\phi_2(u)\psi_2(u)\right\} w({u})d{u}<\infty.\]
Claramente, $\|\varphi\|_{_{\mathcal{H}_1}}^{2}=
\|\varphi_1\|_{_{\mathcal{H}}}^{2}+\|\varphi_2\|_{_{\mathcal{H}}}^{2}$.

Sea \[Y_n^*(u)=\sum_{i=1}^n Y_{ni}^*(u)\]
donde
\[Y_{ni}^*(u)= \frac{1}{\sqrt{n}}\,S(\boldsymbol{X}_i^*;\hat{\theta}_n; u),\ 1\leq i \leq n,\]
\begin{equation}\label{vector-S}
S(\boldsymbol{X}_i^*;\hat{\theta}_n; u)=\left(S_1^0(\boldsymbol{X}_i^*;\hat{\theta}_n; u), S_2^0(\boldsymbol{X}_i^*;\hat{\theta}_n; u)\right),\ 1\leq i \leq n.
\end{equation}

Observar primero que $Y_{ni}^*(u)$ tiene medias cero y momentos segundos finitos, para $1\leq i \leq n$, siempre que $\hat{\theta}_n\in \Theta_0$, lo cual ocurre c.s. porque $\hat{\theta}_n \mathop{\longrightarrow}\limits^{c.s.} \theta$.\vskip .15 cm

Consideremos el n\'ucleo de covarianza $\ K^S_n(u,v)= E_*\!\left\{Y_n^*(u)^\top Y_n^*(v)\right\}$, luego
\[K^S_n(u,v)=\left(E_*\left\{S_i^0(\boldsymbol{X}_1^*;\hat{\theta}_n; u) S_j^0(\boldsymbol{X}_1^*;\hat{\theta}_n; v)\right\}\right), 1\leq i,j \leq 2.\]
Adem\'as, sea $K^S(u,v)=E_{\theta}\left\{S(\boldsymbol{X}_1;{\theta}; u)^{\top} \, S(\boldsymbol{X}_1;{\theta}; v)\right\}$ donde $S$ es como el definido en (\ref{vector-S}), es decir,
\[K^S(u,v)=\Bigl(E_{\theta}\left\{S_i^0(\boldsymbol{X}_1;\theta; u) S_j^0(\boldsymbol{X}_1;\theta; v)\right\}\Bigr), 1\leq i,j \leq 2.\]

Sea $\mathcal{Z}$ un proceso Gaussiano centrado cuyo operador de covarianza $C$ es caracterizado por
\begin{align}
\langle C f, h\rangle_{_{\mathcal{H}_1}}=cov\left(\langle \mathcal{Z}, f\rangle_{_{\!\mathcal{H}_1}}, \langle \mathcal{Z}, h\rangle_{_{\!\mathcal{H}_1}}\right)&=E_{\theta}\left\{\langle \mathcal{Z}, f\rangle_{_{\!\mathcal{H}_1}} \langle \mathcal{Z}, h\rangle_{_{\!\mathcal{H}_1}}\right\}\notag\\[.2 cm]
&=\int_{[0,1]^4} f(u)K^S(u,v)h(v)^{\top} w(u)w(v)dudv.\label{Oper-cov-C-Yn}
\end{align}
Por el Lema \ref{TCL-Hilbert}, $(S_{1n},S_{2n})\stackrel{L}{\longrightarrow} \mathcal{Z}$ en $\mathcal{H}_1$, cuando los datos son iid provenientes del vector aleatorio $\boldsymbol{X}\sim PB(\theta)$, donde $S_{kn}(u)$ es como el definido en el Teorema \ref{ConvDebil-Rnw-Snw}(b), $k=1,2$.

Sea $\{e_k: \, k \geq 0\}$ una base ortonormal de $\mathcal{H}_1$. A continuaci\'on probaremos que se cumplen las condiciones (i)-(iii) del Lema \ref{kundu}.

\begin{itemize}
  \item [(i)] Sea $C_n$ el operador de covarianza  de $Y_n^*$, esto es, sean $f,h \in \mathcal{H}_1$, luego \begin{align} \langle C_n f, h\rangle_{_{\mathcal{H}_1}}&=cov(\langle Y_n^*, f\rangle_{_{\!\mathcal{H}_1}}, \langle Y_n^*, h\rangle_{_{\!\mathcal{H}_1}})=E_{*}\{\langle Y_n^*, f\rangle_{_{\!\mathcal{H}_1}} \langle Y_n^*, h\rangle_{_{\!\mathcal{H}_1}}\}\notag\\[.2 cm] &=\int_{[0,1]^4} f(u)E_*\left\{Y_n^{*}(u)^\top Y_n^*(v)\right\}h(v)^\top w(u)w(v)dudv\notag\\[.2 cm] &=\int_{[0,1]^4} f(u)K^S_n(u,v)h(v)^{\top} w(u)w(v)dudv.\label{def-Cn-Yn} \end{align}
      De (\ref{Oper-cov-C-Yn}), (\ref{def-Cn-Yn}) y de la Proposici\'on \ref{Nucleos-Rnw-Snw}(b),
      \begin{align} \lim_{n\to \infty}\,\langle C_n e_k, e_l\rangle_{_{\mathcal{H}_1}} &=\lim_{n\to \infty}\int_{[0,1]^4} e_k(u)K^S_n(u,v)e_l(v)^{\top}w(u)w(v)du dv\notag\\ &=\int_{[0,1]^4} e_k(u)K^S(u,v)e_l(v)^{\top}w(u)w(v)dudv=\langle C e_k, e_l\rangle_{_{\mathcal{H}_1}} = a_{kl}.\notag \end{align}

\item[(ii)] De (\ref{Oper-cov-C-Yn}), (\ref{def-Cn-Yn}), de la Proposici\'on \ref{Nucleos-Rnw-Snw}(b) y del \'item (i), se obtiene
    \begin{align} \lim_{n\to \infty}\sum_{k=0}^{\infty} \langle C_{n} e_k,e_k\rangle_{_{\mathcal{H}_1}} &= \lim_{n\to \infty}\sum_{k=0}^{\infty}  \int_{[0,1]^4} e_k(u)K^S_n(u,v)e_k(v)^{\top}w(u)w(v)du dv\notag\\ &=\sum_{k=0}^{\infty}  \int_{[0,1]^4} e_k(u)K^S(u,v)e_k(v)^{\top}w(u)w(v)du dv\notag\\ &=\sum_{k=0}^{\infty} \langle Ce_k,e_k\rangle_{_{\mathcal{H}_1}}= \sum_{k=0}^{\infty} a_{kk}<\infty,\notag \end{align}
pues, de la primera ecuaci\'on en (\ref{Oper-cov-C-Yn}) y la igualdad de Parseval
\[\sum_{k=0}^{\infty} a_{kk}=\sum_{k=0}^{\infty} \langle C e_k,e_k\rangle_{_{\mathcal{H}_1}}=\sum_{k=0}^{\infty}  E_{\theta}\!\left\{\langle \mathcal{Z},e_k\rangle^2_{_{\mathcal{H}_1}}\right\}= E_{\theta}\!\left\{\|\mathcal{Z}\|_{_{\mathcal{H}_1}}^{\,2}\right\}< \infty.\]

\item[(iii)] Como $|g(u;\hat{\theta}_n)|\leq 1$, entonces \[|S_k^0(\boldsymbol{X}_i^*;\hat{\theta}_n;u)| \leq X^*_{ki}+\hat{\theta}_{kn}+\hat{\theta}_{3n}+\sqrt{2} \|\boldsymbol{\ell}(\boldsymbol{X}_i^*;\hat{\theta}_n)\|,\ \forall u \in [0,1]^2,\ k=1,2.\] Sea $\, 0<M=\int_{[0,1]^2}w(u)du<\infty$, luego
\[\langle Y_{ni}^*,e_k \rangle^2_{_{\mathcal{H}_1}}\leq\frac{4M}{n} \left(\|\boldsymbol{\ell}(\boldsymbol{X}^*_i;\hat{\theta}_n)\|+ \frac{1}{2}A_{i}\right)^2,\]
donde $A_{i}=(X^*_{1i}+\hat{\theta}_{1n}+\hat{\theta}_{3n})+ (X^*_{2i}+\hat{\theta}_{2n}+\hat{\theta}_{3n})$, siempre que $\hat{\theta}_n \in \Theta_0$, lo cual ocurre c.s. puesto que $\hat{\theta}_n \mathop{\longrightarrow}\limits^{c.s.} \theta$.

Ahora, como
\[\|\boldsymbol{\ell}(\boldsymbol{X}^*_i;\hat{\theta}_n)\|+ \frac{1}{2}A_i\leq
\left\{
\begin{array}{ll}
2\|\boldsymbol{\ell}(\boldsymbol{X}^*_i;\hat{\theta}_n)\|, & \hbox{si}\ \, A_i\leq 2\|\boldsymbol{\ell}(\boldsymbol{X}^*_i;\hat{\theta}_n)\|\\[.25 cm]
A_i, & \hbox{si}\ \, 2\|\boldsymbol{\ell}(\boldsymbol{X}^*_i;\hat{\theta}_n)\|< A_i
  \end{array}
\right.,
\]
entonces
\[\varepsilon< \left|\langle Y_{ni}^*,e_k \rangle_{_{\mathcal{H}_1}}\right| \leq \frac{2\sqrt{M}}{\sqrt{n}}\left(\|\boldsymbol{\ell}(\boldsymbol{X}^*_i; \hat{\theta}_n)\|+ \frac{1}{2}A_i\right),\]
implica que
\begin{align}
I\left\{\left|\langle Y_{ni}^*,e_k \rangle_{_{\mathcal{H}_1}}\right|>\varepsilon\right\} &\leq I\left\{\|\boldsymbol{\ell}(\boldsymbol{X}^*_i;\hat{\theta}_n)\|+ \frac{1}{2}A_i>\frac{\sqrt{n}\,\varepsilon}{2\sqrt{M}}\right\}\notag\\[.2 cm]
&= I\left\{\|\boldsymbol{\ell}(\boldsymbol{X}^*_i;\hat{\theta}_n)\|> \gamma, \ A_i\leq 2\|\boldsymbol{\ell}(\boldsymbol{X}^*_i;\hat{\theta}_n)\| \right\}\notag\\[.2 cm] &\qquad +I\left\{A_i>\frac{\sqrt{n}\,\varepsilon}{2\sqrt{M}},\ A_i> 2\|\boldsymbol{\ell}(\boldsymbol{X}^*_i;\hat{\theta}_n)\| \right\},\notag
\end{align}
donde $\gamma=\gamma(n,\varepsilon)=\frac{\sqrt{n}\,\varepsilon}{4\sqrt{M}}$ es tal que $\gamma\to \infty\,$ cuando $\,n\to \infty$, pues $\,0<M< \infty$. \vskip .3 cm

Por lo tanto,
\begin{align}
&\langle Y_{ni}^*,e_k\rangle^2_{_{\mathcal{H}_1}} I\left\{\left|\langle Y_{ni}^*,e_k \rangle_{_{\mathcal{H}_1}}\right|>\varepsilon\right\}\notag\\[.2 cm]
&\hspace{30 mm}\leq \frac{16M}{n} \ \|\boldsymbol{\ell}(\boldsymbol{X}^*_i;\hat{\theta}_n)\|^2 \,I\left\{\|\boldsymbol{\ell}(\boldsymbol{X}^*_i;\hat{\theta}_n)\|> \gamma,\ A_i\leq 2\|\boldsymbol{\ell}(\boldsymbol{X}^*_i;\hat{\theta}_n)\| \right\}\notag\\[.2 cm]
&\hspace{30 mm}\qquad+\frac{4M}{n}A_i^2\,I\left\{A_i>\frac{\sqrt{n}\,\varepsilon}{2\sqrt{M}}, \ A_i> 2\|\boldsymbol{\ell}(\boldsymbol{X}^*_i;\hat{\theta}_n)\| \right\}.\notag
\end{align}

En consecuencia,
\[L_n(\varepsilon, e_k)=\sum_{i=1}^n E_*\!\left(\langle Y_{ni}^*,e_k\rangle^2_{_{\mathcal{H}_1}} I{\left\{\left|\langle Y_{ni}^*,e_k \rangle_{_{\mathcal{H}_1}}\right|>\varepsilon\right\}}\right)\leq L_{1n}(\varepsilon, e_k)+L_{2n}(\varepsilon, e_k),\]
donde,
\begin{align} L_{1n}(\varepsilon, e_k)&=\frac{16M}{n}\sum_{i=1}^n E_*\!\left[\|\boldsymbol{\ell}(\boldsymbol{X}^*_1;\hat{\theta}_n)\|^2 \,I{\left\{\|\boldsymbol{\ell}(\boldsymbol{X}^*_1;\hat{\theta}_n)\|> \gamma,\ A_1\leq 2\|\boldsymbol{\ell}(\boldsymbol{X}^*_1;\hat{\theta}_n)\| \right\}}\right]\notag\\[.2 cm] &=16M E_*\!\left[\|\boldsymbol{\ell}(\boldsymbol{X}^*_1;\hat{\theta}_n)\|^2 \,I{\left\{\|\boldsymbol{\ell}(\boldsymbol{X}^*_1;\hat{\theta}_n)\|> \gamma,\ A_1\leq 2\|\boldsymbol{\ell}(\boldsymbol{X}^*_1;\hat{\theta}_n)\| \right\}}\right]\notag\\[.2 cm] &\leq 16M \sup_{\hat{\theta}_n\in \Theta_0}E_{\hat{\theta}_n}\!\!\left[\|\boldsymbol{\ell} (\boldsymbol{X}^*_1;\hat{\theta}_n)\|^2\, I{\left\{\|\boldsymbol{\ell} (\boldsymbol{X}^*_1;\hat{\theta}_n)\| >\gamma\right\}}\right]\notag \end{align} y como $\gamma\to \infty$ cuando $n\to \infty$, entonces, por el Supuesto \ref{E(l*l)-l-cont}(1) se concluye que
\begin{equation}\label{L1n}\lim_{n\to \infty}L_{1n}(\varepsilon,e_k)=0.\end{equation}

Considerando la desigualdad de H\"{o}lder y el hecho que la funci\'on indicadora es tal que $0\leq I\{C\}\leq 1$, podemos escribir \begin{align} L_{2n}(\varepsilon, e_k)&=\frac{4M}{n}\sum_{i=1}^n E_*\!\left(A_1^2\,I{\left\{A_1>\frac{\sqrt{n}\,\varepsilon}{2\sqrt{M}},\ A_1> 2\|\boldsymbol{\ell}(\boldsymbol{X}^*_1;\hat{\theta}_n)\| \right\}}\right)\notag\\[.2 cm]
&=4M E_*\!\left(A_1^2\,I{\left\{A_1>\frac{\sqrt{n}\,\varepsilon}{2\sqrt{M}},\ A_1> 2\|\boldsymbol{\ell}(\boldsymbol{X}^*_1;\hat{\theta}_n)\| \right\}}\right)\notag\\[.2 cm]
&\leq 4M \left\{E_*\!\left(A_1^3\right)\right\}^{2/3}\left[ E_*\!\left(I{\left\{A_1>\frac{\sqrt{n}\,\varepsilon}{2\sqrt{M}},\ A_1> 2\|\boldsymbol{\ell}(\boldsymbol{X}^*_1;\hat{\theta}_n)\| \right\}}\right)\right]^{1/3}\notag\\[.2 cm]
&= 4M \left\{E_*\!\left(A_1^3\right)\right\}^{2/3}\left\{ P_*\!\left(A_1>t \right)\right\}^{1/3},\notag \end{align} donde $t=t(n,\varepsilon)=\frac{\sqrt{n}\,\varepsilon}{2\sqrt{M}}$ es tal que $t\to \infty$ cuando $n\to \infty$, pues $\,0<M< \infty$.\vskip .2 cm

Puesto que $|A_1|=A_1$ y $E_*\!\left(A_1^3\right)< \infty$, el Corolario 1.14 (ii) en Serfling (1980, p. 47) implica que $P_*\!\left(A_1>t \right)\to 0$ cuando $t\to \infty$. Adem\'as, como $\, 0<M< \infty$, entonces
\begin{equation}\label{L2n}
\lim_{n\to \infty}L_{2n}(\varepsilon,e_k)=0.\end{equation}

De (\ref{L1n}) y (\ref{L2n}), concluimos que
 \[\lim_{n\to \infty}L_n(\varepsilon,e_k)=0,\]
de donde  se cumple la condici\'on (iii) de Kundu et al. (2000).
\end{itemize}
As\'i, del Lema \ref{kundu}, tenemos que
$Y_n^*\mathop{\longrightarrow} \limits^{L} \mathcal{Z}$, c.s.,  en $\mathcal{H}_1$. Como hab\'iamos mencionado antes, $\mathcal{Z}$ es tambi\'en el l\'imite d\'ebil de $(S_{1n},S_{2n})$  cuando los datos  son iid del vector aleatorio $\boldsymbol{X}\sim PB(\theta)$. Finalmente, el resultado sigue del teorema de la aplicaci\'on continua. $\square$

\begin{observacion}
Es importante observar que el resultado del Teorema \ref{Cons-boot-Rnw-Snw} se cumple si $H_0$ es verdadera o no.

Si de hecho, $H_0$ es verdadera, el Teorema \ref{Cons-boot-Rnw-Snw} implica que la distribuci\'on condicional de $R^*_{n,w}(\hat{\theta}^{\,*}_n)$ est\'a cercana c.s. a la distribuci\'on nula de $R_{n,w}(\hat{\theta}_n)$.

Si $H_0$ es falsa, entonces el Teorema \ref{Cons-boot-Rnw-Snw} nos dice que la distribuci\'on condicional de $R^*_{n,w}(\hat{\theta}^{\,*}_n)$ y la distribuci\'on de $R_{n,w}(\hat{\theta}_n)$, cuando la muestra es tomada de una poblaci\'on con distribuci\'on $PB(\theta)$, est\'an c.s. cercanas, donde $\theta$ es el l\'imite c.s. de $\hat{\theta}_n$.
\end{observacion}
Sea
$$r^*_{n,w,\alpha}=\inf\!\left\{x:P_*(R^*_{n,w}(\hat{\theta}^*_n)\geq x)\leq \alpha\right\}$$
el percentil superior $\alpha$ de la distribuci\'on bootstrap de $R_{n,w}(\hat{\theta}_n)$.

Del Teorema \ref{Cons-boot-Rnw-Snw}, la funci\'on test
$$\Psi^*_R=\left\{
\begin{array}{ll}
1, & \text{si}\ R_{n,w}(\hat{\theta}_n)\geq r^*_{n,w,\alpha}, \\[.15 cm]
0, & \text{en caso contrario},
\end{array}
\right.$$
o equivalentemente, el test que rechaza $H_0$ cuando
$$p_{_R}^*=P_*\!\left(R^*_{n,w}(\hat{\theta}^{\,*}_n) \geq R_{obs}\right)\leq \alpha,$$
es asint\'oticamente correcto en el sentido que, cuando $H_0$ es cierta
$$\lim_{n\to \infty} P_{\theta}\!\left(\Psi^*_R=1\right)=\alpha,$$
donde $R_{obs}$ es el valor observado del test estad\'istico $R_{n,w} (\hat{\theta}_n)$.\\

Similarmente, sea
$$s^*_{n,w,\alpha}=\inf\!\left\{x:P_*(S^*_{n,w}(\hat{\theta}^*_n)\geq x)\leq \alpha\right\}$$
el percentil superior $\alpha$ de la distribuci\'on bootstrap de $S_{n,w}(\hat{\theta}_n)$.

Del Teorema \ref{Cons-boot-Rnw-Snw}, la funci\'on test
$$\Psi^*_S=\left\{
\begin{array}{ll}
1, & \text{si}\ S_{n,w}(\hat{\theta}_n)\geq s^*_{n,w,\alpha}, \\[.15 cm]
0, & \text{en caso contrario},
\end{array}
\right.$$
o equivalentemente, el test que rechaza $H_0$ cuando
$$p_{_S}^*=P_*\!\left(S^*_{n,w}(\hat{\theta}^{\,*}_n) \geq S_{obs}\right)\leq \alpha,$$
es asint\'oticamente correcto en el sentido que, cuando $H_0$ es cierta
$$\lim_{n\to \infty} P_{\theta}\!\left(\Psi^*_S=1\right)=\alpha,$$
donde $S_{obs}$ es el valor observado del test estad\'istico $S_{n,w} (\hat{\theta}_n)$.\\

Como es usual, en la pr\'actica, $r^*_{n,w,\alpha}\,$ o $\,p_{_R}^*$ deben ser aproximados por simulaci\'on. Dados $\boldsymbol{X}_1, \boldsymbol{X}_2,\ldots,\boldsymbol{X}_n$ vectores aleatorios iid de $\boldsymbol{X}=(X_1,X_2)\in\mathbb{N}_0^2$, siguiremos los siguientes pasos:
\begin{itemize}
\item [$(1)$] Calcular el estimador $\hat{\theta}_n$ y calcular $R_{obs}$, el valor de $R_{n,w}(\hat{\theta}_n)$ para la muestra original.
\item [$(2)$] Generar una muestra bootstrap, digamos, $\boldsymbol{X}_1^*, \boldsymbol{X}_2^*,\ldots,\boldsymbol{X}_n^*$ iid desde la distribuci\'on $PB(\hat{\theta}_n)$.
\item [$(3)$] Sobre la base del bootstrap, calcular el estimador $\hat{\theta}^{\,*}_n$ y el valor del test estad\'istico, digamos $R^*$.
\item [$(4)$] Calcular $R_{n,w}(\hat{\theta}^*_n)$ para cada muestra bootstrap y denotar por $R^*_b,\ b=1,2,\ldots,B$, el respectivo valor resultante.
\item [$(5)$] Aproximar el $p-$valor bootstrap, $p_{_R}^*$, por medio de la expresi\'on $$\hat{p}_{_R}=\frac{card\{b:R^*_b\geq R_{obs}\}}{B},$$ o aproximar el punto cr\'itico, $r^*_{n,w,\alpha}$, por $R^*_{a:B}$, donde $a=[(1-\alpha)B]+1$, $[x]$ es la parte entera de $x$, y $R^*_{1:B}, R^*_{2:B},\ldots,R^*_{B:B}$ son los valores $R^*_b,\ b=1,2,\ldots,B$, en orden creciente.
\end{itemize}

\begin{observacion}\label{Boot-Snw} Para aproximar $s^*_{n,w,\alpha}$ se sigue el mismo procedimiento anterior, con los cambios obvios.
\end{observacion}

\section{Alternativas} \label{Alternativas_Rnw-Srw}
En esta secci\'on estudiaremos el comportamiento de los tests propuestos $\Psi^*_R$ y $\Psi^*_S$ bajo alternativas fijas y locales.

\subsection{Alternativas fijas} \label{Alternativas-fijas}
Como uno de nuestros objetivos es la consistencia de los tests de bondad de ajuste, lo pr\'oximo que haremos es estudiar este t\'opico para los tests que proponemos. Con este prop\'osito, primero obtendremos el l\'imite c.s. de  $\frac{1}{n} R_{n,w}$ y $\frac{1}{n} S_{n,w}$.

\begin{teorema}\label{Cons-Rnw-Snw}
Sean $\boldsymbol{X}_1, \boldsymbol{X}_2,\ldots,\boldsymbol{X}_n$ vectores aleatorios iid de $\boldsymbol{X}=(X_1,X_2)\in\mathbb{N}_0^2$ con fgp $g(u)$. Si $\hat{{\theta}}_n \mathop{\longrightarrow}\limits^{c.s.} \theta$, para alg\'un ${\theta}\in \mathbb{R}^3$, entonces
\begin{itemize}
\item[(a)] $\displaystyle\frac{1}{n}R_{n,w}(\hat{{\theta}}_n)\ \mathop{\longrightarrow}\limits^{c.s.} \ \int_0^1\int_0^1\Bigl\{g(u)- g(u;{\theta})\Bigr\}^2 w(u) du =\eta(g;{\theta}).$
\item[(b)] Si $g(u) \in G_2$,
$\displaystyle\frac{1}{n} S_{n,w}(\hat{\theta}_n) \ \mathop{\longrightarrow}\limits^{c.s.}\int_0^1\int_0^1 \left\{D^2_1(u;\theta)+D^2_2(u;\theta)\right\}w(u)du=\xi(g;\theta).$
\end{itemize}
\end{teorema}

\nt {\bf Demostraci\'on} \hspace{2pt} Solamente presentaremos la demostraci\'on de la parte (b), pues la demostraci\'on de la parte (a) sigue pasos similares.

Por definici\'on
\[\frac{1}{n}S_{n,w}(\hat{\theta}_n)=\|D_{1n}\|^2_{_\mathcal{H}}+ \|D_{2n}\|^2_{_\mathcal{H}},\]
donde $D_{kn}=D_{kn}(u;\hat{\theta}_n ), k=1,2$, est\'an definidas en (\ref{EDPs-empiricas}).

De las ecuaciones (\ref{EDPs-fgp}) y (\ref{EDPs-empiricas}), obtenemos
\begin{align}
\left|D_{1n}(u;\hat{\theta}_n )-D_1(u;\theta)\right|& \leq r_1+|\theta_1-\hat{\theta}_{1n}| + |\theta_3-\hat{\theta}_{3n}|+ |\hat{\theta}_{1n}+\hat{\theta}_{3n}|r_0, \notag\\
\left|D_{2n}(u;\hat{\theta}_n )-D_2(u;\theta)\right|&
\leq r_2+|\theta_2-\hat{\theta}_{2n}| + |\theta_3-\hat{\theta}_{3n}|+ |\hat{\theta}_{2n}+\hat{\theta}_{3n}|r_0,\notag
\end{align}
$\forall u \in [0,1]^2$, donde
\[r_0=\sup_{u\in [0,1]^2}\left|g_n(u)- g(u)\right|\]
y
\[r_i=\sup_{u\in [0,1]^2}\left|\frac{\partial } {\partial u_i}g_n(u)-\frac{\partial }{\partial u_i}g(u)\right|,\ i=1,2.\]

De la Proposici\'on \ref{Conv-FuncGenProbBiv}, $r_i=o(1)$, $i=0,1,2$. El hecho que $|\theta_i-\hat{\theta}_{in}| =o(1)$, $i=1,2,3$, junto con (\ref{int-funcion-peso}), implican que
\[\frac{1}{n}S_{n,w}(\hat{\theta}_n)=\|D_{1}\|^2_{_\mathcal{H}}+ \|D_{2}\|^2_{_\mathcal{H}}+o(1),\]
lo que demuestra el resultado porque
$\|D_{1}\|^2_{_\mathcal{H}}+ \|D_{2}\|^2_{_\mathcal{H}}=\xi(g;\theta)$. $\square$\\

Como una consecuencia de los Teoremas \ref{ConvDebil-Rnw-Snw}, \ref{Cons-boot-Rnw-Snw} y \ref{Cons-Rnw-Snw}, el siguiente resultado da la consistencia de los tests $\Psi^*_R$ y $\Psi^*_S$.

\begin{corolario}\label{Altern-fijas-Rnw-Snw}
Sean $\boldsymbol{X}_1, \boldsymbol{X}_2,\ldots,\boldsymbol{X}_n$ vectores aleatorios iid de $\boldsymbol{X}\in\mathbb{N}_0^2$ con fgp $g(u)$. Supongamos que se cumplen las hip\'otesis de los Teoremas \ref{ConvDebil-Rnw-Snw} y \ref{Cons-boot-Rnw-Snw}.
\begin{itemize}
\item[(a)] Si  $\eta(g;\theta)>0$, entonces
$P(\Psi^*_R=1) \to 1.$
\item[(b)] Si $g(u) \in G_2$ y $\xi(g;\theta)>0$, entonces $P(\Psi^*_S=1) \to 1.$
\end{itemize}\end{corolario}

\begin{observacion}
Notar que $\eta(g;{\theta})\geq 0$ ($\xi(g;\theta) \geq 0$). Si $w>0$ en casi todo $[0,1]^2$, entonces $\eta(g;\theta)= 0$ ($\xi(g;\theta)= 0$) s\'i y s\'olo si $H_0$ es cierta.

Por lo tanto, la consistencia de los tests propuestos est\'a garantizada simplemente tomando una funci\'on de peso que sea positiva en casi todo $[0,1]^2$.
\end{observacion}

\subsection{Alternativas contiguas} \label{Alternativas-contiguas-Rnw-Swn}
El resultado de consistencia dado en el Corolario \ref{Altern-fijas-Rnw-Snw} no distingue entre medidas de probabilidad alternativas de $P$. Una mejor discriminaci\'on se obtiene considerando alternativas para las cuales la potencia tiende a valores menores o iguales que 1. Esto se logra reemplazando una alternativa fija por una sucesi\'on de alternativas que converge a la nula a una cierta velocidad, que se suelen denominar alternativas contiguas.\vskip .2 cm

Con este objetivo consideremos ahora un arreglo triangular $\boldsymbol{X}_{n,1}, \boldsymbol{X}_{n,2},\ldots,\boldsymbol{X}_{n,n}$ de vectores aleatorios bivariantes independientes por filas que toman valores en $\mathbb{N}_0^2$ y funci\'on de probabilidad conjunta $P_n(x_1,x_2)$ dada por
\begin{equation}\label{P(n)=P(theta)+b}
P_n(x_1,x_2)=P_{\theta}(x_1,x_2)\left\{1+\frac{1}{\sqrt{n}} \, b_n(x_1,x_2)\right\},
\end{equation}
donde $P_{\theta}(x_1,x_2)$ es la funci\'on de probabilidad del vector aleatorio bivariante que tiene una distribuci\'on $PB(\theta)$, para alg\'un $\theta \in \Theta$, y $b_n(x_1,x_2)$ satisface las siguientes condiciones.
\begin{supuesto}\label{b(x,y)}
\begin{enumerate}
  \item [$(1)$] $E_{\theta}\{b_n(X_1,X_2)\}=0$, $\forall n$.
  \item [$(2)$] $b_n(x_1,x_2)\to b(x_1,x_2)$, $\forall (x_1,x_2)\in \mathbb{N}_0^2$.
  \item [$(3)$] $\displaystyle \sup_{n} E_{\theta}\!\left\{b_n(X_1,X_2)^4\right\} <\infty$.
\end{enumerate}
\end{supuesto}

El Supuesto \ref{b(x,y)}(1) asegura que $\sum_{x_1,\,x_2 \in \mathbb{N}_0}P_n(x_1,x_2)=1$; los Supuestos \ref{b(x,y)}(2) y (3) contienen condiciones de tipo t\'ecnico que usaremos en las demostraciones.\vskip .2 cm

El Teorema \ref{ConvDebil-Rnw-Snw} establece que, cuando $H_0$ es cierta, $R_{n,w}(\hat{\theta}_n)$ converge en distribuci\'on a una combinaci\'on lineal de variables $\chi^2$ independientes con 1 grado de libertad, donde los pesos son los autovalores del operador $C_R(\theta)$ dado en (\ref{Operador-C}). Sea $\{\phi_j^R\}$ el conjunto de autofunciones ortonormales correspondiente a los autovalores $\{\lambda_j^R\}$ de $C_R(\theta)$.\vskip .2 cm

Observar que lo expuesto anteriormente para $R_{n,w}(\hat{\theta}_n)$, tambi\'en es v\'alido para $S_{n,w}(\hat{\theta}_n)$, considerando el operador $C_S(\theta)$ con sus conjuntos $\{\phi_j^S\}$ y $\{\lambda_j^S\}$ de autofunciones ortonormales y de autovalores, respectivamente.\vskip .2 cm

El siguiente Teorema da la ley l\'imite de estos estad\'isticos bajo las alternativas $P_n(x_1,x_2)$ en (\ref{P(n)=P(theta)+b}).

\begin{teorema}\label{Altern-contiguas-S_n,w}
Sea $\boldsymbol{X}_{n,1}, \boldsymbol{X}_{n,2},\ldots,\boldsymbol{X}_{n,n}$ un arreglo triangular de vectores aleatorios bivariantes que son independientes por filas y que toman valores en $\mathbb{N}_0^2$, con funci\'on de probabilidad dada por $P_n(x_1,x_2)$ definida en (\ref{P(n)=P(theta)+b}). Supongamos que se cumplen los Supuestos \ref{hat(theta)-theta} y \ref{b(x,y)}. Entonces
\begin{itemize}
\item[(a)] $R_{n,w}(\hat{\theta}_n)\, \mathop{\longrightarrow} \limits^{L} \, \displaystyle\sum_{k=1}^{\infty} \lambda_k^R\!\left(Z_k+c_k^R\right)^2,$
donde $\,c_k^R=\mathop{\sum}\limits_{x_1,x_2}b(x_1,x_2)\,\phi_k^R(x_1,x_2)$ y $Z_1,Z_2,\ldots\,$ son variables normales est\'andar independientes.
\item[(b)] \vskip .3 cm $S_{n,w}(\hat{\theta}_n)\ \mathop{\longrightarrow} \limits^{L} \ \displaystyle\sum_{k=1}^{\infty} \lambda_k^S\!\left(Z_k+c_k^S\right)^2,$
donde $\,c_k^S=\mathop{\sum}\limits_{x_1,x_2}b(x_1,x_2)\,\phi_k^S(x_1,x_2)\,$ y $Z_1,Z_2,\ldots\,$ son variables normales est\'andar independientes.
\end{itemize}
\end{teorema}
\nt {\bf Demostraci\'on} \hspace{2pt}
Sea $A_n \subseteq \mathbb{N}_0^2$ tal que $P_{\theta}(A_n)\to 0$, entonces
\[ P_n(A_n)=P_{\theta}(A_n)+\frac{1}{\sqrt{n}}\sum_{(x_1,x_2)\in A_n}P_{\theta}(x_1,x_2)b_n(x_1,x_2).\]
De la desigualdad de  H\"older y del Supuesto \ref{b(x,y)}(3) se sigue que
\begin{align}
&\sum_{(x_1,x_2)\in A_n}P_{\theta}(x_1,x_2)b_n(x_1,x_2)\notag\\
&\hspace{35mm}\leq \left\{\sum_{(x_1,x_2)\in A_n}P_{\theta}(x_1,x_2) \right\}^{3/4}
\left\{\sum_{(x_1,x_2)\in A_n}P_{\theta}(x_1,x_2)b_n(x_1,x_2)^4 \right\}^{1/4} \notag\\[.25 cm]
&\hspace{35mm}\leq \sup_{n} E_{\theta}\left\{b_n(X_1,X_2)^4\right\} <\infty,\notag
\end{align}
y por tanto $P_n(A_n)\to 0$, esto es, $P_n$ es contigua a $P_{\theta}$.

A continuaci\'on  veremos que, bajo $P_n$,  $\sqrt{n}(\hat{\theta}_n-\theta)$ converge en ley a una distribuci\'on normal. 
Con este objetivo, sean $\boldsymbol{X}_1, \boldsymbol{X}_2, \ldots, \boldsymbol{X}_n$ vectores aleatorios iid con funci\'on de probabilidad com\'un $P_{\theta}$ y sea
\[l_n=\log \frac{P_n(\boldsymbol{X}_1)P_n(\boldsymbol{X}_2)\ldots P_n(\boldsymbol{X}_n)}
{P_{\theta}(\boldsymbol{X}_1)P_{\theta}(\boldsymbol{X}_2)\ldots P_{\theta}(\boldsymbol{X}_n)}.\]
Sea $Z_{n,i}=b_n(\boldsymbol{X}_i)$, $1\leq i \leq n$. Entonces, por desarrollo en serie de Taylor,
\begin{equation} \label{ln}
l_n=\sum_{i=1}^n \log\left(1+\frac{1}{\sqrt{n}}Z_{n,i} \right)=\frac{1}{\sqrt{n}}\sum_{i=1}^n Z_{n,i}-\frac{1}{2n}\sum_{i=1}^n Z_{n,i}^2+\varrho_n,
\end{equation}
donde $\varrho_n=\varrho_n(Z_{n,1},Z_{n,2}, \ldots, Z_{n,n})$,
\[\varrho_n=\sum_{i=1}^n \frac{2}{3!\,n\sqrt{n}\,a^3_{in}} Z_{n,i}^3,\]
con $a_{in}=a_{in}(Z_{n,i})$
\[a_{in}=1+\frac{\alpha_i}{\sqrt{n}}\,Z_{n,i},\]
para alg\'un $0<\alpha_i<1$, $1\leq i \leq n$.
A continuaci\'on estudiamos el l\'imite de cada uno de los sumandos en el lado derecho de (\ref{ln}).

Sea $\varepsilon$ una constante positiva. Adem\'as, sea $M_n=\sqrt{n}/2$ y consideremos $\tilde{\varrho}_n=\varrho_n(Z_{n,1}I\{|Z_{n,1}|\leq M_n\}, \ldots, Z_{n,n}I\{|Z_{n,n}|\leq M_n\})$.

Notemos que $\tilde{a}_{in}=a_{in}(Z_{n,i}I\{|Z_{n,i}|\leq M_n\})\geq \frac{1}{2}, 1\leq i \leq n$, y por lo tanto
\[E_{\theta}(|\tilde{\varrho}_n|)\leq \frac{2^3}{3}\frac{1}{\sqrt{n}} \,\displaystyle \sup_m E_{\theta}\! \left\{|b_m(\boldsymbol{X}_1)|^3 \right\},\]
lo cual implica que
\begin{equation} \label{resto1}
P_{\theta}(|\tilde{\varrho}_n|>\varepsilon )\to 0.
\end{equation}
Como
\begin{align}
P_{\theta}\left(|\tilde{\varrho}_{n} -\varrho_{n}|>\varepsilon\right) &\leq P_{\theta}\left(\bigcup_{i=1}^n\left\{|Z_{n,i}|> M_n \right\}\right)\notag\\
&\leq \sum_{i=1}^n P_{\theta}\left(|Z_{n,i}|> M_n \right)\notag\\[.15 cm]
&\leq \frac{n}{M_n^4}\,\sup_m E_{\theta}\!\left\{b_m(\boldsymbol{X}_1)^4 \right\}\to 0.\label{resto2}
\end{align}
De (\ref{resto1}) y (\ref{resto2}), concluimos que
\begin{equation} \label{resto}
P_{\theta}\left(|\varrho_{n}|>\varepsilon\right)\to 0.
\end{equation}

Por el Supuesto \ref{b(x,y)},
$E_{\theta}(Z_{n,i})=0,$
y
\[\sigma_n^2=Var_{\theta}(Z_{n,i}) \to \sigma^2=E_{\theta}\left\{b(X_1,X_2)^2\right\}<\infty.\]
Adem\'as,
\begin{align}
E_{\theta}\left[Z_{n,i}^2\,I\{|Z_{n,i}|>na\} \right] &\leq E_{\theta}^{1/2}\left(Z_{n,i}^4 \right)P^{1/2}_{\theta}\!
\left(|Z_{n,i}|>na\right)\notag\\[.2 cm]
&\leq\frac{1}{(na)^4}\left[\sup_n E_{\theta}\left\{b_n(X_1,X_2)^4 \right\} \right]^2\to 0, \quad \forall a>0.\notag
\end{align}
Por el teorema central del l\'imite para arreglos triangulares (Teorema 1.9.3 en Serfling, 1980, pp. 31-32, \cite{Ser80}) se sigue que
\begin{equation}\label{Zi-conv-N}
\frac{1}{\sqrt{n}}\sum_{i=1}^n Z_{n,i}\ \mathop{\longrightarrow} \limits^{L}\ N(0,\sigma^2).
\end{equation}

Se tiene que
\[E_{\theta}\left(\frac{1}{n}\sum_{i=1}^n Z_{n,i}^2\right)=\sigma_n^2\to \sigma^2,\]
\[Var_{\theta}\left(\frac{1}{n}\sum_{i=1}^n Z_{n,i}^2\right)=\frac{Var_{\theta}(Z_{n,i}^2)}{n}\leq \frac{1}{n}\sup_n E_{\theta}\left\{b_n(X_1,X_2)^4\right\}\to 0,\]
de donde
\begin{equation}\label{Zi^2-conv-sigma^2}
\frac{1}{2n}\sum_{i=1}^n Z_{n,i}^2 \mathop{\longrightarrow} \limits^{P}\  \frac{1}{2}\sigma^2.
\end{equation}
Ahora, del Supuesto \ref{hat(theta)-theta}, (\ref{ln}) y (\ref{resto})--(\ref{Zi^2-conv-sigma^2}), la sucesi\'on
\[\left(  \sqrt{n}(\hat{\theta}_n-\theta), l_n \right)\]
converge en ley a una distribuci\'on normal multivariante de dimensi\'on 4, cuando los datos provienen de $P_{\theta}$. As\'i, por el tercer Lema de Le Cam (Corolario 12.3.2 en Lehmann y Romano (2005) \cite{LehRom05}), concluimos que, cuando los datos provienen de $P_n$, $\sqrt{n}(\hat{\theta}_n-\theta)$ converge en distribuci\'on a una ley normal, lo cual implica que es acotada en probabilidad.\vskip .2 cm

Siguiendo pasos similares a los dados en la demostraci\'on del Teorema \ref{ConvDebil-Rnw-Snw} (b), podemos probar que (\ref{Snw-Sn}) tambi\'en se cumple cuando los datos provienen de $P_n$, con $P_n(|s_n|>\varepsilon)\to 0$, $\forall \varepsilon>0$.

As\'i, cuando los datos tienen la funci\'on de probabilidad $P_n$, aplicando el Teorema 2.3 en Gregory (1977) \cite{Gre77}, obtenemos que
\[\|S_{1n}\|_{_\mathcal{H}}^2+\|S_{2n}\|_{_\mathcal{H}}^2 \mathop{\longrightarrow} \limits^{L} \ \sum_{k=1}^{\infty} \lambda_k^S\left(Z_k+c_k^S\right)^2,\]
y con ello se consigue el resultado en (b). La demostraci\'on del resultado en (a) sigue los mismos pasos. $\square$\vskip .2 cm

Del Teorema \ref{Altern-contiguas-S_n,w}, concluimos que el test $\Psi^*_R$ ($\Psi^*_S$) es capaz de detectar alternativas como las establecidas en (\ref{P(n)=P(theta)+b}), que convergen a la DPB a una raz\'on de $n^{-1/2}$.


\chapter{Estad\'istico $\boldsymbol{W_n(\hat{\theta}_n)}$}\label{Estadistico-Wn}

\section{Definici\'on del test estad\'istico} \label{Def-Test-estadistico}

En este cap\'itulo proponemos otro test para contrastar la hip\'otesis nula, cuyo test estad\'istico lo deduciremos de la Proposici\'on \ref{Soluc-sistema-de-dos-EDP}. Para ello consideremos que se verifican las hip\'otesis en dicha proposici\'on.

Sea $(X_1,X_2)\in \mathbb{N}^2_0$ un vector aleatorio y sea $g(u_1,u_2)=E\left(u_1^{X_1} u_2^{X_2}\right)$ su fgp, luego, por definici\'on
\[g(u)=\sum_{r,s\geq 0}u_1^r u_2^s P(r,s),\]
donde $P(r,s)=P(X_1=r,X_2=s)$.\vskip .2 cm

De la ecuaci\'on anterior y de las definiciones de $D_1(u;\theta)$ y $D_2(u;\theta)$ dadas en la Proposici\'on \ref{Soluc-sistema-de-dos-EDP}, podemos escribir
\begin{align}
D_1(u;\theta)&=\sum_{r\geq 0}\sum_{s\geq 0}\left\{(r+1)P(r+1,s)-(\theta_1-\theta_3)P(r,s)-\theta_3P(r,s-1)\right\}u_1^r u_2^s,\notag \\[.2 cm]
D_2(u;\theta) &=\sum_{r\geq 0}\sum_{s\geq 0}\left\{(s+1)P(r,s+1)-(\theta_2-\theta_3)P(r,s)-\theta_3P(r-1,s)\right\}u_1^r u_2^s.\notag
\end{align}

Consideremos ahora las versiones emp\'iricas de las ecuaciones anteriores, esto es, consideremos $D_{1n}(u;\hat{\theta}_n)$ y $D_{2n}(u;\hat{\theta}_n)$ definidas en (\ref{EDPs-empiricas}). Si $H_0$ fuera cierta entonces $D_{1n}(u;\hat{\theta}_n)$ y $D_{2n}(u;\hat{\theta}_n)$ deber\'ian ser pr\'oximas a 0, $\forall u \in [0,1]^2$. Esta proximidad puede interpretarse de varias formas. En el cap\'itulo anterior ya vimos que esto era equivalente a
$\int \{D_{1n}(u;\hat{\theta}_n)^2+D_{2n}(u;\hat{\theta}_n)^2\}w(u)du\approx 0$.\vskip .2 cm

Veamos otra interpretaci\'on siguiendo un razonamiento similar al hecho en Nakamura y P\'erez-Abreu (1993) \cite{NaPe93} para el caso univariante, que presentamos al final de la Secci\'on \ref{Test-bondad-ajuste-dim-1}. Para ello observemos que
\begin{align}
D_{1n}(u;\hat{\theta}_n)&=\sum_{r\geq 0}\sum_{s\geq 0}  d_1(r,s;\hat{\theta}_n) u_1^ru_2^s,\notag\\[.2 cm]
D_{2n}(u;\hat{\theta}_n)&=\sum_{r\geq 0}\sum_{s\geq 0} d_2(r,s;\hat{\theta}_n) u_1^ru_2^s,\notag
\end{align}
donde
\begin{align}
d_1(r,s;\hat{\theta}_n)&=(r+1)p_n(r+1,s)-(\hat{\theta}_{1n}-\hat{\theta}_{3n}) p_n(r,s)-\hat{\theta}_{3n} p_n(r,s-1),\notag\\[.2 cm]
d_2(r,s;\hat{\theta}_n)&=(s+1)p_n(r,s+1)-(\hat{\theta}_{2n}-\hat{\theta}_{3n}) p_n(r,s)-\hat{\theta}_{3n} p_n(r-1,s),\notag
\end{align}
y
\[p_n(r,s)=\frac{1}{n}\sum_{k=1}^n I{\{X_{1k}=r,X_{2k}=s\}}\]
es la frecuencia relativa emp\'irica del par $(r,s)$. Por tanto, $D_{in}(u;\hat{\theta}_n)= 0$,
$\forall u \in [0,1]^2$, $i=1,2$, s\'i y s\'olo si los coeficientes de $u_1^ru_2^s$ en las expansiones anteriores son  nulos $\forall r,s \geq 0$. Esto nos lleva a considerar el siguiente estad\'istico para contrastar $H_0$:
\begin{equation}\label{Estad-Wn-bivariante}
W_n(\hat{\theta}_n)=\sum_{r\geq 0}\sum_{s\geq 0}\{d_1(r,s;\hat{\theta}_n)^2+d_2(r,s;\hat{\theta}_n)^2\}=
\sum_{r,s = 0}^M\{d_1(r,s;\hat{\theta}_n)^2+d_2(r,s;\hat{\theta}_n)^2\},
\end{equation}
donde $M=\max \{X_{1(n)}, X_{2(n)}\}$, $X_{k(n)}=\max_{1\leq i\leq n}X_{ki}$, $k=1,2$.

Teniendo en cuenta que
\[d_k(r,s;\hat{\theta}_n)=\frac{1}{n} \sum_{i=1}^n \phi_{krs}(\boldsymbol{X}_i; \hat{\theta}_n),\quad k=1,2,\]
con
\begin{equation}\label{psi_1rs-2rs}
\begin{array}{l}
\phi_{1rs}(x;{\theta})=(r+1)I{\{x_1=r+1,x_2=s\}}-(\theta_1-\theta_3) I{\{x_1=r,x_2=s\}}\\[.3 cm]
\hspace{85mm}-\theta_3I{\{x_1=r,x_2=s-1\}},\\[.5 cm]
\phi_{2rs}(x;{\theta})=(s+1)I{\{x_1=r,x_2=s+1\}}-(\theta_2-\theta_3) I{\{x_1=r,x_2=s\}}\\[.3 cm]
\hspace{85mm}-\theta_3I{\{x_1=r-1,x_2=s\}},
\end{array}
\end{equation}
donde $x=(x_1,x_2)$, entonces el estad\'istico $W_n(\hat{\theta}_n)$ puede ser expresado como sigue:
\[W_n(\hat{\theta}_n)=\frac{1}{n^2}\sum_{i,j=1}^n h(\boldsymbol{X}_i,\boldsymbol{X}_j; \hat{\theta}_n),\]
donde
\begin{equation}\label{definicion-h}
\begin{array}{l}
h(x,y; {\theta})=h_1(x,y; {\theta})+h_2(x,y; {\theta}),\\[.35 cm]
h_k(x,y; {\theta})=\displaystyle\sum_{r \geq 0}\sum_{s \geq 0} \phi_{krs}(x; {\theta}) \phi_{krs}(y; {\theta}),\quad k=1,2,
\end{array}
\end{equation}
con $x=(x_1,x_2)$ e $y=(y_1,y_2)$.\\

Un test razonable para contrastar $H_0$ deber\'ia rechazar la hip\'otesis nula para valores grandes de $W_n(\hat{\theta}_n)$. Ahora, para determinar qu\'e son los valores grandes, debemos calcular la distribuci\'on nula de $W_n(\hat{\theta}_n)$ o al menos una aproximaci\'on de ella.

Como la distribuci\'on nula de $W_n(\hat{\theta}_n)$ es desconocida, trataremos de estimarla empleando el modo cl\'asico, esto es, aproximaremos la distribuci\'on nula mediante la distribuci\'on asint\'otica nula. En la siguiente secci\'on estudiaremos esta situaci\'on.

\section[Aproximaci\'on de la distribuci\'on nula de $W_n(\hat{\theta}_n)$]{Aproximaci\'on de la distribuci\'on nula de $\boldsymbol{W_n(\hat{\theta}_n)}$}\label{Distribucion-nula-Wn}

\subsection{Distribuci\'on asint\'otica nula}\label{Distribucion-asintotica-nula-Wn}
El siguiente resultado proporciona la distribuci\'on asint\'otica nula de $W_n(\hat{\theta}_n)$.
\begin{teorema}\label{TeoConvDebil-Wn}
Sean  $\boldsymbol{X}_{1},\boldsymbol{X}_{2},\ldots, \boldsymbol{X}_{n}$ v.a. iid de $\boldsymbol{X}=(X_{1},X_{2})\sim PB(\theta_{1},\theta_{2},\theta_{3})$. Supongamos que se cumple el Supuesto \ref{hat(theta)-theta}. Entonces
\[nW_n(\hat{\theta}_n)=\frac{1}{n}\sum_{i,j=1}^n h_d(\boldsymbol{X}_i,\boldsymbol{X}_j; \theta)+\rho_n,\]
donde $P_{{\theta}}(|\rho_n|>\varepsilon)\to 0$, $\forall \varepsilon>0$,
\begin{equation}\label{nucleo_de_Wn_aproximado}
h_d(x,y;\theta)=h(x,y;\theta)+\boldsymbol{\ell}(x;\theta) S \boldsymbol{\ell}(y;\theta)^{\top},
\end{equation}
con $h(x,y;\theta)$ definido en (\ref{definicion-h}), $x=(x_1,x_2)$, $y=(y_1,y_2)$, $S=\sum_{r,s\geq 0}A_{rs}$ y $A_{rs}$ es la matriz sim\'etrica dada por
\begin{equation}\label{Matriz-Ars}
A_{rs}=\left(
\begin{array}{ccc}
a^2 & 0 & a(b-a) \\
0 & a^2 & a(c-a) \\
a(b-a) & a(c-a) & (b-a)^2+(c-a)^2
\end{array}
\right),
\end{equation}
con $\,a=P(r,s;\theta),\ b=P(r,s-1;\theta),\ c=P(r-1,s;\theta)$. Adem\'as
\[nW_n(\hat{\theta}_n)\ \mathop{\longrightarrow} \limits^{\!L}\ \sum_{j\geq 1}\lambda_j\chi^2_{1j},\]
donde $\chi^2_{11},\chi^2_{12},\ldots$ son v.a. independientes $\chi^2$ con 1 grado de libertad y el conjunto $\{\lambda_j\}$ son los autovalores no nulos del operador $C(\theta)$ definido sobre el espacio de funciones $\{\tau:\mathbb{N}_0^2\to \mathbb{R},  \text{tal que} \ E_{\theta}\!\left\{\tau^2(\boldsymbol{X})\right\}<\infty,\forall \theta\in\Theta\}$, como sigue
\begin{equation}\label{Operador_C_para_Wn}
C(\theta) \tau(x)= E_{\theta}\{h_d(x,\boldsymbol{Y};\theta) \tau(\boldsymbol{Y})\}.
\end{equation}
\end{teorema}
\nt {\bf Demostraci\'on} \hspace{2pt} Por definici\'on
\[W_n(\hat{\theta}_n)=\frac{1}{n^2}\sum_{i,j=1}^n h(\boldsymbol{X}_i,\boldsymbol{X}_j; \hat{\theta}_n).\]
Por desarrollo en serie de Taylor
\begin{align}
&W_n(\hat{\theta}_n)\notag\\[.2 cm]
&=\!\frac{1}{n^2}\!\sum_{i,j=1}^n\! \left\{\!h(\boldsymbol{X}_i,\boldsymbol{X}_j; \theta)\!+Q^{(1)}(\boldsymbol{X}_i,\boldsymbol{X}_j;\theta) (\hat{\theta}_n\!-\theta)^{\top}\!\!+\frac{1}{2}(\hat{\theta}_n\!-\theta)  Q^{(2)}(\boldsymbol{X}_i,\boldsymbol{X}_j;\theta) (\hat{\theta}_n\!-\theta)^{\top}\!\right\}\notag\\[.2 cm]
&\hspace{4mm}\left.+\frac{1}{3!\,n^2}\sum_{i,j=1}^n\, \sum_{k_1,k_2,k_3=1}^3 \! \frac{\partial^3} {\partial \vartheta_{k_1}\partial \vartheta_{k_2}\partial \vartheta_{k_3}}h(\boldsymbol{X}_i,\boldsymbol{X}_j; \vartheta)\right|_{\vartheta=\tilde{\theta}} \!\! (\hat{\theta}_{k_1n}\!-\theta_{k_1})(\hat{\theta}_{k_2n}\!-\theta_{k_2}) (\hat{\theta}_{k_3n}\!-\theta_{k_3}),\label{Wn-aprox}
\end{align}
donde $Q^{(k)}(x,y;\theta)$ representan las $k-$\'esimas derivadas de $h(x,y;\vartheta)$ respecto de $\vartheta$ evaluadas en $\theta$, $k=1,2$, y $\tilde{\theta}=\alpha \hat{\theta}_n+(1-\alpha)\theta$, para alg\'un $\,0<\alpha<1$.

Las derivadas de $h(x,y;\theta)$ est\'an dadas por
\begin{align}
\frac{\partial}{\partial \vartheta_{1}}h(x,y;\theta)&=-\sum_{r,s\geq 0}\Bigl[ I{\{x_1=r,x_2=s\}}\phi_{1rs}(y;\theta)+I{\{y_1=r,y_2=s\}} \phi_{1rs}(x;\theta)\Bigr],\notag\\[.13 cm]
\frac{\partial}{\partial \vartheta_{2}}h(x,y;\theta)&=-\sum_{r,s\geq 0}\Bigl[ I{\{x_1=r,x_2=s\}}\phi_{2rs}(y;\theta)+I{\{y_1=r,y_2=s\}}\phi_{2rs}(x;\theta) \Bigr],\notag\\[.13 cm]
\frac{\partial}{\partial \vartheta_{3}}h(x,y;\theta)&=\sum_{r,s\geq 0}\Bigl(I{\{x_1=r,x_2=s\}}-I{\{x_1=r,x_2=s-1\}}\Bigr) \phi_{1rs}(y;\theta)\notag\\ &\hspace{12mm}+\sum_{r,s\geq 0}\Bigl(I{\{y_1=r,y_2=s\}}-I{\{y_1=r,y_2=s-1\}}\Bigr) \phi_{1rs}(x;\theta)\notag\\
&\hspace{12mm}+\sum_{r,s\geq 0}\Bigl(I{\{x_1=r,x_2=s\}}-I{\{x_1=r-1,x_2=s\}}\Bigr) \phi_{2rs}(y;\theta)\notag\\ &\hspace{12mm}+\sum_{r,s\geq 0}\Bigl(I{\{y_1=r,y_2=s\}}-I{\{y_1=r-1,y_2=s\}}\Bigr) \phi_{2rs}(x;\theta),\notag\\[.1 cm]
\frac{\partial^2}{\partial \vartheta_{1}^2}h(x,y;\theta)&=2\sum_{r,s\geq 0} I{\{x_1=r,x_2=s\}}I{\{y_1=r,y_2=s\}},\notag\\
\frac{\partial^2}{\partial \vartheta_{1}\partial\vartheta_{2}}h(x,y;\theta)&=0,\notag\\[.13 cm]
\frac{\partial^2}{\partial \vartheta_{1}\partial\vartheta_{3}}h(x,y;\theta)&=-\!\sum_{r,s\geq 0} \Bigl[I{\{x_1=r,x_2=s\}}\bigl(I{\{y_1=r,y_2=s\}}-I{\{y_1=r,y_2=s-1\}}\bigr) \Bigr.\notag\\
&\hspace{10mm}\Bigl.+ I\{y_1=r,y_2=s\}\bigl(I{\{x_1=r,x_2=s\}}-I{\{x_1=r,x_2=s-1\}}\bigr) \Bigr],\notag\\[.1 cm]
\frac{\partial^2}{\partial \vartheta_{2}^2}h(x,y;\theta)&=2\sum_{r,s\geq 0} I{\{x_1=r,x_2=s\}}I{\{y_1=r,y_2=s\}},\notag\\
\frac{\partial^2}{\partial \vartheta_{2}\partial\vartheta_{3}}h(x,y;\theta)&=-\!\sum_{r,s\geq 0} \Bigl[I{\{x_1=r,x_2=s\}}(I{\{y_1=r,y_2=s\}}-I{\{y_1=r-1,y_2=s\}}) \Bigr.\notag\\
&\hspace{10mm}\Bigl. + I{\{y_1=r,y_2=s\}} \bigl(I{\{x_1=r,x_2=s\}}-I{\{x_1=r-1,x_2=s\}}\bigr)\Bigr],\notag\\[.1 cm]
\frac{\partial^2}{\partial \vartheta_{3}^2}h(x,y;\theta)&=2\sum_{r,s\geq 0}\Bigl[\bigl(I{\{x_1=r,x_2=s\}}-I{\{x_1=r,x_2=s-1\}}\bigr)\Bigr. \notag\\ &\hspace{17mm}\Bigl.\times\bigl(I{\{y_1=r,y_2=s\}}-I{\{y_1=r,y_2=s-1\}}\bigr) \Bigr] \notag\\
&\hspace{12mm} +2\sum_{r,s\geq 0}\Bigl[\bigl(I{\{x_1=r,x_2=s\}}- I{\{x_1=r-1,x_2=s\}}\bigr)\Bigr. \notag\\ &\hspace{29mm}\Bigl. \times\bigl(I{\{y_1=r,y_2=s\}}-I{\{y_1=r-1,y_2=s\}}\bigr)\Bigr].\notag
\end{align}
Adem\'as,
\[\frac{\partial^3} {\partial \vartheta_{i}\partial \vartheta_{j}\partial \vartheta_{k}}h(x,y;\theta)=0,\ 1\leq i,j,k\leq 3.\]

De lo anterior se obtiene que
\[\begin{array}{ll}
E_{\theta}\left\{Q^{(1)}(\boldsymbol{X}_1,\boldsymbol{X}_1; {\theta})\right\}&= 2(\theta_1-\theta_3,\theta_2-\theta_3,4\theta_3-\theta_1-\theta_2)<\infty,\\[.2 cm]
E_{\theta}\left\{Q^{(1)}(\boldsymbol{X}_1,\boldsymbol{X}_2; {\theta})\right\}&=\mathbf{0}, \ (\text{vector nulo de } \mathbb{R}^3),\\[.2 cm]
E_{\theta}\left\{Q^{(2)}(\boldsymbol{X}_1,\boldsymbol{X}_1; {\theta})\right\}&= -2 \left(
\begin{array}{rrr}
-1 & 0 & 1 \\
0 & -1 & 1 \\
1 & 1 & -4
\end{array}
\right)<\infty,
\\[.2 cm]
E_{\theta}\left\{Q^{(2)}(\boldsymbol{X}_1,\boldsymbol{X}_2; {\theta})\right\}&= 2 \sum_{r,s\geq 0} A_{rs}<\infty,
\end{array}\]
donde $A_{rs}$ es la matriz definida en (\ref{Matriz-Ars}).

Puesto que
\[\frac{1}{n^2}\sum_{i,j=1}^n Q^{(k)}(\boldsymbol{X}_i,\boldsymbol{X}_j;\theta)=
\frac{1}{n^2}\sum_{i\neq j} Q^{(k)}(\boldsymbol{X}_i,\boldsymbol{X}_j; {\theta})+
\frac{1}{n^2}\sum_{i=1}^n Q^{(k)}(\boldsymbol{X}_i,\boldsymbol{X}_i; {\theta}),\ k=1,2,\]
y por la ley fuerte de los grandes n\'umeros para U-estad\'isticos (v\'ease por ejemplo Teorema 5.4 en Serfling (1980) \cite{Ser80}),
\[\frac{1}{n^2}\sum_{i\neq j} Q^{(k)}(\boldsymbol{X}_i,\boldsymbol{X}_j; {\theta})\mathop{\longrightarrow}\limits^{c.s.}
E_{\theta}\left\{Q^{(k)}(\boldsymbol{X}_1,\boldsymbol{X}_2; {\theta})\right\}<\infty,\ k=1,2,\]
y por la ley fuerte de los grandes n\'umeros,
\[\frac{1}{n}\sum_{i=1}^n Q^{(k)}(\boldsymbol{X}_i,\boldsymbol{X}_i; {\theta})\mathop{\longrightarrow}\limits^{c.s.}E_{\theta}\left\{Q^{(k)} (\boldsymbol{X}_1,\boldsymbol{X}_1; {\theta})\right\}<\infty,\ k=1,2.\]
As\'i, (\ref{Wn-aprox}) se puede escribir como
\[W_n(\hat{\theta}_n)=\frac{1}{n^2}\sum_{i,j=1}^n h(\boldsymbol{X}_i,\boldsymbol{X}_j; \theta)+\sqrt{n}(\hat{\theta}_n-\theta) \frac{1}{n}S \sqrt{n}(\hat{\theta}_n-\theta)^{\top}+o_{_P}(1),\]
donde $S$ es la matriz sim\'etrica dada por $S=\frac{1}{2}\,E_{\theta}\!\left\{Q^{(2)}(\boldsymbol{X}_1,\boldsymbol{X}_2; {\theta})\right\}$.\vskip .2 cm

Considerando ahora el Supuesto \ref{hat(theta)-theta}, se obtiene
\[W_n(\hat{\theta}_n)=\frac{1}{n^2}\sum_{i,j=1}^n h_d(\boldsymbol{X}_i,\boldsymbol{X}_j; \theta)+\varepsilon_n,\]
donde $h_d(x,y;\theta)$ es el definido en (\ref{nucleo_de_Wn_aproximado}) y $\varepsilon_n=o_{_P}(n^{-1})$.\vskip .2 cm

Notar que $h_d(x,y;\theta)=h_d(y,x;\theta)$, pues $\left\{\boldsymbol{\ell}(x;\theta) S\boldsymbol{\ell}(y;\theta)^{\top}\right\}^{\top}=\boldsymbol{\ell}(y;\theta) S\boldsymbol{\ell}(x;\theta)^{\top}\in \mathbb{R}$, y bajo $H_0$, resulta
\[E_{\theta}\!\left\{|h_d(\boldsymbol{X}_1,\boldsymbol{X}_1;\theta)|\right\} < \infty \quad \mbox{y} \quad  E_{\theta}\!\left\{h_d(\boldsymbol{X}_1,\boldsymbol{X}_2;\theta)^2\right\}< \infty. \]

Adem\'as, por el Supuesto \ref{hat(theta)-theta} y el hecho que $E_{\theta}\!\left\{h(\boldsymbol{X}_1,\boldsymbol{X}_2;\theta)\right\}=0$, dan como resultado $E_{\theta}\{h_d(\boldsymbol{X}_1,\boldsymbol{X}_2;\theta)\}=0 $.

Por \'ultimo, como $h_d$ es degenerado, $E_{\theta}\!\left\{h_d(\boldsymbol{X}_1,\boldsymbol{X}_2;\theta)/ \boldsymbol{X}_1\right\}=0$, entonces, por el Teorema 6.4.1.B en Serfling (1980) \cite{Ser80},
\[\frac{1}{n}\sum_{i,j=1}^n h_d(\boldsymbol{X}_i,\boldsymbol{X}_j; \theta)
\mathop{\longrightarrow} \limits^{\!L}\ \sum_{j\geq 1}\lambda_j\,\chi^2_{1j},\]
con lo cual se consigue el resultado. $\square$\\

Los mismos comentarios hechos al final de la Subsecci\'on \ref{Distribucion-asintotica-nula} para $R_{n,w}(\hat{\theta}_n)$ y $S_{n,w}(\hat{\theta}_n)$ se pueden hacer para $W_n(\hat{\theta}_n)$, es decir, la distribuci\'on asint\'otica nula no proporciona una estimaci\'on \'util a la distribuci\'on nula de $W_n(\hat{\theta}_n)$. As\'i, en la siguiente secci\'on consideramos otra forma de aproximar la distribuci\'on nula del test estad\'istico, el m\'etodo bootstrap.

\subsection[Estimador bootstrap de la distribuci\'on nula de $W_n(\hat{\theta}_n)$]{Estimador bootstrap de la distribuci\'on nula de $\boldsymbol{W_n(\hat{\theta}_n)}$}\label{Estimador bootstrap}
Como mencionamos en la Secci\'on \ref{Aproximacion-bootstrap}, un modo alternativo de aproximar la distribuci\'on nula es mediante el m\'etodo bootstrap. Para ello, sean $\boldsymbol{X}_{1}, \boldsymbol{X}_{2},\ldots, \boldsymbol{X}_{n}$ vectores aleatorios iid que toman valores en $\mathbb{N}_0^2$ tales que $\hat{\theta}_n= \hat{\theta}_n(\boldsymbol{X}_{\!1}, \boldsymbol{X}_{2},\ldots, \boldsymbol{X}_{n}) \in \Theta$.

Sean $\boldsymbol{X}^*_{1}, \boldsymbol{X}^*_{2},\ldots, \boldsymbol{X}^*_{n}$ v.a. iid de una poblaci\'on que se distribuye seg\'un la ley $PB(\hat{\theta}_{1n},\hat{\theta}_{2n}, \hat{\theta}_{3n})$, dado $\boldsymbol{X}_{1}, \boldsymbol{X}_{2},\ldots, \boldsymbol{X}_{n}$ y sea $W^*_{n}(\hat{\theta}^*_n)$ la versi\'on bootstrap de $W_{n}(\hat{\theta}_n)$ obtenida al reemplazar $\boldsymbol{X}_{\!1}, \boldsymbol{X}_{2},\ldots, \boldsymbol{X}_{n}$ y
$\hat{\theta}_n\!=\! \hat{\theta}_n(\boldsymbol{X}_{\!1}, \boldsymbol{X}_{2},\ldots, \boldsymbol{X}_{n})$ por $\boldsymbol{X}^*_{\!1}, \boldsymbol{X}^*_{2},\ldots, \boldsymbol{X}^*_{n}$ y $\hat{\theta}^*_n= \hat{\theta}_n(\boldsymbol{X}^*_{1}, \boldsymbol{X}^*_{2},\ldots, \boldsymbol{X}^*_{n})$, respectivamente, en la expresi\'on de $W_{n}(\hat{\theta}_n)$.

Como en el Cap\'itulo \ref{Estadisticos-tipoCramer-von-Mises}, sea $P_*$ la ley de probabilidad condicional bootstrap, dado $\boldsymbol{X}_{1},\boldsymbol{X}_{2},\ldots, \boldsymbol{X}_{n}$.

Para mostrar que el m\'etodo bootstrap aproxima consistentemente a la distribuci\'on nula de $W_n(\hat{\theta}_n)$ nos ser\'an \'utiles las siguientes expresiones.

Observar que del segundo sumando de (\ref{nucleo_de_Wn_aproximado}) podemos escribir que
\begin{align}
\boldsymbol{\ell}(x;\theta)A_{rs}\,\boldsymbol{\ell}(y;\theta)^{\top}&= \boldsymbol{\ell}(x;\theta)
\left(\begin{array}{ccc}
  a & 0 & 0 \\
  0 & a & 0 \\
  b-a & c-a & 0
\end{array}\right)
\left(\begin{array}{ccc}
  a & 0 & b-a \\
  0 & a & c-a \\
  0 & 0 & 0
\end{array}\right)\boldsymbol{\ell}(y;\theta)^{\top}\notag\\
&=\left(\boldsymbol{\ell}(x;\theta) (a,0,b-a)^{\top},\boldsymbol{\ell}(x;\theta) (0,a,c-a)^{\top}\right)\left(
\begin{array}{l}
(a,0,b-a)\boldsymbol{\ell}(y;\theta)^{\top}\\[.1 cm]
(0,a,c-a)\boldsymbol{\ell}(y;\theta)^{\top}
\end{array}
\right)\notag\\[.2 cm]
&=\boldsymbol{\ell}(x;\theta)\, (a,0,b-a)^{\top} \boldsymbol{\ell}(y;\theta) \,(a,0,b-a)^{\top}\notag\\[.2 cm]
&\hspace{8mm}+\boldsymbol{\ell}(x;\theta)\,(0,a,c-a)^{\top} \boldsymbol{\ell}(y;\theta)\,(0,a,c-a)^{\top}.\notag
\end{align}
Sean
\begin{equation}\label{psi_3rs-4rs}
\phi_{3rs}(x;\theta)=\boldsymbol{\ell}(x;\theta) (a,0,b-a)^{\top}\quad\text{y}\quad \phi_{4rs}(x;\theta)=\boldsymbol{\ell}(x;\theta)(0,a,c-a)^{\top},
\end{equation}
donde $x=(x_1,x_2)$.

Con esta notaci\'on,
\begin{equation}\label{l-Ars-l}
\boldsymbol{\ell}(x;\theta)A_{rs}\,\boldsymbol{\ell}(y;\theta)^{\top}= \phi_{3rs}(x;\theta)\phi_{3rs}(y;\theta)+ \phi_{4rs}(x;\theta)\phi_{4rs}(y;\theta).
\end{equation}

Tambi\'en nos ser\'a de utilidad el siguiente resultado.

\begin{proposicion}\label{Nucleos-Wn}
Sean  $\boldsymbol{X}_{1},\boldsymbol{X}_{2},\ldots, \boldsymbol{X}_{n}$ vectores aleatorios iid de $\boldsymbol{X}=(X_{1},X_{2}) \in \mathbb{N}_0^2$. Supongamos que se verifica el Supuesto \ref{E(l*l)-l-cont} y que $\hat{\theta}_n \mathop{\longrightarrow}\limits^{c.s.} \theta$, para alg\'un $\theta \in \Theta$. Entonces
\[\sup_{p,q,u,v\in \mathbb{N}_0} \left| K_n(p,q,u,v)-K(p,q,u,v)\right| \stackrel{c.s.}{\longrightarrow}0,\]
donde
\begin{align}
K_n(p,q,u,v)&=\left(E_*\{\phi_{ipq}(\boldsymbol{X}_1^*;\hat{\theta}_n) \phi_{juv}(\boldsymbol{X}_1^*;\hat{\theta}_n)\}\right), 1\leq i,j \leq 4.\notag\\[.2 cm]
K(p,q,u,v)&=\Bigl(E_{\theta}\!\left\{\phi_{ipq}(\boldsymbol{X}_1;\theta) \phi_{juv}(\boldsymbol{X}_1;\theta)\right\}\Bigr), 1\leq i,j \leq 4.\notag
\end{align}
con $\,\phi_{1rs}(\boldsymbol{X}_1;\vartheta)$, $\phi_{2rs}(\boldsymbol{X}_1;\vartheta)$ definidos en (\ref{psi_1rs-2rs}), y $\,\phi_{3rs}(\boldsymbol{X}_1;{\vartheta})$ y $\phi_{4rs}(\boldsymbol{X}_1;{\vartheta})$ definidos en (\ref{psi_3rs-4rs}).
\end{proposicion}
\nt {\bf Demostraci\'on} \hspace{2pt}
El Supuesto \ref{E(l*l)-l-cont} y el hecho de que $\hat{\theta}_n \mathop{\longrightarrow}\limits^{c.s.} \theta$ implican que
\begin{equation}\label{E*-Et}
\sup_{p,q,u,v \in \mathbb{N}_0} \left| E_*\{\phi_{ipq}(\boldsymbol{X}_1^*;\hat{\theta}_n) \phi_{juv}(\boldsymbol{X}_1^*;\hat{\theta}_n)\}- E_{\theta}\{\phi_{ipq}(\boldsymbol{X}_1;\theta) \phi_{juv}(\boldsymbol{X}_1;\theta)\}\right| \stackrel{c.s.}{\longrightarrow}0, \end{equation}
donde $1\leq i,j \leq 4$.

Para probar este resultado consideremos los siguientes casos:
\begin{itemize}
  \item [(a)] $i,j$ tales que $1\leq i,j \leq 2$,
  \item [(b)] $i,j$ tales que $3\leq i,j \leq 4$ y
  \item [(c)] $i,j$ tales que $i=1,2,$ $j=3,4$.
\end{itemize}
Notar primero que por el Teorema del Valor Medio
\[P(r,s;\hat{\theta}_n)-P(r,s;\theta)=\nabla_{\vartheta}P(r,s;\tilde{\theta}) (\hat{\theta}_n-\theta)^{\top},\  \tilde{\theta}=\alpha \hat{\theta}_n+(1-\alpha)\theta, \ \text{para alg\'un} \ 0<\alpha <1,\]
donde $\nabla_{\vartheta}P(r,s;\tilde{\theta})$ denota el vector gradiente de $P(r,s;\vartheta)$ evaluado en $\tilde{\theta}$. Por las relaciones de recurrencia dadas en (\ref{form-rec-derDPB}), obtenemos que
\[\nabla_{\vartheta}P(r,s;\tilde{\theta})=(a-c,b-c,d-a-b+c),\]
donde $a=P(r-1,s;\tilde{\theta}), b=P(r,s-1;\tilde{\theta}), c=P(r,s;\tilde{\theta})$ y $d=P(r-1,s-1;\tilde{\theta})$.

Como $0\leq P(r,s;\tilde{\theta})\leq 1$ y puesto que $\hat{\theta}_n \mathop{\longrightarrow}\limits^{c.s.} \theta$, resulta
\begin{equation}\label{P-o(1)}
\sup_{r,s \in \mathbb{N}_0} \left|P(r,s;\hat{\theta}_n)-P(r,s;\theta)\right|=o(1).
\end{equation}

\begin{itemize}
  \item En el caso (a), una situaci\'on es la siguiente
\begin{align}
E_*&\{\phi_{1pq}(\boldsymbol{X}_1^*;\hat{\theta}_n) \phi_{1uv}(\boldsymbol{X}_1^*;\hat{\theta}_n)\}\notag\\[.18 cm]
&\hspace{5mm} =\left\{
\begin{array}{ll}
\begin{array}{l}
(p+1)^2P(p+1,q;\hat{\theta}_n)+\hat{\theta}_{3n}^2P(p,q-1;\hat{\theta}_n)\\[.27 cm]
\hspace{33mm}+(\hat{\theta}_{1n}-\hat{\theta}_{3n})^2 P(p,q;\hat{\theta}_n)
\end{array}
 & \hbox{si } u=p, v=q,\\[.7 cm]
-(p+1)(\hat{\theta}_{1n}-\hat{\theta}_{3n}) P(p+1,q;\hat{\theta}_n), & \hbox{si } u=p+1, v=q, \\[.3 cm]
-(p+1)\hat{\theta}_{3n} P(p+1,q;\hat{\theta}_n), & \hbox{si } u=p+1, v=q+1, \\[.3 cm]
-p(\hat{\theta}_{1n}-\hat{\theta}_{3n}) P(p,q;\hat{\theta}_n), & \hbox{si }  u=p-1, v=q, \\[.3 cm]
-p\hat{\theta}_{3n} P(p,q-1;\hat{\theta}_n), & \hbox{si } u=p-1, v=q-1,\\[.3 cm]
\hat{\theta}_{3n}(\hat{\theta}_{1n}-\hat{\theta}_{3n}) P(p,q;\hat{\theta}_n), & \hbox{si }  u=p, v=q+1, \\[.3 cm]
\hat{\theta}_{3n}(\hat{\theta}_{1n}-\hat{\theta}_{3n}) P(p,q-1;\hat{\theta}_n), & \hbox{si }  u=p, v=q-1,
\end{array}
\right.\notag
\end{align}

\nt y an\'alogo para $E_{\theta}\left\{\phi_{1pq}(\boldsymbol{X}_1;\theta) \phi_{1uv}(\boldsymbol{X}_1;\theta)\right\}$, con las modificaciones obvias.\vskip .25 cm

Al hacer la resta de las esperanzas anteriores nos encontramos con las siguientes tres situaciones diferentes
\begin{itemize}
   \item [(a.1)] $t(\hat{\theta}_n) P(r,s;\hat{\theta}_n)-t(\theta)P(r,s;\theta)$,
   \item [(a.2)] $r\!\left\{t(\hat{\theta}_n) P(r,s;\hat{\theta}_n)-t(\theta)P(r,s;\theta)\!\right\}$,
   \item [(a.3)] $r^2\!\left\{P(r,s;\hat{\theta}_n)-P(r,s;\theta)\!\right\}$,
 \end{itemize}
donde $t(\vartheta)=\vartheta^k_3$, $t(\vartheta)=(\vartheta_1-\vartheta_3)^k$, para $k=1,2$, o bien $t(\vartheta)=\vartheta_3(\vartheta_1-\vartheta_3)$.\vskip .3 cm

Para la situaci\'on (a.1), puesto que $t(\vartheta)$ es continua como funci\'on de $\vartheta\in \Theta_0$ y como $\hat{\theta}_n \mathop{\longrightarrow}\limits^{c.s.} \theta$, entonces $t(\hat{\theta}_n)=t(\theta)+o(1)$, por lo tanto
\begin{equation}\label{t*P}
t(\hat{\theta}_n) P(r,s;\hat{\theta}_n)-t(\theta)P(r,s;\theta)= t(\theta)\{P(r,s;\hat{\theta}_n)-P(r,s;\theta)\}+o(1)P(r,s;\hat{\theta}_n),
\end{equation}
como $t(\theta)< \infty$, de (\ref{P-o(1)}) resulta
\[\sup_{r,s\in \mathbb{N}_0}\left|t(\hat{\theta}_n) P(r,s;\hat{\theta}_n)-t(\theta)P(r,s;\theta)\right|=o(1).\]

Para la situaci\'on (a.3), por desarrollo en serie de Taylor
\[r^2\!\left\{P(r,s;\hat{\theta}_n)-P(r,s;\theta)\right\}= \sum_{i=1}^3r^2\frac{\partial}{\partial \theta_i}P(r,s;\tilde{\theta})
(\hat{\theta}_{in}-{\theta}_i),\]
donde $\tilde{\theta}=\alpha \hat{\theta}_n+(1-\alpha)\theta$, para alg\'un $0<\alpha <1$.\vskip .2 cm

Veamos el primer sumando y an\'alogo el resto:

Como $\frac{\partial}{\partial \theta_1}P(r,s;\tilde{\theta})= P(r-1,s;\tilde{\theta})-P(r,s;\tilde{\theta})$ y $E_{\vartheta}(X^2_1)< \infty, \forall \vartheta\in \Theta_0$, entonces
\begin{align}
r^2\frac{\partial}{\partial \theta_1}P(r,s;\tilde{\theta})
(\hat{\theta}_{1n}-{\theta}_1)&=(\hat{\theta}_{1n}-{\theta}_1)r^2 \{P(r-1,s;\tilde{\theta})-P(r,s;\tilde{\theta})\}\notag\\
&\leq |\hat{\theta}_{1n}-{\theta}_1|\sum_{r,s\geq 0}r^2\{P(r-1,s;\tilde{\theta})+P(r,s;\tilde{\theta})\}\notag\\[.1 cm]
&\leq|\hat{\theta}_{1n}-{\theta}_1|\left[E_{\tilde{\theta}}\{(X_1+1)^2\}+ E_{\tilde{\theta}}(X_1^2)\right]=o(1).\notag
\end{align}
De donde
\begin{align}
&r^2\!\left\{P(r,s;\hat{\theta}_n)-P(r,s;\theta)\right\}\notag\\[.1 cm]
&\hspace{5mm}\leq \!\left\{|\hat{\theta}_{1n}\!-{\theta}_1|+2|\hat{\theta}_{3n}\! -{\theta}_3|\right\} \left[E_{\tilde{\theta}}\left\{(X_1+1)^2\right\}+ E_{\tilde{\theta}}\!\left(X_1^2\right)\right] +2|\hat{\theta}_{2n}\!-{\theta}_2|E_{\tilde{\theta}}\!\left(X_1^2\right).\notag
\end{align}
Por lo tanto
\[\sup_{r,s\in \mathbb{N}_0}\left|r^2\!\left\{P(r,s;\hat{\theta}_n)-P(r,s;\theta)\right\}\right|=o(1).\]

Para la situaci\'on (a.2) y considerando (\ref{t*P}), obtenemos
\begin{align}
&r\left\{t(\hat{\theta}_n) P(r,s;\hat{\theta}_n)-t(\theta)P(r,s;\theta)\right\} \notag \\
&\hspace{48mm}=r\, t(\theta) \left\{P(r,s;\hat{\theta}_n)-P(r,s;\theta)\right\}+rP(r,s;\hat{\theta}_n)\, o(1).\notag
\end{align}

Puesto que $t(\theta)< \infty$, $E_{\vartheta}(X_1)< \infty, \forall \vartheta\in \Theta_0$ y siguiendo pasos similares a los dados en la situaci\'on (a.3) anterior se consigue que
\[\sup_{r,s\in \mathbb{N}_0}\left|r\left\{t(\hat{\theta}_n) P(r,s;\hat{\theta}_n)-t(\theta)P(r,s;\theta)\!\right\}\right|=o(1).\]

  \item Para el caso (b), tenemos, por ejemplo 
\begin{align}
E_*\!\left\{\phi_{3pq}(\boldsymbol{X}_1^*;\hat{\theta}_n) \phi_{3uv}(\boldsymbol{X}_1^*;\hat{\theta}_n)\!\right\}&= E_*\!\left\{\boldsymbol{\ell} (\boldsymbol{X}_1^*;\hat{\theta}_n) B(p,q;\hat{\theta}_n)^{\top} \boldsymbol{\ell}(\boldsymbol{X}_1^*;\hat{\theta}_n) B(u,v;\hat{\theta}_n)^{\top}\!\right\} \notag\\[.1 cm] &= B(p,q;\hat{\theta}_n)J(\hat{\theta}_n) B(u,v;\hat{\theta}_n)^{\top},\notag\\[.2 cm]
E_{\theta}\!\left\{\phi_{3pq}(\boldsymbol{X}_1;\theta) \phi_{3uv}(\boldsymbol{X}_1;\theta)\right\}&=B(p,q;\theta)J(\theta) B(u,v;\theta)^{\top}, \notag
\end{align}
donde
\begin{equation}\label{B(r,s)}
B(r,s;\vartheta)=(P(r,s;\vartheta),0, P(r,s-1;\vartheta)-P(r,s;\vartheta)).
\end{equation}

Como, por el Supuesto \ref{E(l*l)-l-cont}, $J(\vartheta)$ es continua como funci\'on de $\vartheta\in \Theta_0$ y $\hat{\theta}_n \mathop{\longrightarrow}\limits^{c.s.} \theta$, entonces $J(\hat{\theta}_n)=J(\theta)+o(1)$. Adem\'as, por el Teorema del Valor Medio
\begin{equation}\label{Bhat(t)-Bt}
B(r,s;\hat{\theta}_n)^{\top}-B(r,s;\theta)^{\top}= grad(P) (\hat{\theta}_n-\theta)^{\top}=\mathbf{o}(1),
\end{equation}
pues
\[grad(P)=\left(\begin{array}{c}
\nabla_{\theta}P(r,s;\tilde{\theta}) \\
\mathbf{0} \\
\nabla_{\theta}P(r,s-1;\tilde{\theta})- \nabla_{\theta}P(r,s;\tilde{\theta})
\end{array}\right)
< \infty,\]
donde $\tilde{\theta}=\alpha \hat{\theta}_n+(1-\alpha)\theta$, para alg\'un $0<\alpha <1$.

Por tanto
\begin{align}
&\left|E_*\!\left\{\phi_{3pq}(\boldsymbol{X}_1^*;\hat{\theta}_n) \phi_{3uv}(\boldsymbol{X}_1^*;\hat{\theta}_n)\!\right\}- E_{\theta}\!\left\{\phi_{3pq}(\boldsymbol{X}_1;\theta) \phi_{3uv}(\boldsymbol{X}_1;\theta)\right\}\right|\notag\\[.2 cm]
&\hspace{42mm}\leq \left|\left\{B(p,q;\hat{\theta}_n)-B(p,q;\theta)\right\} J(\theta)B(u,v;\hat{\theta}_n)^{\top}\right| \notag\\[.2 cm]
&\hspace{47mm}+ \left|B(p,q;\theta)J(\theta)\!\left\{B(u,v;\hat{\theta}_n)^{\top}- B(u,v;\theta)^{\top}\right\}\right|+o(1).\notag
\end{align}

Puesto que $J(\vartheta)<\infty$ y $B(r,s;\vartheta)<\infty$, $\forall \vartheta\in\Theta_0, \forall r,s\in \mathbb{N}_0$, entonces se logra
\[\sup_{p,q,u,v\in \mathbb{N}_0}\left|E_*\!\left\{\phi_{3pq}(\boldsymbol{X}_1^*;\hat{\theta}_n) \phi_{3uv}(\boldsymbol{X}_1^*;\hat{\theta}_n)\!\right\}- E_{\theta}\!\left\{\phi_{3pq}(\boldsymbol{X}_1;\theta) \phi_{3uv}(\boldsymbol{X}_1;\theta)\right\}\right| \stackrel{c.s.}{\longrightarrow}0.\]

Aplicando el mismo procedimiento se demuestra (\ref{E*-Et}) para $3\leq i,j\leq 4$.

  \item \vskip .4 cm En el caso (c), una situaci\'on es
\[E_*\!\left\{\phi_{1pq}(\boldsymbol{X}_1^*;\hat{\theta}_n) \phi_{3uv}(\boldsymbol{X}_1^*;\hat{\theta}_n)\right\}- E_{\theta}\{\phi_{1pq}(\boldsymbol{X}_1;\theta) \phi_{3uv}(\boldsymbol{X}_1;\theta)\}=A_1+A_2+A_3,\]
donde
\begin{align}
A_1&=(p+1)P(p+1,q;\hat{\theta}_n) \boldsymbol{\ell}(p+1,q;\hat{\theta}_n)B(u,v;\hat{\theta}_n)^{\top}\notag\\[.1 cm] &\hspace{50 mm}-(p+1)P(p+1,q;\theta) \boldsymbol{\ell}(p+1,q;\theta)B(u,v;\theta)^{\top},\notag\\[.2 cm]
A_2&=(\theta_1-\theta_3)P(p,q;\theta) \boldsymbol{\ell}(p,q;\theta)B(u,v;\theta)^{\top}\notag\\[.1 cm]
&\hspace{50 mm} -(\hat{\theta}_{1n}-\hat{\theta}_{3n})P(p,q;\hat{\theta}_n) \boldsymbol{\ell}(p,q;\hat{\theta}_n)B(u,v;\hat{\theta}_n)^{\top},\notag \\[.2 cm]
A_3&=\theta_3 P(p,q-1;\theta) \boldsymbol{\ell}(p,q-1;\theta)B(u,v;\theta)^{\top}\notag\\[.1 cm]
&\hspace{50 mm} -\hat{\theta}_{3n} P(p,q-1;\hat{\theta}_n) \boldsymbol{\ell}(p,q-1;\hat{\theta}_n)B(u,v;\hat{\theta}_n)^{\top} \notag
\end{align}
y $B$ es definido como en (\ref{B(r,s)}).\vskip .2 cm

Veamos primero que $\displaystyle\sup_{r,s,u,v\in \mathbb{N}_0}\left|A_3\right|=o(1)$. De manera an\'aloga se demuestra que $\displaystyle\sup_{r,s,u,v\in \mathbb{N}_0}\left|A_2\right|=o(1)$.

\begin{align}
\hat{\theta}_{3n} P(r,s;\hat{\theta}_n) \boldsymbol{\ell}&(r,s;\hat{\theta}_n)B(u,v;\hat{\theta}_n)^{\top}
-\theta_3 P(r,s;\theta) \boldsymbol{\ell}(r,s;\theta)B(u,v;\theta)^{\top}\notag\\[.15 cm]
&=(\hat{\theta}_{3n}-\theta_3)P(r,s;\hat{\theta}_n) \boldsymbol{\ell}(r,s;\hat{\theta}_n) B(u,v;\hat{\theta}_n)^{\top}\notag\\[.15 cm]
&\ \ \ +\theta_3P(r,s;\theta)\boldsymbol{\ell}(r,s;\theta) \left\{B(u,v;\hat{\theta}_n)^{\top}- B(u,v;\theta)^{\top}\right\}\notag\\[.15 cm]
&\ \ \ +\theta_3\left\{P(r,s;\hat{\theta}_n) \boldsymbol{\ell}(r,s;\hat{\theta}_n)-P(r,s;\theta) \boldsymbol{\ell}(r,s;\theta)\right\} B(u,v;\hat{\theta}_n)^{\top}.\label{Tercer-suma}
\end{align}

Como $J(\vartheta)$ es finita $\forall \vartheta \in \Theta_0$, esto implica que $|P(r,s;\vartheta)\boldsymbol{\ell}_i(r,s;\vartheta)|\leq L$, para cierta constante positiva $L>0$, $\forall r,s\in \mathbb{N}_0$,
$\forall \vartheta \in \Theta_0$, $i=1,2,3$.\vskip .2 cm

Puesto que $\hat{\theta}_n \mathop{\longrightarrow}\limits^{c.s.} \theta$, de (\ref{Bhat(t)-Bt}) resulta
\begin{equation}\label{Bi_hat(theta)-Bi(theta)}
\sup_{u,v\in \mathbb{N}_0}\left|B_i(u,v;\hat{\theta}_n)-B_i(u,v;\theta)\right|=o(1),\ i=1,2,3.
\end{equation}
Por lo tanto
\[\sup_{r,s,u,v\in \mathbb{N}_0}\left|(\hat{\theta}_{3n}-\theta_3)P(r,s;\hat{\theta}_n) \boldsymbol{\ell}(r,s;\hat{\theta}_n) B(u,v;\hat{\theta}_n)^{\top}\right|=o(1),\]
\[\sup_{r,s,u,v\in \mathbb{N}_0}\left|\theta_3P(r,s;\theta)\boldsymbol{\ell}(r,s;\theta) \left\{B(u,v;\hat{\theta}_n)^{\top}- B(u,v;\theta)^{\top}\right\}\right|=o(1).\]

Para el tercer sumando de (\ref{Tercer-suma}), el Lema \ref{P(hat(t))*l(hat(t))-P(t)*l(t)} junto con que $\hat{\theta}_n\in\Theta_0$ y el hecho que $|B_i(u,v;\hat{\theta}_n)|\leq 1$, $i=1,2,3$, implican
\[\sup_{r,s,u,v\in \mathbb{N}_0}\left|\theta_3\left\{P(r,s;\hat{\theta}_n) \boldsymbol{\ell}(r,s;\hat{\theta}_n)-P(r,s;\theta) \boldsymbol{\ell}(r,s;\theta)\right\} B(u,v;\hat{\theta}_n)^{\top}
\right|=o(1).\]

Veamos ahora que $\ \sup_{r,s,u,v\in \mathbb{N}_0}\left|A_1\right|=o(1)$. Para ello, notar que
\begin{align}
rP(r,s;\hat{\theta}_n) \boldsymbol{\ell}(r,s;\hat{\theta}_n) & B(u,v;\hat{\theta}_n)^{\top}-rP(r,s;\theta) \boldsymbol{\ell}(r,s;\theta)B(u,v;\theta)^{\top}\notag\\
&=r\left\{P(r,s;\hat{\theta}_n)\boldsymbol{\ell}(r,s;\hat{\theta}_n)- P(r,s;\theta) \boldsymbol{\ell}(r,s;\theta) \right\}B(u,v;\hat{\theta}_n)^{\top}\notag\\
&\hspace{18mm}+r P(r,s;\theta) \boldsymbol{\ell}(r,s;\theta)\left\{B(u,v;\hat{\theta}_n)^{\top}\!- B(u,v;\theta)^{\top} \right\}.\notag
\end{align}
Usando el Lema \ref{P(hat(t))*l(hat(t))-P(t)*l(t)}(a) junto con que $\hat{\theta}_n\in\Theta_0$, obtenemos
\begin{align}
&\sup_{r,s\in\mathbb{N}_0}r \left|\boldsymbol{\ell}_i(r,s;\hat{\theta}_n) P(r,s;\hat{\theta}_n)-\boldsymbol{\ell}_i(r,s;\theta) P(r,s;\theta)\right|\notag\\
&\hspace{17mm}\leq\sum_{r,s \geq 0} r\left|\boldsymbol{\ell}_i(r,s;\hat{\theta}_n) P(r,s;\hat{\theta}_n)- \boldsymbol{\ell}_i(r,s;\theta) P(r,s;\theta)\right|=o(1),\ i=1,2,3.\notag
\end{align}
De esta \'ultima relaci\'on y puesto que $|B_i(u,v;\hat{\theta}_n)|\leq 1$, $i=1,2,3$, nos resulta
\[\sup_{r,s,u,v\in \mathbb{N}_0}\left|r\left\{P(r,s;\hat{\theta}_n) \boldsymbol{\ell}(r,s;\hat{\theta}_n)-P(r,s;\theta) \boldsymbol{\ell}(r,s;\theta)\right\} B(u,v;\hat{\theta}_n)^{\top}
\right|=o(1).\]

Por la desigualdad de Cauchy-Schwarz y puesto que $J(\theta)< \infty$, entonces
\begin{align}
\sup_{r,s\in\mathbb{N}_0}r|\boldsymbol{\ell}_i(r,s;\theta)| P(r,s;\theta) &\leq \sum_{r,s\geq 0}r|\boldsymbol{\ell}_i(r,s;\theta)| P(r,s;\theta)\notag\\[.1 cm]
&\leq \left\{E_{\theta}\!\left(X_1^2\right)\right\}^{\!1/2} \left\{ E_{\theta}\!\left(\|\boldsymbol{\ell}(\boldsymbol{X}_1;\theta)\|^2\right)\right\}^{\!\!1/2}< \infty,\ i=1,2,3.\notag
\end{align}

Esta \'ultima desigualdad junto con (\ref{Bi_hat(theta)-Bi(theta)}) implican \[\sup_{r,s,u,v\in \mathbb{N}_0}\left|r P(r,s;\theta) \boldsymbol{\ell}(r,s;\theta)\left\{B(u,v;\hat{\theta}_n)^{\top}- B(u,v;\theta)^{\top} \right\}
\right|=o(1).\]

As\'i, concluimos que $\ \sup_{r,s,u,v\in \mathbb{N}_0}\left|A_1\right|=o(1)$.
\end{itemize}

Los casos (a), (b) y (c) muestran que se verifica (\ref{E*-Et}) para $1\leq i,j\leq 4$, con lo cual se demuestra la proposici\'on. $\square$

Ahora probaremos que el m\'etodo bootstrap estima consistentemente la distribuci\'on nula de $W_n(\hat{\theta}_n)$.

\begin{teorema}\label{Cons-boot-Wn}
Sean  $\boldsymbol{X}_{1},\boldsymbol{X}_{2},\ldots, \boldsymbol{X}_{n}$ vectores aleatorios iid de $\boldsymbol{X}=(X_{1},X_{2}) \in \mathbb{N}_0^2$. Supongamos que se verifica el Supuesto \ref{E(l*l)-l-cont} y que $\hat{\theta}_n \mathop{\longrightarrow}\limits^{c.s.} \theta$, para alg\'un $\theta \in \Theta$. Entonces
\[\sup_{x\in\mathbb{R}}\!\left|P_*\left(nW^*_{n}(\hat{\theta}^*_n) \leq x\right)-P_{\theta}\!\left(nW_{n}(\hat{\theta}_n)\leq x\right)\right|\ \mathop{\longrightarrow}\limits^{c.s.}\ 0.\]
\end{teorema}
\nt {\bf Demostraci\'on} \hspace{2pt} Por definici\'on
\[W^*_n(\hat{\theta}^*_n)=\frac{1}{n^2}\sum_{i,j=1}^n h^*(\boldsymbol{X}^*_i,\boldsymbol{X}^*_j;\hat{\theta}^*_n).\]
Siguiendo pasos similares a los dados en la demostraci\'on del Teorema \ref{TeoConvDebil-Wn} se puede ver que
\[W^*_n(\hat{\theta}^*_n)=W^*_{1n}(\hat{\theta}^*_n),\]
donde
\[W^*_{1n}(\hat{\theta}^*_n)=\frac{1}{n^2}\sum_{i,j=1}^n h^*_d(\boldsymbol{X}^*_i,\boldsymbol{X}^*_j;\hat{\theta}_n)+\xi^*,\]
con $\xi^*=o_{_{P_*}}(n^{-1})$ c.s. y $h^*_d$ es definido como $h_d$ en (\ref{nucleo_de_Wn_aproximado}) con $\boldsymbol{X}_i$, $\boldsymbol{X}_j$ y $\theta$ reemplazados por $\boldsymbol{X}^*_i$,$\boldsymbol{X}^*_j$ y $\hat{\theta}_n$, respectivamente.

De (\ref{definicion-h}) y (\ref{l-Ars-l}), $h^*_d$ se puede escribir como
\begin{align}
h^*_d(\boldsymbol{X}^*_i,\boldsymbol{X}^*_j;\hat{\theta}_n)&= h^*_1(\boldsymbol{X}^*_i,\boldsymbol{X}^*_j;\hat{\theta}_n)+ h^*_2(\boldsymbol{X}^*_i,\boldsymbol{X}^*_j;\hat{\theta}_n)+ \sum_{r,s\geq 0}\boldsymbol{\ell}(\boldsymbol{X}^*_i;\hat{\theta}_n) A^*_{rs}\boldsymbol{\ell}(\boldsymbol{X}^*_j;\hat{\theta}_n)^{\top}\notag\\[.1 cm]
&=\sum_{1\leq k\leq 4} \sum_{r,s\geq 0} \phi_{krs}(\boldsymbol{X}^*_i;\hat{\theta}_n) \phi_{krs}(\boldsymbol{X}^*_j;\hat{\theta}_n),\notag
\end{align}
con lo cual
\[W^*_{1n}(\hat{\theta}^*_n)=\sum_{i,j=1}^n \frac{1}{n^2} \sum_{1\leq k\leq 4} \sum_{r,s\geq 0} \phi_{krs}(\boldsymbol{X}^*_i;\hat{\theta}_n) \phi_{krs}(\boldsymbol{X}^*_j;\hat{\theta}_n).\]

La demostraci\'on se completar\'a verificando las condiciones (i)-(iii) del Lema \ref{kundu} para $W^*_{1n}(\hat{\theta}^*_n)$. Para ello, consideremos el siguiente espacio de Hilbert separable
\[\mathcal{H}_2=\left\{z=z(r,s)=(z_{1rs},z_{2rs},z_{3rs},z_{4rs})_{r,s\geq 0}\ \ \text{tal que}\ \ \|z\|_{_{\mathcal{H}_2}}^2=\sum_{1\leq k\leq 4}\,\sum_{r,s\geq 0} \!z_{krs}^2<\infty\right\}.\]
con producto escalar $\,\langle x,y\rangle_{_{\!\mathcal{H}_2}}=\displaystyle\sum_{1\leq k\leq 4} \sum_{r,s\geq 0}x_{krs}y_{krs}<\infty$.

En este espacio de Hilbert se tiene que
\[nW^*_{1n}(\hat{\theta}^*_n)=\sum_{1\leq k\leq 4}\sum_{r,s\geq 0}\left\{\frac{1}{\sqrt{n}}\sum_{1\leq i\leq n} \phi_{krs}(\boldsymbol{X}^*_i;\hat{\theta}_n)\right\}^2 \equiv\|\Phi_n^*\|_{_{\mathcal{H}_2}}^2,\]
donde $\Phi_n^*$ est\'a dado por
\[\Phi_n^*(r,s)=\sum_{i=1}^n \Phi_{ni}^*(r,s),\] \[\Phi_{ni}^*(r,s)=\frac{1}{\sqrt{n}}\Phi(\boldsymbol{X}^*_i;\hat{\theta}_n;r,s), 1\leq i\leq n,\]
\begin{equation}\label{vector-Phi}
\Phi(\boldsymbol{X}^*_i;\hat{\theta}_n;r,s)= \left(\phi_{1rs}(\boldsymbol{X}^*_i;\hat{\theta}_n), \phi_{2rs}(\boldsymbol{X}^*_i;\hat{\theta}_n), \phi_{3rs}(\boldsymbol{X}^*_i;\hat{\theta}_n), \phi_{4rs}(\boldsymbol{X}^*_i;\hat{\theta}_n)\right), 1\leq i\leq n,
\end{equation}

\nt con $\phi_{1rs}(x;{\theta})$ y $\phi_{2rs}(x;{\theta})$ definidos en (\ref{psi_1rs-2rs}), $\phi_{3rs}(x;{\theta})$ y $\phi_{4rs}(x;{\theta})$ definidos en (\ref{psi_3rs-4rs}), donde $x$ y $\theta$ son reemplazados por $\boldsymbol{X}^*_i$ y $\hat{\theta}_n$, respectivamente.\vskip .3 cm

Por el Supuesto \ref{E(l*l)-l-cont} y por (\ref{form-rec-DPB}), $\Phi_{ni}^*(r,s)$ tiene medias nulas y momentos segundos finitos, $1\leq i \leq n$, pues $\hat{\theta}_n \mathop{\longrightarrow}\limits^{c.s.} \theta$.

Sean $(p,q),(u,v)\in \mathbb{N}_0^2$ y consideremos el n\'ucleo de covarianza $K_n(p,q,u,v)=E_*\left\{\Phi_n^*(p,q)^\top \Phi_n^*(u,v)\right\}$, luego
\[K_n(p,q,u,v)=\left(E_*\{\phi_{ipq}(\boldsymbol{X}_1^*;\hat{\theta}_n) \phi_{juv}(\boldsymbol{X}_1^*;\hat{\theta}_n)\}\right), 1\leq i,j \leq 4.\]

Adem\'as, sea $K(p,q,u,v)=E_{\theta}\left\{\Phi(\boldsymbol{X}_1;\theta;p,q)^{\top} \Phi(\boldsymbol{X}_1;\theta;u,v)\right\}$ donde $\Phi$ es como el definido en (\ref{vector-Phi}), es decir,
\[K(p,q,u,v)=\Bigl(E_{\theta}\!\left\{\phi_{ipq}(\boldsymbol{X}_1;\theta) \phi_{juv}(\boldsymbol{X}_1;\theta)\right\}\Bigr), 1\leq i,j \leq 4.\]

Sea $\mathcal{Z}$ la sucesi\'on Gaussiana centrada cuyo operador de covarianza $C$ est\'a caracterizado por
\begin{align}
\langle C x,y\rangle_{_{\!\mathcal{H}_2}}=cov\left(\langle \mathcal{Z}, x\rangle_{_{\!\mathcal{H}_2}}, \langle \mathcal{Z}, y\rangle_{_{\!\mathcal{H}_2}}\right)&=E_{\theta}\left\{\langle \mathcal{Z}, x\rangle_{_{\!\mathcal{H}_2}} \langle \mathcal{Z}, y\rangle_{_{\!\mathcal{H}_2}}\right\}\notag\\[.2 cm]
&=\sum_{p,q,u,v\geq 0} x(p,q)K(p,q,u,v)y(u,v)^{\top}.\label{Oper-cov-C}
\end{align}

Por el Lema \ref{TCL-Hilbert}, $(\Phi_{1n},\Phi_{2n},\Phi_{3n},\Phi_{4n})\stackrel{L}{\longrightarrow} \mathcal{Z}$ en $\mathcal{H}_2$, cuando los datos son iid provenientes de $\boldsymbol{X}\sim PB(\theta)$, donde \[\Phi_{kn}(r,s)=\frac{1}{\sqrt{n}}\sum_{1\leq i\leq n}\phi_{krs}(\boldsymbol{X}_i;\theta), 1\leq k\leq 4.\]

Para nuestro caso una base ortonormal conveniente en $\mathcal{H}_2$ es la dada por el conjunto $\left\{e^k_{r_0s_0}: r_0,s_0\geq 0, k=1,2,3,4\right\}$, donde
\[\begin{array}{c}
e^1_{r_0s_0}(r,s)=\bigl(I{\{r=r_0,s=s_0\}},0,0,0\bigr)_{r,s\geq 0},\\[.35 cm]
e^2_{r_0s_0}(r,s)=\bigl(0,I{\{r=r_0,s=s_0\}},0,0\bigr)_{r,s\geq 0},\\[.35 cm]
e^3_{r_0s_0}(r,s)=\bigl(0,0,I{\{r=r_0,s=s_0\}},0\bigr)_{r,s\geq 0},\\[.35 cm]
e^4_{r_0s_0}(r,s)=\bigl(0,0,0,I{\{r=r_0,s=s_0\}}\bigr)_{r,s\geq 0}.
\end{array}\]

Ahora mostraremos que se verifican las condiciones (i)-(iii) del Lema \ref{kundu}.
\begin{itemize}
  \item [(i)] Sea $C_n$ el operador de covarianza  de $\Phi_n^*$, esto es, $\forall x,y \in \mathcal{H}_2$ \begin{align} \langle C_n x,y\rangle_{_{\!\mathcal{H}_2}}&=cov(\langle \Phi_n^*, x\rangle_{_{\!\mathcal{H}_2}}, \langle \Phi_n^*, y\rangle_{_{\!\mathcal{H}_2}})=E_{*}\!\left\{\langle \Phi_n^*, x\rangle_{_{\!\mathcal{H}_2}} \langle \Phi_n^*, y\rangle_{_{\!\mathcal{H}_2}}\right\}\notag\\[.2 cm] &=\sum_{p,q,u,v\geq 0} x(p,q)E_*\!\left\{\Phi_n^*(p,q)^\top \Phi_n^*(u,v)\right\}y(u,v)^{\top}\notag\\[.2 cm] &=\sum_{p,q,u,v\geq 0} x(p,q)K_n(p,q,u,v)y(u,v)^{\top}.\label{def-Cn} \end{align}

      De (\ref{Oper-cov-C}), (\ref{def-Cn}) y de la Proposici\'on \ref{Nucleos-Wn}
      \begin{align} \lim_{n\to \infty}\langle C_n e^k_{p_0q_0}, e^l_{u_0v_0}\rangle_{_{\!\mathcal{H}_2}} &=\lim_{n\to \infty} \sum_{p,q,u,v\geq 0} e^k_{p_0q_0}(p,q)K_n(p,q,u,v) e^l_{u_0v_0}(u,v)^{\top}\notag\\[.2 cm] &=\sum_{p,q,u,v\geq 0} e^k_{p_0q_0}(p,q)K(p,q,u,v)e^l_{u_0v_0}(u,v)^{\top}\notag\\[.2 cm]
      &=\langle C e^k_{p_0q_0}, e^l_{u_0v_0}\rangle_{_{\!\mathcal{H}_2}}= a^{kl}_{p_0q_0u_0v_0}.\notag 
      \end{align}

  \item [(ii)] De (\ref{Oper-cov-C}), (\ref{def-Cn}), de la Proposici\'on \ref{Nucleos-Wn} y del \'item (i), se obtiene
\begin{align} \lim_{n\to \infty} \sum_{1\leq k\leq 4}\sum_{p_0,q_0\geq 0} &\langle C_{n} e^k_{p_0q_0},e^k_{p_0q_0}\rangle_{_{\!\mathcal{H}_2}}\notag\\[.2 cm]
&= \lim_{n\to \infty}\sum_{1\leq k\leq 4}\sum_{p_0,q_0\geq 0} \sum_{p,q,p,q\geq 0} e^k_{p_0q_0}(p,q)K_n(p,q,p,q) e^l_{p_0q_0}(p,q)^{\top}\notag\\[.2 cm]
&= \sum_{1\leq k\leq 4}\sum_{p_0,q_0\geq 0} \sum_{p,q,p,q\geq 0} e^k_{p_0q_0}(p,q)K(p,q,p,q) e^l_{p_0q_0}(p,q)^{\top}\notag\\[.2 cm]
&=\sum_{1\leq k\leq 4}\sum_{p_0,q_0\geq 0} \langle C e^k_{p_0q_0},e^k_{p_0q_0}\rangle_{_{\!\mathcal{H}_2}}\notag\\[.2 cm]
&= \sum_{1\leq k\leq 4}\sum_{p_0,q_0\geq 0} a^{kk}_{p_0q_0p_0q_0}<\infty,\notag \end{align}
pues, de la primera ecuaci\'on en (\ref{Oper-cov-C}) y la igualdad de Parseval
\begin{align}
\sum_{1\leq k\leq 4}\sum_{p_0,q_0\geq 0} a^{kk}_{p_0q_0p_0q_0}&=\sum_{1\leq k\leq 4}\sum_{p_0,q_0\geq 0} \langle C e^k_{p_0q_0},e^k_{p_0q_0}\rangle_{_{\!\mathcal{H}_2}}\notag\\[.2 cm]
&=\sum_{1\leq k\leq 4}\sum_{p_0,q_0\geq 0}  E_{\theta}\!\left\{\langle \mathcal{Z},e^k_{p_0q_0}\rangle_{_{\!\mathcal{H}_2}}^2\right\}= E_{\theta}\!\left\{\|\mathcal{Z}\|_{_{\mathcal{H}_2}}^{\,2}\right\}< \infty.\notag
\end{align}

  \item [(iii)] Sean $r_0,s_0\in \mathbb{N}_0$, fijos,
\begin{equation}\label{PI-Psi-e1}
\langle \Phi^*_{ni},e^k_{r_0s_0}\rangle_{_{\mathcal{H}_2}}^2=\frac{1}{n} \phi^2_{kr_0s_0}(\boldsymbol{X}_i^*;\hat{\theta}_n), k=1,2,3,4.
\end{equation}
Para $k=1$, tenemos
\[|\langle \Phi^*_{ni},e^1_{r_0s_0}\rangle_{_{\mathcal{H}_2}}|=\frac{1}{\sqrt{n}} |\phi_{1r_0s_0}(\boldsymbol{X}_i^*;\hat{\theta}_n)|\leq r_0+1+|\hat{\theta}_{1n}- \hat{\theta}_{3n}|+\hat{\theta}_{3n}.\]
Por tanto, dado $\varepsilon >0$, existir\'a $n_0=n_0(r_0,\varepsilon)\in \mathbb{N}_0$ tal que
\[\frac{1}{\sqrt{n}} |\phi_{1r_0s_0}(\boldsymbol{X}_i^*;\hat{\theta}_n)|\leq \frac{1}{\sqrt{n}}(r_0+1+|\hat{\theta}_{1n}- \hat{\theta}_{3n}|+\hat{\theta}_{3n})\leq \varepsilon, \forall n> n_0.\]
De donde,
\begin{equation}\label{I-varepsilon}
I{\left\{\frac{1}{\sqrt{n}} |\phi_{1r_0s_0}(\boldsymbol{X}_i^*;\hat{\theta}_n)|> \varepsilon\right\}}=
\left\{
  \begin{array}{ll}
    1, & \hbox{si  } n\leq n_0,\\
    0, & \hbox{si  } n>n_0.
  \end{array}
\right.
\end{equation}

Se tiene que
\[L_n(\varepsilon,e^1_{r_0s_0})=\frac{1}{n}\sum_{i=1}^n E_*\!\left[\phi^2_{1r_0s_0}(\boldsymbol{X}_i^*;\hat{\theta}_n) I{\left\{\frac{1}{\sqrt{n}} |\phi_{1r_0s_0}(\boldsymbol{X}_i^*;\hat{\theta}_n)|> \varepsilon\right\}}\right].\]
Adem\'as, de (\ref{psi_1rs-2rs})
\begin{align}
\phi^2_{1r_0s_0}(\boldsymbol{X}_i^*;\hat{\theta}_n)&=(r_0+1)^2 \, I{\{X_{1i}^*=r_0+1,X_{2i}^*=s_0\}}\notag\\[.2 cm]
&\hspace{-4mm}+(\hat{\theta}_{1n}-\hat{\theta}_{3n})^2\, I{\{X_{1i}^*=r_0,X_{2i}^*=s_0\}}+\hat{\theta}_{3n}^2 I{\{X_{1i}^*=r_0,X_{2i}^*=s_0-1\}}.\notag
\end{align}

Con esta \'ultima igualdad y (\ref{I-varepsilon})
\begin{align}
&\sum_{i=1}^nE_*\!\left[\phi^2_{1r_0s_0}(\boldsymbol{X}_i^*;\hat{\theta}_n) I{\left\{\frac{1}{\sqrt{n}} |\phi_{1r_0s_0}(\boldsymbol{X}_i^*;\hat{\theta}_n)|> \varepsilon\right\}}\right]\notag\\
&=\sum_{i=1}^n\sum_{r,s\geq 0}\left[(r_0+1)^2 \, I{\{r=r_0+1,s=s_0\}}+(\hat{\theta}_{1n}-\hat{\theta}_{3n})^2 \, I{\{r=r_0,s=s_0\}}\right.\notag\\
&\hspace{22mm}\left.+\hat{\theta}_{3n}^2 \, I{\{r=r_0,s=s_0-1\}}\right]I{\left\{\frac{1}{\sqrt{n}} |\phi_{1r_0s_0}(r,s;\hat{\theta}_n)|> \varepsilon\right\}}\,P(r,s;\hat{\theta}_n)\notag\\
&\leq\sum_{i=1}^{n_0}\sum_{r,s\geq 0}\left[(r_0+1)^2 \, I{\{r=r_0+1,s=s_0\}}+(\hat{\theta}_{1n}-\hat{\theta}_{3n})^2 \, I{\{r=r_0,s=s_0\}}\right.\notag\\
&\hspace{22mm}\left.+\hat{\theta}_{3n}^2 \, I{\{r=r_0,s=s_0-1\}}\right]\,P(r,s;\hat{\theta}_n)\notag\\
& =\sum_{i=1}^{n_0}\!\left\{(r_0+1)^2P(r_0+1,s_0;\hat{\theta}_n) +(\hat{\theta}_{1n}\!-\hat{\theta}_{3n})^2P(r_0,s_0;\hat{\theta}_n)+ \hat{\theta}_{3n}^2P(r_0,s_0\!-1;\hat{\theta}_n)\!
\right\}\notag\\
& \leq n_0\left\{(r_0+1)^2 +(\hat{\theta}_{1n}-\hat{\theta}_{3n})^2+ \hat{\theta}_{3n}^2
\right\}.\notag
\end{align}

Por tanto
\[L_n(\varepsilon,e^1_{r_0s_0})\leq \frac{n_0}{n}\!\left\{(r_0+1)^2 +(\hat{\theta}_{1n}-\hat{\theta}_{3n})^2+ \hat{\theta}_{3n}^2
\right\}.\]

Puesto que $r_0$ y $\varepsilon$ son fijos y $n_0=n_0(r_0,\varepsilon)$, entonces esta \'ultima desigualdad converge a cero. As\'i
\[\lim_{n\to \infty}L_n(\varepsilon,e^1_{r_0s_0})=0.\]

An\'alogamente se logra que $\lim_{n\to \infty}L_n(\varepsilon,e^2_{r_0s_0})=0$.\vskip .2 cm

De (\ref{PI-Psi-e1}), para $k=3$ y de (\ref{psi_3rs-4rs})
\[\langle \Phi^*_{ni},e^3_{r_0s_0}\rangle_{_{\mathcal{H}_2}}=\frac{1}{\sqrt{n}} \boldsymbol{\ell}(\boldsymbol{X}_i^*;\hat{\theta}_n) d(r_0,s_0;\hat{\theta}_n)^{\top},\]
donde $d(r,s;\hat{\theta}_n)=(P(r,s;\hat{\theta}_n),0, P(r,s-1;\hat{\theta}_n)-P(r,s;\hat{\theta}_n))$ y por tanto
\begin{align}
&L_n(\varepsilon,e^3_{r_0s_0})\notag\\
&\hspace{4mm}=\frac{1}{n}\sum_{i=1}^n E_*\!\left[\left\{ \boldsymbol{\ell}(\boldsymbol{X}_i^*;\hat{\theta}_n) d(r_0,s_0;\hat{\theta}_n)^{\top}\right\}^2 I{\left\{\frac{1}{\sqrt{n}} \left| \boldsymbol{\ell}(\boldsymbol{X}_i^*;\hat{\theta}_n) d(r_0,s_0;\hat{\theta}_n)^{\top}\right|> \varepsilon\right\}}\right].\notag
\end{align}

Ahora, como
\begin{equation}\label{(l*d)^2}
\left|\boldsymbol{\ell} (\boldsymbol{X}_i^*;\hat{\theta}_n)d(r,s;\hat{\theta}_n)^{\top}\right| \leq \|\boldsymbol{\ell} (\boldsymbol{X}_i^*;\hat{\theta}_n)\| \|d(r,s;\hat{\theta}_n)\| \leq \sqrt{2}\|\boldsymbol{\ell} (\boldsymbol{X}_i^*;\hat{\theta}_n)\|,
\end{equation}
entonces
\[\varepsilon < \frac{1}{\sqrt{n}} | \boldsymbol{\ell} (\boldsymbol{X}_i^*;\hat{\theta}_n)d(r_0,s_0;\hat{\theta}_n)^{\top}| \leq \frac{\sqrt{2}}{\sqrt{n}}\|\boldsymbol{\ell} (\boldsymbol{X}_i^*;\hat{\theta}_n)\|,\]
por tanto
\begin{align}
I{\left\{\frac{1}{\sqrt{n}} \left| \boldsymbol{\ell}(\boldsymbol{X}_i^*;\hat{\theta}_n) d(r_0,s_0;\hat{\theta}_n)^{\top}\right|> \varepsilon\right\}} &\leq I{\left\{\frac{\sqrt{2}}{\sqrt{n}}\,\|\boldsymbol{\ell} (\boldsymbol{X}_i^*;\hat{\theta}_n)\|> \varepsilon\right\}}\notag\\[.2 cm]
&=I{\left\{\|\boldsymbol{\ell} (\boldsymbol{X}_i^*;\hat{\theta}_n)\| >\gamma\right\}},\notag
\end{align}
donde $\gamma=\gamma(n)=\varepsilon \frac{\sqrt{n}}{\sqrt{2}}$ es tal que $\gamma\to \infty$ cuando $n\to \infty$.\vskip .2 cm

Adem\'as, de (\ref{(l*d)^2}), se tiene que $\,\left( \boldsymbol{\ell}(\boldsymbol{X}_i^*;\hat{\theta}_n) d(r_0,s_0;\hat{\theta}_n)^{\top}\right)^2\!\!\leq 2 \|\boldsymbol{\ell} (\boldsymbol{X}_i^*;\hat{\theta}_n)\|^2$ y como $\boldsymbol{X}_1^*,\boldsymbol{X}_2^*,\ldots,\boldsymbol{X}_n^*$ son vectores aleatorios iid de  $\boldsymbol{X}^*\sim PB(\hat{\theta}_n)$, entonces
\begin{align}
L_n(\varepsilon,e^3_{r_0s_0})&\leq\frac{2}{n}\sum_{i=1}^n E_*\!\left(\|\boldsymbol{\ell} (\boldsymbol{X}_i^*;\hat{\theta}_n)\|^2 I{ \left\{\|\boldsymbol{\ell} (\boldsymbol{X}_i^*;\hat{\theta}_n)\| >\gamma\right\}}\right)\notag\\[.1 cm]
&\leq 2 \sup_{\hat{\theta}_n\in \Theta_0}E_{\hat{\theta}_n}\!\!\left(\|\boldsymbol{\ell} (\boldsymbol{X}^*;\hat{\theta}_n)\|^2 I{ \left\{\|\boldsymbol{\ell} (\boldsymbol{X}^*;\hat{\theta}_n)\| >\gamma\right\}}\right),\notag
\end{align}

y como $\gamma\to \infty$ cuando $n\to \infty$, entonces, por el Supuesto \ref{E(l*l)-l-cont} (1) se concluye que
\[\lim_{n\to \infty}L_n(\varepsilon,e^3_{r_0s_0})=0.\]

De la misma forma se prueba que $\lim_{n\to \infty}L_n(\varepsilon,e^4_{r_0s_0})=0$.
\end{itemize}

As\'i, por el Lema \ref{kundu}, tenemos que
$\Phi_n^*\mathop{\longrightarrow} \limits^{L} \mathcal{Z}$, c.s., en $\mathcal{H}_2$. Como ya hab\'iamos mencionado, $\mathcal{Z}$ es tambi\'en el l\'imite d\'ebil de $(\Phi_{1n},\Phi_{2n},\Phi_{3n},\Phi_{4n})$  cuando los datos son iid provenientes de $\boldsymbol{X}\sim PB(\theta)$. Finalmente, el resultado se obtiene por el Teorema de la aplicaci\'on continua. $\square$\\

Similarmente a lo expuesto en la parte final de la Secci\'on \ref{Aproximacion-distribucion-nula}, sea
$$w^*_{n,\alpha}=\inf\!\left\{x:P_*(W^*_n(\hat{\theta}^*_n)\geq x)\leq \alpha\right\}$$
el percentil superior $\alpha$ de la distribuci\'on bootstrap de $W_n(\hat{\theta}_n)$.

Del Teorema \ref{Cons-boot-Wn}, la funci\'on test
$$\Psi^*_W=\left\{
\begin{array}{ll}
1, & \text{si}\ W_n(\hat{\theta}_n)\geq w^*_{n,\alpha}, \\[.15 cm]
0, & \text{en caso contrario},
\end{array}
\right.$$
o equivalentemente, el test que rechaza $H_0$ cuando
$$p_{_W}^*=P_*\!\left(W^*_n(\hat{\theta}^{\,*}_n) \geq W_{obs}\right)\leq \alpha,$$
es asint\'oticamente correcto en el sentido que, cuando $H_0$ es cierta
$$\lim_{n\to \infty} P_{\theta}\!\left(\Psi^*_W=1\right)=\alpha,$$
donde $W_{obs}$ es el valor observado del test estad\'istico $W_n (\hat{\theta}_n)$.

\begin{observacion}\label{Boot-Wn} Para aproximar $w^*_{n,\alpha}$ se sigue el mismo procedimiento presentado al final de la Secci\'on \ref{Aproximacion-distribucion-nula}, con los cambios obvios.
\end{observacion}

\newpage
\section{Alternativas} \label{Alternativas-Wn}
En esta secci\'on estudiaremos el comportamiento del test $\Psi^*_W$ bajo alternativas fijas y locales.

\subsection{Alternativas fijas}\label{Alternativas-fijas-Wn}
Como una de nuestras metas es obtener tests de bondad de ajuste que sean consistentes, lo pr\'oximo que haremos es estudiar este t\'opico para el nuevo test que proponemos. Con este fin, primero obtendremos el l\'imite c.s. de $W_n(\hat{\theta}_n)$.

\begin{teorema}\label{Convergencia-de-W_n(theta-hat)}
Sean $\boldsymbol{X}_1, \boldsymbol{X}_2,\ldots,\boldsymbol{X}_n$ vectores aleatorios iid de $\boldsymbol{X}=(X_1,X_2)\in\mathbb{N}_0^2$,
con $E(X_k^2)<\infty$, $k=1,2$. Sea $p(r,s)=P(X_1=r,X_2=s)$. Si
$\hat{\theta}_n \mathop{\longrightarrow}\limits^{c.s.} \theta$, para alg\'un $\theta\in \mathbb{R}^3$, entonces
\[W_n(\hat{\theta}_n)\ \mathop{\longrightarrow}\limits^{c.s.}\sum_{r,s \geq 0}\left\{a_1(r,s;\theta)^2+a_2(r,s;\theta)^2\right\}=\nu(P;\theta),\]
donde
\begin{align}
a_1(r,s;\theta)&=(r+1)p(r+1,s)-(\theta_1-\theta_3)p(r,s)-\theta_3p(r,s-1), \notag\\[.2 cm]
a_2(r,s;\theta)&=(s+1)p(r,s+1)-(\theta_2-\theta_3)p(r,s)-\theta_3p(r-1,s).\notag
\end{align}
\end{teorema}

\begin{observacion}\label{eta(P;theta) geq 0}
Notar que $\nu(P;\theta)\geq 0$. Adem\'as, de la Proposici\'on \ref{Soluc-sistema-de-dos-EDP} obtenemos que, $\nu(P;\theta)=0$ s\'i y s\'olo si $H_0$ es cierta.
\end{observacion}

\nt {\bf Demostraci\'on del Teorema \ref{Convergencia-de-W_n(theta-hat)}} \hspace{2pt}
Se tiene que
\begin{equation}\label{daux1}
d_1(r,s;\hat{\theta}_n)=d_1(r,s;{\theta})-(\hat{\theta}_{1n}-\theta_1)p_n(r,s)+ (\hat{\theta}_{3n}-\theta_3)\{p_n(r,s)-p_n(r,s-1)\}.
\end{equation}
Tambi\'en se tiene que
\[\sum_{r,s\geq 0}d_1(r,s;{\theta})^2=\frac{1}{n^2}\sum_{i\neq j} h_1(\boldsymbol{X}_i,\boldsymbol{X}_j; {\theta})+
\frac{1}{n^2}\sum_{i=1}^n h_1(\boldsymbol{X}_i,\boldsymbol{X}_i; {\theta}).\]
Por la ley fuerte de los grandes n\'umeros para U-estad\'isticos (v\'ease por ejemplo Teorema 5.4 en Serfling (1980) \cite{Ser80}),
\[\frac{1}{n^2}\sum_{i\neq j} h_1(\boldsymbol{X}_i,\boldsymbol{X}_j; {\theta})\mathop{\longrightarrow}\limits^{c.s.}
E\{h_1(\boldsymbol{X}_1,\boldsymbol{X}_2; {\theta})\}=\sum_{r,s \geq 0}a_1(r,s;\theta)^2.\]
Como $E(X_1^2)<\infty$, por la ley fuerte de los grandes n\'umeros,
\[\frac{1}{n}\sum_{i=1}^n h_1(\boldsymbol{X}_i,\boldsymbol{X}_i; {\theta})\mathop{\longrightarrow}\limits^{c.s.} E\{h_1(\boldsymbol{X}_1,\boldsymbol{X}_1; {\theta})\}=E\left\{
\sum_{r,s \geq 0}\phi_{1rs}(\boldsymbol{X}_1; {\theta})^2\right\}<\infty.\]
Por tanto,
\begin{equation}\label{daux2}
\sum_{r,s\geq 0}d_1(r,s;{\theta})^2\mathop{\longrightarrow}\limits^{c.s.}
\sum_{r,s \geq 0}a_1(r,s;\theta)^2.
\end{equation}

Como $p_n(r,s)^2\leq p_n(r,s)$, $\forall r,s \geq 0$, y  $\sum_{r,s\geq 0}p_n(r,s)=1$, se tiene que
\begin{equation}\label{daux3}
(\hat{\theta}_{1n}-\theta_1)^2 \sum_{r,s\geq 0} p_n(r,s)^2 \leq (\hat{\theta}_{1n}-\theta_1)^2 =o(1),
\end{equation}
y an\'alogamente,
\begin{equation}\label{daux4}
(\hat{\theta}_{3n}-\theta_3)^2\sum_{r,s\geq 0}\{p_n(r,s)-p_n(r,s-1)\}^2=o(1).
\end{equation}
De (\ref{daux1})--(\ref{daux4}),
\begin{equation}\label{daux5}
\sum_{r,s\geq 0}d_1(r,s;\hat{\theta}_n)^2\mathop{\longrightarrow}\limits^{c.s.}
\sum_{r,s \geq 0}a_1(r,s;\theta)^2.
\end{equation}
Procediendo similarmente
\begin{equation}\label{daux6}
\sum_{r,s\geq 0}d_2(r,s;\hat{\theta}_n)^2\mathop{\longrightarrow}\limits^{c.s.}
\sum_{r,s \geq 0}a_2(r,s;\theta)^2.
\end{equation}
Finalmente, el resultado se tiene de (\ref{daux5}) y (\ref{daux6}). $\square$\\

Como una consecuencia de los Teoremas \ref{TeoConvDebil-Wn}, \ref{Cons-boot-Wn} y \ref{Convergencia-de-W_n(theta-hat)}, y la Observaci\'on \ref{eta(P;theta) geq 0}, el siguiente resultado da la consistencia del test $\Psi^*_W$.

\begin{corolario}\label{Altern-fijas-Wn}
Sean $\boldsymbol{X}_1, \boldsymbol{X}_2,\ldots,\boldsymbol{X}_n$ vectores aleatorios iid de $\boldsymbol{X}\in\mathbb{N}_0^2$. Supongamos que se cumplen las hip\'otesis de los Teoremas \ref{TeoConvDebil-Wn} y \ref{Cons-boot-Wn}.

Si $H_0$ no es cierta, entonces $P(\Psi^*_W=1) \to 1.$
\end{corolario}

\subsection{Alternativas contiguas} \label{Alternativas-contiguas-Wn}
Como se estableci\'o en el Teorema \ref{TeoConvDebil-Wn}, cuando $H_0$ es cierta, $nW_n(\hat{\theta}_n)$ converge d\'ebilmente a una combinaci\'on lineal de v.a. $\chi^2$ independientes con 1 grado de libertad, donde los pesos son los autovalores del operator $C(\theta)$ dado en (\ref{Operador_C_para_Wn}).
Sea $\{\phi_j\}$ el conjunto ortonormal de autofunciones correspondientes a los autovalores $\{\lambda_j\}$ de $C(\theta)$.

El siguiente Teorema da la ley l\'imite del test estad\'istico bajo las alternativas $P_n(x_1,x_2)$ en (\ref{P(n)=P(theta)+b}).

\begin{teorema}\label{Altern-contiguas-Wn}
Sea $\boldsymbol{X}_{n,1}, \boldsymbol{X}_{n,2},\ldots,\boldsymbol{X}_{n,n}$ un arreglo triangular de vectores aleatorios bivariantes que son independientes por filas y que toman valores en $\mathbb{N}_0^2$, con funci\'on de probabilidad dada por $P_n(x_1,x_2)$ definida en (\ref{P(n)=P(theta)+b}). Supongamos que se cumplen los Supuestos \ref{hat(theta)-theta} y \ref{b(x,y)}. Entonces
\[nW_n(\hat{\theta}_n)\ \mathop{\longrightarrow} \limits^{L} \ \sum_{j=1}^{\infty} \lambda_j\left(Z_j+c_j\right)^2,\]
donde $c_j=\mathop{\sum}\limits_{x_1,x_2}b(x_1,x_2)\phi_j(x_1,x_2)$ y $Z_1,Z_2,\ldots$ son variables normales est\'andar independientes.
\end{teorema}
\nt {\bf Demostraci\'on} \hspace{2pt} Sigue los mismos pasos que la demostraci\'on del Teorema \ref{Altern-contiguas-S_n,w}, por lo tanto la omitiremos. $\square$\vskip .2 cm

Del Teorema \ref{Altern-contiguas-Wn}, concluimos que el test $\Psi^*_W$ es capaz de detectar alternativas como las establecidas en (\ref{P(n)=P(theta)+b}), que convergen a la DPB a una raz\'on de $n^{-1/2}$.

\newpage

\[\begin{array}{c}\\  \\ \end{array}\]

\newpage


\chapter{Resultados num\'ericos} \label{Resultados-numericos}
Las propiedades estudiadas en las Secciones \ref{Aproximacion-distribucion-nula}, \ref{Alternativas_Rnw-Srw}, \ref{Distribucion-nula-Wn} y \ref{Alternativas-Wn} son asint\'oticas, esto es, tales propiedades describen el comportamiento de los tests propuestos para muestras de tama\~no grande. Para estudiar la bondad de la aproximaci\'on bootstrap para muestras de tama\~no finito, as\'i como tambi\'en para comparar la potencia de los tests propuestos tanto entre ellos como con otros tests, hemos llevado a cabo un experimento de simulaci\'on. En esta secci\'on describimos dicho experimento y damos un resumen de los resultados que hemos obtenido.

Todos los c\'alculos computacionales realizados en este texto se llevaron a cabo mediante el uso de programas escritos en el lenguaje R \cite{R}.

Como competidores de los tests estad\'isticos propuestos, hemos considerado los tests dados por Crockett (1979) \cite{Cro79} (denotado por $T$), Loukas y Kemp (1986) \cite{LoKe86} (denotado por $I_B$), y Rayner y Best (1995) \cite{RaBe95} (denotado por  $NI_B$). La expresi\'on de cada test estad\'istico, as\'i como la regiones cr\'iticas correspondientes ya las presentamos en el Cap\'itulo \ref{Resultados previos y definiciones}.

Para calcular $R_{n,w}(\hat{\theta}_n)$ y $S_{n,w}(\hat{\theta}_n)$ es necesario dar una forma expl\'icita a la funci\'on de peso $w$. En el caso univariante, Baringhaus et al. (2000) \cite{BaGuHe00} consideraron la funci\'on de peso $w(u)=u^a$, con $a\geq 0$. Varias extensiones son posibles. Aqu\'i hemos tenido en cuenta la siguiente
\begin{equation}\label{funcion-peso-explicita}
w(u;a_1,a_2)=u_1^{a_1}u_2^{a_2}.
\end{equation}

Observar que las \'unicas restricciones que hemos impuesto a la funci\'on de peso son que $w$ sea positiva casi en todas partes en $[0,1]^2$ y la establecida en (\ref{int-funcion-peso}). La funci\'on $w(u;a_1,a_2)$ dada en (\ref{funcion-peso-explicita}) cumple dichas condiciones siempre que $a_i>-1$, $i=1,2$.

En el Cap\'itulo \ref{Expresiones-matematicas} damos las expresiones matem\'aticas de los estad\'isticos $R_{n,w}(\hat{\theta}_n)$ y $S_{n,w}(\hat{\theta}_n)$ para la funci\'on de peso dada en (\ref{funcion-peso-explicita}). Tambi\'en damos la expresi\'on para el estad\'istico $W_n(\hat{\theta}_n)$.

Adem\'as, mostramos los resultados que obtuvimos al aplicar los tests propuestos a dos conjuntos de datos reales.

\section{Datos simulados}\label{Simulacion-datos}

Para estudiar la bondad de la aproximaci\'on bootstrap en muestras de tama\~no finito tanto para los tests propuestos, como  para las aproximaciones a la distribuci\'on nula de los tests estad\'isticos $T$, $I_B$ y $NI_B$, que est\'an basados en sus distribuciones asint\'oticas nulas, generamos muestras de tama\~no $n=30(20)70$ de la distribuci\'on $PB(\theta_1,\theta_2,\theta_3)$, con $\theta_1=\theta_2=$1.0 y $\theta_3$ tal que el coeficiente de correlaci\'on, $\rho=\theta_3/\sqrt{\theta_1\,\theta_2}$ sea igual a 0.25, 0.50 y 0.75, con el fin de examinar la bondad de las aproximaciones para datos con una correlaci\'on baja, media y alta, respectivamente.

Para estimar el par\'ametro $\theta$ empleamos el m\'etodo de m\'axima verosimilitud como se describe en Kocherlakota y Kocherlakota (1992, pp. 103--105) \cite{KoKo92}. Luego, aproximamos los $p-$valores bootstrap, $p^*$, de los tests propuestos, para ello, en el caso de $R_{n,w}(\hat{\theta}_n)$ y $S_{n,w}(\hat{\theta}_n)$ usamos la funci\'on de peso dada en (\ref{funcion-peso-explicita}) para $(a_1,a_2)\in \{(0,0), (1,0)\}$ y generamos $B=500$ muestras bootstrap. Tambi\'en, calculamos los $p-$valores (asint\'oticos) asociados a los tests estad\'isticos $T$, $I_B$ y $NI_B$.

El procedimiento anterior lo repetimos $1000$ veces y calculamos la fracci\'on de los $p-$valores estimados que resultaron ser menores o iguales que 0.05 y 0.10, que son las estimaciones de las probabilidades del error tipo I para $\alpha=$ 0.05 y 0.10 (en las tablas, denotamos esto como f05 y f10), respectivamente.

Si las aproximaciones consideradas fuesen exactas, entonces los $p-$valores calculados deber\'ian ser una muestra aleatoria de una distribuci\'on uniforme en el intervalo $(0,1)$. Por lo tanto, para medir la bondad de las aproximaciones consideradas, calculamos el $p-$valor del test estad\'istico de uniformidad de Kolmogorov-Smirnov ($KS$) para cada conjunto de 1000 $p-$valores obtenido para cada uno de los tests estad\'isticos.

Repetimos el experimento anterior para $\theta_1=$ 1.5, $\theta_2=$ 1.0 y $\theta_3$ de manera que el coeficiente de correlaci\'on,
$\rho=\theta_3/\sqrt{\theta_1\,\theta_2}$, fuera aproximadamente igual a 0.25, 0.50 y 0.75. En este caso, ya que $\theta_1\neq \theta_2$,
adem\'as de $(a_1,a_2)\in \{(0,0), (1,0)\}$ consideramos
$(a_1,a_2)=(0,1)$ para $R_{n,a}$ y $S_{n,a}$, con el fin de examinar el efecto de dar un peso diferente a cada uno de los componentes cuando estos tienen distintas esperanzas.

Las Tablas \ref{Resultados-igual-E-1-simulacion-error-I}-- \ref{Resultados-diferente-E-3-simulacion-error-I} resumen los resultados obtenidos. En dichas tablas, denotamos por $R_{n,a}$ y $S_{n,a}$ a los tests estad\'isticos $R_{n,w}(\hat{\theta}_n)$ y $S_{n,w}(\hat{\theta}_n)$, respectivamente, cuando la funci\'on de peso $w$ toma la forma dada por (\ref{funcion-peso-explicita}), para alg\'un $a=(a_1,a_2)$.

Observando los valores dados en las tablas, concluimos que la aproximaci\'on asint\'otica de los $p-$valores de los tests estad\'isticos $T$, $I_B$ y $NI_B$ no da resultados satisfactorios para los casos tratados. Por el contrario, el bootstrap proporciona una aproximaci\'on precisa a la distribuci\'on nula de $R_{n,w}(\hat{\theta}_n)$ y $S_{n,w}(\hat{\theta}_n)$ en todos los casos tratados. La aproximaci\'on bootstrap a la distribuci\'on nula de $W_n(\hat{\theta}_n)$ es adecuada para $n\geq 50$.

\begin{landscape}
\begin{table}
\caption{Resultados de simulaci\'on para la probabilidad del error tipo I. Caso $E(X_1)=E(X_2)$, $\rho=$ 0.25} \label{Resultados-igual-E-1-simulacion-error-I}
\begin{center}
{\normalsize 
\begin{tabular}{|l|cc|c|cc|c|cc|c|}
\cline{2-10}\multicolumn{1}{c|}{}
&\multicolumn{3}{c|}{$n=30$}& \multicolumn{3}{c|}{$n=50$} & \multicolumn{3}{c|}{$n=70$}\\ \hline
\begin{tabular}{l}
$\theta=$ (1.0,1.0,0.25) \\
$\rho=$ 0.25
\end{tabular}
& f05 & f10 & $KS$ & f05 & f10 & $KS$ & f05 & f10 & $KS$\\
\hline
$R_{n,(0,0)}$ & 0.035 & 0.091 & 0.818621 & 0.047 & 0.100 & 0.459543 & 0.041 & 0.083 & 0.329116\\
$S_{n,(0,0)}$ & 0.046 & 0.090 & 0.902243 & 0.049 & 0.097 & 0.818621 & 0.041 & 0.089 & 0.769894\\
\hline
$R_{n,(1,0)}$ & 0.035 & 0.093 & 0.257432 & 0.050 & 0.100 & 0.769894 & 0.042 & 0.086 & 0.413150\\
$S_{n,(1,0)}$ & 0.039 & 0.094 & 0.665399 & 0.048 & 0.098 & 0.995881 & 0.042 & 0.093 & 0.559560\\
\hline
$W_{n}$       & 0.022 & 0.056 & 1.00e-05 & 0.033 & 0.078 & 0.111356 & 0.038 & 0.090 & 0.612128\\
\hline
$T$           & 0.011 & 0.031 & $<$ 2.2e-16 & 0.046 & 0.092 & 0.060937 & 0.013 & 0.038 & $<$ 2.2e-16 \\
$I_B$         & 0.027 & 0.061 & $<$ 2.2e-16 & 0.098 & 0.144 & 0.001642 & 0.022 & 0.054 & $<$ 2.2e-16 \\
$NI_B$        & 0.010 & 0.034 & $<$ 2.2e-16 & 0.068 & 0.111 & 0.003452 & 0.013 & 0.033 & $<$ 2.2e-16 \\
\hline
\end{tabular}}
\end{center}
\end{table}
\end{landscape}

\begin{landscape}
\begin{table}
\caption{Resultados de simulaci\'on para la probabilidad del error tipo I. Caso $E(X_1)=E(X_2)$, $\rho=$ 0.50} \label{Resultados-igual-E-2-simulacion-error-I}
\begin{center}
{\normalsize 
\begin{tabular}{|l|cc|c|cc|c|cc|c|}
\cline{2-10}\multicolumn{1}{c|}{}
&\multicolumn{3}{c|}{$n=30$}& \multicolumn{3}{c|}{$n=50$} & \multicolumn{3}{c|}{$n=70$}\\ \hline
\begin{tabular}{l}
$\theta=$ (1.0,1.0,0.50) \\
$\rho=$ 0.50
\end{tabular}
& f05 & f10 & $KS$ & f05 & f10 & $KS$ & f05 & f10 & $KS$\\
\hline
$R_{n,(0,0)}$ & 0.047 & 0.091 & 0.508494 & 0.044 & 0.101 & 0.129364 & 0.048 & 0.100 & 0.413150\\
$S_{n,(0,0)}$ & 0.049 & 0.094 & 0.863178 & 0.045 & 0.098 & 0.329116 & 0.046 & 0.100 & 0.459543\\
\hline
$R_{n,(1,0)}$ & 0.049 & 0.097 & 0.665399 & 0.039 & 0.096 & 0.863178 & 0.054 & 0.097 & 0.129364\\
$S_{n,(1,0)}$ & 0.045 & 0.092 & 0.459543 & 0.043 & 0.093 & 0.291736 & 0.053 & 0.096 & 0.459543\\
\hline
$W_{n}$       & 0.022 & 0.061 & 0.013476 & 0.032 & 0.077 & 0.111356 & 0.037 & 0.081 & 0.111356\\
\hline
$T$           & 0.064 & 0.095 & 0.018402 & 0.024 & 0.039 & $<$ 2.2e-16 & 0.021 & 0.053 & 0.000179 \\
$I_B$         & 0.152 & 0.186 & $<$ 2.2e-16 & 0.073 & 0.119 & $<$ 2.2e-16 & 0.051 & 0.081 & $<$ 2.2e-16 \\
$NI_B$        & 0.080 & 0.132 & 0.054004 & 0.018 & 0.049 & $<$ 2.2e-16 & 0.007 & 0.035 & $<$ 2.2e-16 \\
\hline
\end{tabular}}
\end{center}
\end{table}
\end{landscape}

\begin{landscape}
\begin{table}
\caption{Resultados de simulaci\'on para la probabilidad del error tipo I. Caso $E(X_1)=E(X_2)$, $\rho=$ 0.75} \label{Resultados-igual-E-3-simulacion-error-I}
\begin{center}
{\normalsize 
\begin{tabular}{|l|cc|c|cc|c|cc|c|}
\cline{2-10}\multicolumn{1}{c|}{}
&\multicolumn{3}{c|}{$n=30$}& \multicolumn{3}{c|}{$n=50$} & \multicolumn{3}{c|}{$n=70$}\\ \hline
\begin{tabular}{l}
$\theta=$ (1.0,1.0,0.75) \\
$\rho=$ 0.75
\end{tabular}
& f05 & f10 & $KS$ & f05 & f10 & $KS$ & f05 & f10 & $KS$\\
\hline
$R_{n,(0,0)}$ & 0.045 & 0.097 & 0.769894 & 0.060 & 0.107 & 0.995881 & 0.056 & 0.107 & 0.508494\\
$S_{n,(0,0)}$ & 0.053 & 0.092 & 0.369615 & 0.061 & 0.106 & 0.769894 & 0.050 & 0.099 & 0.902243\\
\hline
$R_{n,(1,0)}$ & 0.052 & 0.096 & 0.995881 & 0.059 & 0.106 & 0.960002 & 0.050 & 0.109 & 0.769894\\
$S_{n,(1,0)}$ & 0.050 & 0.096 & 0.818621 & 0.059 & 0.104 & 0.934732 & 0.048 & 0.106 & 0.329116\\
\hline
$W_{n}$       & 0.029 & 0.076 & 0.024117 & 0.036 & 0.085 & 0.111356 & 0.038 & 0.088 & 0.129364\\
\hline
$T$           & 0.025 & 0.049 & $<$ 2.2e-16 & 0.034 & 0.065 & 1.00e-07 & 0.024 & 0.058 & 5.30e-06 \\
$I_B$         & 0.116 & 0.140 & $<$ 2.2e-16 & 0.141 & 0.162 & $<$ 2.2e-16 & 0.129 & 0.153 & $<$ 2.2e-16 \\
$NI_B$        & 0.045 & 0.074 & 6.10e-06 & 0.033 & 0.081 & $<$ 2.2e-16 & 0.029 & 0.063 & $<$ 2.2e-16 \\
\hline
\end{tabular}}
\end{center}
\end{table}
\end{landscape}

\begin{landscape}
\begin{table}
\caption{Resultados de simulaci\'on para la probabilidad del error tipo I. Caso $E(X_1)\neq E(X_2)$, $\rho\approx$ 0.25} \label{Resultados-diferente-E-1-simulacion-error-I}
\begin{center}
{\normalsize 
\begin{tabular}{|l|cc|c|cc|c|cc|c|}
\cline{2-10}\multicolumn{1}{c|}{}
&\multicolumn{3}{c|}{$n=30$}& \multicolumn{3}{c|}{$n=50$} & \multicolumn{3}{c|}{$n=70$}\\ \hline
\begin{tabular}{l}
$\theta=$ (1.5,1.0,0.31) \\
$\rho=$ 0.25311
\end{tabular}
& f05 & f10 & $KS$ & f05 & f10 & $KS$ & f05 & f10 & $KS$\\
\hline
$R_{n,(0,0)}$ & 0.048 & 0.101 & 0.257432 & 0.053 & 0.110 & 0.508494 & 0.047 & 0.100 & 0.989545\\
$S_{n,(0,0)}$ & 0.049 & 0.104 & 0.769894 & 0.052 & 0.104 & 0.172476 & 0.045 & 0.092 & 0.508494\\
\hline
$R_{n,(1,0)}$ & 0.046 & 0.097 & 0.413150 & 0.053& 0.104 & 0.863178 & 0.046 & 0.104 & 0.559560\\
$S_{n,(1,0)}$ & 0.042 & 0.098 & 0.291736 & 0.049 & 0.104 & 0.459543 & 0.051 & 0.104 & 0.508494\\
\hline
$R_{n,(0,1)}$ & 0.055 & 0.106 & 0.718379 & 0.061 & 0.112 & 0.902243 & 0.048 & 0.090 & 0.863178\\
$S_{n,(0,1)}$ & 0.049 & 0.103 & 0.459543 & 0.051 & 0.102 & 0.329116 & 0.048 & 0.089 & 0.612128\\
\hline
$W_{n}$       & 0.022 & 0.066 & 0.041633 & 0.036 & 0.076 & 0.111356 & 0.037 & 0.082 & 0.111356\\
\hline
$T$           & 0.018 & 0.046 & 1.00e-07 & 0.021 & 0.060 & 0.000318 & 0.053 & 0.093 & 0.099411 \\
$I_B$         & 0.031 & 0.060 & $<$ 2.2e-16 & 0.013 & 0.028 & $<$ 2.2e-16 & 0.108 & 0.161 & 0.000589 \\
$NI_B$        & 0.016 & 0.041 & $<$ 2.2e-16 & 0.010 & 0.018 & $<$ 2.2e-16 & 0.076 & 0.136 & 0.048751 \\
\hline
\end{tabular}}
\end{center}
\end{table}
\end{landscape}

\begin{landscape}
\begin{table}
\caption{Resultados de simulaci\'on para la probabilidad del error tipo I. Caso $E(X_1)\neq E(X_2)$, $\rho\approx$ 0.50} \label{Resultados-diferente-E-2-simulacion-error-I}
\begin{center}
{\normalsize 
\begin{tabular}{|l|cc|c|cc|c|cc|c|}
\cline{2-10}\multicolumn{1}{c|}{}
&\multicolumn{3}{c|}{$n=30$}& \multicolumn{3}{c|}{$n=50$} & \multicolumn{3}{c|}{$n=70$}\\ \hline
\begin{tabular}{l}
$\theta=$ (1.5,1.0,0.62) \\
$\rho=$ 0.50623
\end{tabular}
& f05 & f10 & $KS$ & f05 & f10 & $KS$ & f05 & f10 & $KS$\\
\hline
$R_{n,(0,0)}$ & 0.050 & 0.094 & 0.329116 & 0.048 & 0.093 & 0.902243 & 0.054 & 0.110 & 0.665399\\
$S_{n,(0,0)}$ & 0.046 & 0.100 & 0.769894 & 0.048 & 0.091 & 0.769894 & 0.045 & 0.094 & 0.226206\\
\hline
$R_{n,(1,0)}$ & 0.049 & 0.093 & 0.718379 & 0.046 & 0.092 & 0.769894 & 0.050 & 0.104 & 0.718379\\
$S_{n,(1,0)}$ & 0.043 & 0.095 & 0.612128 & 0.048 & 0.087 & 0.818621 & 0.050 & 0.097 & 0.508494\\
\hline
$R_{n,(0,1)}$ & 0.050 & 0.089 & 0.612128 & 0.043 & 0.100 & 0.718379 & 0.054 & 0.100 & 0.665399\\
$S_{n,(0,1)}$ & 0.052 & 0.091 & 0.459543 & 0.043 & 0.099 & 0.459543 & 0.052 & 0.089 & 0.508494\\
\hline
$W_{n}$       & 0.026 & 0.055 & 0.003013 & 0.037 & 0.071 & 0.111356 & 0.039 & 0.079 & 0.111356\\
\hline
$T$           & 0.056 & 0.088 & 0.000526 & 0.050 & 0.104 & 0.011917 & 0.049 & 0.096 & 0.001109 \\
$I_B$         & 0.147 & 0.201 & $<$ 2.2e-16 & 0.169 & 0.223 & $<$ 2.2e-16 & 0.147 & 0.196 & $<$ 2.2e-16 \\
$NI_B$        & 0.094 & 0.152 & 0.000622 & 0.082 & 0.145 & 0.006666 & 0.076 & 0.120 & 0.078967 \\
\hline
\end{tabular}}
\end{center}
\end{table}
\end{landscape}

\begin{landscape}
\begin{table}
\caption{Resultados de simulaci\'on para la probabilidad del error tipo I. Caso $E(X_1)\neq E(X_2)$, $\rho\approx$ 0.75} \label{Resultados-diferente-E-3-simulacion-error-I}
\begin{center}
{\normalsize 
\begin{tabular}{|l|cc|c|cc|c|cc|c|}
\cline{2-10}\multicolumn{1}{c|}{}
&\multicolumn{3}{c|}{$n=30$}& \multicolumn{3}{c|}{$n=50$} & \multicolumn{3}{c|}{$n=70$}\\ \hline
\begin{tabular}{l}
$\theta=$ (1.5,1.0,0.92) \\
$\rho=$ 0.75118
\end{tabular}
& f05 & f10 & $KS$ & f05 & f10 & $KS$ & f05 & f10 & $KS$\\
\hline
$R_{n,(0,0)}$ & 0.059 & 0.102 & 0.369615 & 0.053 & 0.094 & 0.508494 & 0.046 & 0.091 & 0.960002\\
$S_{n,(0,0)}$ & 0.055 & 0.105 & 0.612128 & 0.049 & 0.092 & 0.769894 & 0.048 & 0.088 & 0.508494\\
\hline
$R_{n,(1,0)}$ & 0.051 & 0.096 & 0.413150 & 0.051 & 0.099 & 0.459543 & 0.050 & 0.091 & 0.863178\\
$S_{n,(1,0)}$ & 0.053 & 0.101 & 0.459543 & 0.051 & 0.092 & 0.769894 & 0.045 & 0.086 & 0.665399\\
\hline
$R_{n,(0,1)}$ & 0.058 & 0.104 & 0.718379 & 0.049 & 0.098 & 0.459543 & 0.046 & 0.089 & 0.863178\\
$S_{n,(0,1)}$ & 0.058 & 0.108 & 0.863178 & 0.050 & 0.091 & 0.559560 & 0.046 & 0.094 & 0.329116\\
\hline
$W_{n}$       & 0.037 & 0.081 & 0.000714 & 0.042 & 0.079 & 0.111356 & 0.037 & 0.083 & 0.149677\\
\hline
$T$           & 0.029 & 0.059 & 1.70e-06 & 0.057 & 0.094 & 0.008821 & 0.078 & 0.109 & 0.065401\\
$I_B$         & 0.091 & 0.116 & $<$ 2.2e-16 & 0.209 & 0.239 & $<$ 2.2e-16 & 0.196 & 0.220 & $<$ 2.2e-16\\
$NI_B$        & 0.021 & 0.051 & $<$ 2.2e-16 & 0.089 & 0.152 & 0.001554 & 0.094 & 0.149 & 0.003483\\
\hline
\end{tabular}}
\end{center}
\end{table}
\end{landscape}

Para estudiar la potencia repetimos el experimento anterior para muestras de tama\~no $n=50$ para varias alternativas: la distribuci\'on binomial bivariante $BB(m; p_1, p_2, p_3)$, donde $p_1+p_2-p_3\leq 1,\, p_1,p_2\geq p_3>0$;  la distribuci\'on binomial negativa bivariante $BNB(\nu; \gamma_0, \gamma_1, \gamma_2)$, donde $\nu\!\in\! \mathbb{N}, \gamma_0,\gamma_1\!>\!\gamma_2\!>\!0$;  mixturas de la DPB de la forma $pPB(\theta)+(1\!-\!p)PB(\lambda), 0<p<1$, denotada por $PPB(p;\theta,\lambda)$;  la distribuci\'on Neyman tipo A bivariante  $NTAB(\lambda;\lambda_1,\lambda_2,\lambda_3)$, donde $0\!<\!\lambda_1+\lambda_2+\lambda_3\!\leq 1$; y la distribuci\'on serie logar\'itmica bivariante $SLB(\lambda_1,\lambda_2,\lambda_3)$, donde $0\!<\!\lambda_1+\lambda_2+\lambda_3\!<\!1$ (ver, por ejemplo, Kocherlakota y Kocherlakota \cite{KoKo92} para una descripci\'on de estas distribuciones).

Para generar las muestras aleatorias de las distribuciones bivariantes que usamos como alternativas, implementamos algoritmos computacionales siguiendo los procedimientos de simulaci\'on dados en Kocherlakota y Kocherlakota \cite{KoKo92}.

La distribuci\'on binomial bivariante y la distribuci\'on Neyman tipo A bivariante han sido utilizadas como alternativas en otros art\'iculos relacionados (v\'ease, por ejemplo, Loukas y Kemp (1986) \cite{LoKe86}, Rayner y Best (1995) \cite{RaBe95}). Las otras distribuciones alternativas empleadas son distribuciones an\'alogas a las utilizadas por G\"urtler y Henze (2000) \cite{GuHe00} en el caso univariante.

Los par\'ametros para estas alternativas fueron elegidos de manera que el \'indice de dispersi\'on de cada componente del vector aleatorio estuvieran cercanos de 1, concretamente, de tal manera que $|var(X_i)/E(X_i)-1| <1$, $i=1,2$. En este sentido, las alternativas consideradas est\'an ``cerca'' de la DPB.

Por otra parte, consideramos $a_1=a_2=0$, porque para esta elecci\'on de $a_1$ y $a_2$ los test estad\'isticos $R_{n,w}(\hat{\theta}_n)$ y $S_{n,w}(\hat{\theta}_n)$ que resultan ocupan menos tiempo computacional que para otras combinaciones de $a$.

Las Tablas \ref{Resultados-simulacion-potencia-1} y \ref{Resultados-simulacion-potencia-2} muestran las alternativas consideradas y la potencia estimada para el nivel de significaci\'on nominal $\alpha=$ 0.05. Estas tablas tambi\'en muestran el \'indice de dispersi\'on de cada componente del vector aleatorio, as\'i como el coeficiente de correlaci\'on. Al examinar los resultados en estas tablas llegamos a la conclusi\'on de que los tests que hemos propuesto son capaces de detectar todas las alternativas tratadas, mientras que, en general, los otros tests no pueden detectar la mayor\'ia de las alternativas, especialmente los tests basados en $I_B$ y $NI_B$.

De acuerdo a los resultados de las Tablas \ref{Resultados-simulacion-potencia-1} y \ref{Resultados-simulacion-potencia-2} podemos decir que las potencias de $R_{n,w}(\hat{\theta}_n)$ y $S_{n,w}(\hat{\theta}_n)$ est\'an muy cerca, siendo $R_{n,w}(\hat{\theta}_n)$ un poco m\'as potente que $S_{n,w}(\hat{\theta}_n)$ en la mayor\'ia de los casos tratados. El test basado en $W_n(\hat{\theta}_n)$ es algo menos potente que los otros dos tests propuestos, en las alternativas ensayadas.

\begin{landscape}
\begin{table}
\caption{Resultados de simulaci\'on para la potencia. Alternativas: $BB(m; p_1, p_2, p_3)$, $BNB(\nu; \gamma_0, \gamma_1, \gamma_2)$ y $PPB(p;\theta,\lambda)$ }\label{Resultados-simulacion-potencia-1}
\begin{center}
{\normalsize 
\begin{tabular}{|l|ccc|cccccc|}
\hline
Alternativa & $\frac{var(X_1)}{E(X_1)}$ & $\frac{var(X_2)}{E(X_2)}$& $\rho$ & $R_{n,(0,0)}$ & $S_{n,(0,0)}$ & $W_{n}$ & $T$ & $I_B$ &  $NI_B$\\
\hline\hline
  $BB(1;0$\text{.}$41,0$\text{.}$02,0$\text{.}$01)$ & 0.590 & 0.980 & 0.026 & 0.863 & 0.881 & 0.829 &  0.111 & 0.000 & 0.000\\
  $BB(1;0$\text{.}$41,0$\text{.}$03,0$\text{.}$02)$ & 0.590 & 0.970 & 0.092 & 0.845 & 0.866 & 0.779 &  0.115 & 0.000 & 0.000\\
  $BB(2;0$\text{.}$61,0$\text{.}$01,0$\text{.}$01)$ & 0.390 & 0.990 & 0.080 & 0.988 & 0.953 & 0.948 & 0.931 & 0.004 & 0.004\\
  $BB(1;0$\text{.}$61,0$\text{.}$03,0$\text{.}$02)$ & 0.390 & 0.970 & 0.020 & 1.000 & 1.000 & 0.999 & 0.945 & 0.000 & 0.000\\
  $BB(2;0$\text{.}$71,0$\text{.}$01,0$\text{.}$01)$ & 0.290 & 0.990 & 0.064 & 1.000 & 0.996 & 1.000 & 1.000 & 0.000 & 0.000\\
  \hline\hline
  $BNB(1;0$\text{.}$92,0$\text{.}$97,0$\text{.}$01)$  & 1.920 & 1.970 & 0.486 & 0.928 & 0.890 & 0.524 & 0.843 & 0.622 & 0.974\\  
  $BNB(1;0$\text{.}$97,0$\text{.}$97,0$\text{.}$01)$  & 1.970 & 1.970 & 0.493 & 0.933 & 0.884 & 0.526 & 0.860 & 0.633 & 0.975\\
  $BNB(1;0$\text{.}$97,0$\text{.}$97,0$\text{.}$02)$  & 1.970 & 1.970 & 0.493 & 0.936 & 0.901 & 0.518 & 0.846 & 0.616 & 0.975\\ 
  $BNB(1;0$\text{.}$98,0$\text{.}$98,0$\text{.}$01)$  & 1.980 & 1.980 & 0.495 & 0.944 & 0.906 & 0.530 & 0.855 & 0.607 & 0.980\\
  $BNB(1;0$\text{.}$99,0$\text{.}$99,0$\text{.}$01)$  & 1.990 & 1.990 & 0.498 & 0.932 & 0.893 & 0.510 & 0.850 & 0.585 & 0.973\\
\hline\hline
  $PPB(0$\text{.}$35;(0$\text{.}$2,0$\text{.}$2,0$\text{.}$1);(0$\text{.}$9,1$\text{.}$0,0$\text{.}$6))$ & 1.170 & 1.202 & 0.762 & 0.850 & 0.845 & 0.622 & 0.528 & 0.000 & 0.000\\
  $PPB(0$\text{.}$37;(0$\text{.}$2,0$\text{.}$2,0$\text{.}$1);(0$\text{.}$9,0$\text{.}$9,0$\text{.}$8))$ & 1.178 & 1.178 & 0.956 & 0.748 & 0.754 & 0.642 & 0.458 & 0.001 & 0.001\\
  $PPB(0$\text{.}$37;(0$\text{.}$2,0$\text{.}$2,0$\text{.}$1);(0$\text{.}$9,1$\text{.}$0,0$\text{.}$2))$ & 1.178 & 1.212 & 0.461 & 0.920 & 0.901 & 0.582 & 0.653 & 0.000 & 0.000\\
  $PPB(0$\text{.}$40;(0$\text{.}$2,0$\text{.}$2,0$\text{.}$1);(0$\text{.}$9,1$\text{.}$0,0$\text{.}$3))$ & 1.190 & 1.226 & 0.566 & 0.964 & 0.941 & 0.556 & 0.759 & 0.000 & 0.000\\
  $PPB(0$\text{.}$40;(0$\text{.}$2,0$\text{.}$3,0$\text{.}$1);(1$\text{.}$0,0$\text{.}$9,0$\text{.}$1))$ & 1.226 & 1.131 & 0.370 & 0.957 & 0.918 & 0.516 & 0.765 & 0.000 & 0.000 \\
  \hline
\end{tabular}}
\end{center}
\end{table}
\end{landscape}

\begin{landscape}
\begin{table}
\caption{Resultados de simulaci\'on para la potencia. Alternativas: $NTAB(\lambda;\lambda_1,\lambda_2,\lambda_3)$ y $SLB(\lambda_1,\lambda_2,\lambda_3)$ }\label{Resultados-simulacion-potencia-2}
\begin{center}
{\normalsize
\begin{tabular}{|l|ccc|cccccc|}
\hline
Alternativa & $\frac{var(X_1)}{E(X_1)}$ & $\frac{var(X_2)}{E(X_2)}$& $\rho$ & $R_{n,(0,0)}$ & $S_{n,(0,0)}$ & $W_{n}$ & $T$ & $I_B$ &  $NI_B$\\
\hline\hline
  $NTAB(0$\text{.}$41;0$\text{.}$01,0$\text{.}$01,0$\text{.}$98)$ & 1.990 & 1.990 & 0.995 & 0.934 & 0.922 & 0.853 & 0.681 & 0.001 & 0.880\\
  $NTAB(0$\text{.}$50;0$\text{.}$01,0$\text{.}$01,0$\text{.}$98)$ & 1.990 & 1.990 & 0.995 & 0.931 & 0.916 & 0.831 & 0.685 & 0.001 & 0.875\\
  $NTAB(0$\text{.}$70;0$\text{.}$01,0$\text{.}$01,0$\text{.}$98)$ & 1.990 & 1.990 & 0.995 & 0.937 & 0.928 & 0.777 & 0.727 & 0.001 & 0.893\\
  $NTAB(0$\text{.}$71;0$\text{.}$01,0$\text{.}$01,0$\text{.}$98)$ & 1.990 & 1.990 & 0.995 & 0.945 & 0.943 & 0.796 & 0.736 & 0.000 & 0.903\\
  $NTAB(0$\text{.}$75;0$\text{.}$01,0$\text{.}$01,0$\text{.}$98)$ & 1.990 & 1.990 & 0.995 & 0.954 & 0.937 & 0.776 & 0.732 & 0.000 & 0.911\\
  \hline\hline
  $SLB(0$\text{.}$25,0$\text{.}$15,0$\text{.}$10)$   & 0.690 & 0.779 & 0.104 & 1.000 & 1.000 & 0.995 & 0.354 & 0.034 & 0.031\\
  $SLB(5d/7,d/7,d/7)^*$   & 1.000 & 1.000 & 0.289 & 0.945 & 0.999 & 0.973 & 0.258 & 0.177 & 0.171\\
  $SLB(3d/4,d/8,d/8)^*$   & 1.000 & 1.000 & 0.267 & 0.950 & 1.000 & 0.981 &  0.274 & 0.183 & 0.164\\
  $SLB(7d/9,d/9,d/9)^*$   & 1.000 & 1.000 & 0.250 & 0.948 & 0.999 & 0.979 &  0.270 & 0.169 & 0.166\\
  $SLB(0$\text{.}$51,0$\text{.}$01,0$\text{.}$02)$   & 0.668 & 0.981 & 0.098 & 1.000 & 1.000 & 0.999 & 0.429 & 0.046 & 0.041\\
  \hline
  \multicolumn{10}{l}{\normalsize $^*\, d=1-\exp(-1)\approx$ 0.63212.}
\end{tabular}}
\end{center}
\end{table}
\end{landscape}

Tambi\'en comparamos los tests estad\'isticos $R_{n,w}(\hat{\theta}_n)$, $S_{n,w}(\hat{\theta}_n)$ y $W_n(\hat{\theta}_n)$ desde un punto de vista computacional. La Tabla \ref{Tiempo-CPU} muestra el tiempo de CPU promedio consumido por cada uno de los tests estad\'isticos. En este aspecto, $W_n(\hat{\theta}_n)$ es mucho m\'as eficiente que $R_{n,w}(\hat{\theta}_n)$ y $S_{n,w}(\hat{\theta}_n)$. Los c\'alculos fueron realizados en un servidor que tiene las siguientes caracter\'isticas t\'ecnicas: Intel(R) Xeon(R) CPU X3430 @ 2.40GHz.

\begin{table}
\caption{Tiempo de CPU promedio (en segundos).}\label{Tiempo-CPU}
\begin{center}
\begin{tabular}{|l|r|r|r|}
\cline{2-4}\multicolumn{1}{c|}{} & \multicolumn{1}{c|}{\ \ \ $n=30$\ \ \ } &\multicolumn{1}{c|}{\ \ \ $n=50$\ \ \ } & \multicolumn{1}{c|}{\ \ \ $n=70$\ \ \ } \\
  \hline
  $R_{n,(0,0)}$ & 39678.35\ \ & 42275.29\ \ & 44567.67\ \ \\
  $S_{n,(0,0)}$ &  3019.81\ \ &  7296.59\ \ & 14202.57\ \ \\
  $W_n$         &  1452.07\ \ &  1807.28\ \ &  2142.86\ \ \\
   \hline
\end{tabular}
\end{center}
\end{table}

\section{Conjuntos de datos reales}\label{Conjuntos-datos-reales}
Ahora, aplicaremos los tests propuestos a dos conjuntos de datos reales. El primer conjunto de datos comprende el n\'umero de plantas de las especies \emph{Lacistema aggregatum} y \emph{Protium guianense} en cada uno de 100 cuadrantes contiguos. El primer autor que analiz\'o estos datos y los present\'o en detalle, fue Holgate (1966) \cite{Hol66}, despu\'es fueron analizados por Gillings (1974) \cite{Gil74}, Crockett (1979) \cite{Cro79}, Loukas y Kemp (1986) \cite{LoKe86}, Kocherlakota y Kocherlakota (1992) \cite{KoKo92} y tambi\'en por Rayner y Best (1995) \cite{RaBe95}. Crockett (1979) \cite{Cro79}, Loukas y Kemp (1986) \cite{LoKe86} y Rayner y Best (1995) \cite{RaBe95} examinaron dichos datos para averiguar si correspond\'ian a un modelo Poisson bivariante, concluyeron que la DPB no modelaba correctamente los datos mencionados.

El segundo conjunto de datos se refiere a la demanda de atenci\'on en sanidad en Australia, que fueron analizados por Karlis y Tsiamyrtzis (2008) \cite{KaTs08}. Los datos se refieren a la encuesta de Salud en Australia en el periodo 1977--1978. El tama\~no de la muestra es bastante grande ($n=5190$). Estos autores utilizaron dos variables: el n\'umero de consultas con un m\'edico o un especialista ($X_1$) y el n\'umero total de medicamentos prescritos y no prescritos utilizados en los \'ultimos dos d\'ias previos a la encuesta ($X_2$). Los datos se analizaron bajo el supuesto de que $(X_1,X_2)$ ten\'ia una DPB.

La Tabla \ref{Resultados-datos-reales} muestra los $p-$valores para contrastar la bondad de ajuste a una DPB para los dos conjuntos de datos reales empleando los tres tests estad\'isticos que hemos propuesto. Para ello, en el caso de $R_{n,w}(\hat{\theta}_n)$ y $S_{n,w}(\hat{\theta}_n)$, usamos
$(a_1,a_2)\in \{(0,0), (1,0), (0,1)\}$. Para los datos de las plantas, todos los tests rechazaron la hip\'otesis nula, esto es, que los datos no son bien modelados por una DPB. Este resultado coincide con los obtenidos por los investigadores que analizaron previamente este conjunto de datos. A la misma conclusi\'on se llega al emplear los datos de salud.

\begin{table}[h]
\caption{
Resultados para los conjuntos de datos reales.}\label{Resultados-datos-reales}
\begin{center}
\begin{tabular}{|l|c|c|}
  \cline{2-3} \multicolumn{1}{c|}{}
  &   \multicolumn{1}{c|}{Plantas} &   \multicolumn{1}{c|}{Salud}\\ \hline
$R_{n,(0,0)}$ &  0.002 & 0.000\\
$R_{n,(1,0)}$ &  0.004 & 0.000\\
$R_{n,(0,1)}$ &  0.004 & 0.000\\ \hline\hline
$S_{n,(0,0)}$ &  0.002 & 0.002\\
$S_{n,(1,0)}$ &  0.006 & 0.000\\
$S_{n,(0,1)}$ &  0.008 & 0.000\\ \hline\hline
$W_n$ &  0.042 & 0.000\\ \hline\hline
\multicolumn{1}{|c|}{$\hat{\theta}_n$} & (0.64000,\,0.94000,\,0.19852) &
 (0.30173,\,1.21830,\,0.12518) \\ \hline
\end{tabular}
\end{center}
\end{table}


\chapter{Expresiones matem\'aticas de los tests y algunos aspectos computacionales}\label{Expresiones-matematicas}
Las expresiones matem\'aticas en las siguientes secciones son necesarias para implementar computacionalmente los tests estad\'isticos que hemos propuesto.

\section[C\'alculo del test estad\'istico $R_{n,w}(\hat{\theta}_n)$]{C\'alculo del test estad\'istico $\boldsymbol{R_{n,w}(\hat{\theta}_n)}$ } \label{Calculo-estadistico-Rnw}
Para $w$ definido como en (\ref{funcion-peso-explicita}) y para  $a_1, a_2\in \mathbb{N}_0$, de (\ref{Estad-Rnw}), obtenemos
\begin{align}
R_{n,w}(\hat{\theta}_n)&=\frac{1}{n}\int_0^1\int_0^1  \left[\sum_{i=1}^n\left[ u_1^{X_{1i}}u_2^{X_{2i}}- e^{\hat{\theta}_{1n}(u_1-1)+\hat{\theta}_{2n} (u_2-1) +\hat{\theta}_{3n}(u_1-1)(u_2-1)}\right]\right]^{2} u_1^{a_1}\, u_2^{a_2} \, du\notag\\[.2 cm]
&=\frac{1}{n}\sum_{i=1}^n \sum_{j=1}^n\int_0^1\int_0^1 u_1^{X_{1i}+X_{1j}+a_1}u_2^{X_{2i}+ X_{2j}+a_2}\,du\notag\\[.2 cm]
&\ \ \ - 2\,e^{\hat{\theta}_{3n}-\hat{\theta}_{1n}- \hat{\theta}_{2n}}  \!\sum_{i=1}^n\!\int_0^1 \!u_1^{X_{1i}+a_1} \,e^{\left\{\left(\hat{\theta}_{1n}- \hat{\theta}_{3n}\right)u_1\!\right\}}\!\int_0^1\! u_2^{X_{2i}+a_2} \,e^{\left\{\left(\hat{\theta}_{2n}-\hat{\theta}_{3n}+ \hat{\theta}_{3n}u_1\right)u_2\!\right\}}\,du\notag\\[.2 cm]
&\ \ \ +n\,e^{\left\{2\left(\hat{\theta}_{3n}-\hat{\theta}_{1n}- \hat{\theta}_{2n}\right)\right\}} \int_0^1u_1^{a_1}\,e^{\left\{2\left(\hat{\theta}_{1n}- \hat{\theta}_{3n}\right)u_1\right\}}\int_0^1u_2^{a_2}\, e^{\left\{2\left(\hat{\theta}_{2n}- \hat{\theta}_{3n}+\hat{\theta}_{3n}u_1\right) u_2\right\}}\, du.\label{Rnw-1}
\end{align}
Para las integrales de (\ref{Rnw-1}) que son de la forma $\displaystyle\int_0^1 t^k\exp(rt)\,dt,\,$ con $\,k\in \mathbb{N}_0, \ r>0$, integrando por partes, resulta
\begin{equation}\label{Rnw-2}
\int_0^1 t^k\,e^{rt}\,dt=\sum_{i=0}^{k} \frac{(-1)^i\, k!\exp(r)}{(k-i)!\,r^{i+1}}+\frac{(-1)^{k+1}\,k!}{r^{k+1}}\,,\ \forall\, k\!\in\mathbb{N}_0,\ r>0.
\end{equation}
Desarrollando (\ref{Rnw-1}) y empleando (\ref{Rnw-2}) con $t=u_2$, y para $k$ y $r$ apropiados, obtenemos
\begin{align}
&R_{n,w}(\hat{\theta}_n)=\frac{1}{n}\sum_{i=1}^n \sum_{j=1}^n \frac{1}{(X_{1i} +X_{1j}+a_1+1)(X_{2i}+X_{2j}+a_2+1)}\notag\\[.4 cm]
&\hspace{4mm}-2\exp(\hat{\theta}_{3n}-\hat{\theta}_{1n}- \hat{\theta}_{2n})\sum_{i=1}^n(X_{2i}+a_2)!\left\{\sum_{k=0}^{X_{2i}+ a_2}\! \frac{(-1)^k\, \exp(\hat{\theta}_{2n}- \hat{\theta}_{3n})}{(X_{2i}+a_2-k)!} \right.\notag\\[.35 cm]
&\hspace{9mm}\left.\times\int_0^1\! \frac{u_1^{X_{1i}+a_1}\exp\{\hat{\theta}_{1n}u_1\}\ du_{1}} {(\hat{\theta}_{2n}-\hat{\theta}_{3n}+\hat{\theta}_{3n}u_1 )^{k+1}} +(-1)^{X_{2i}+a_2+1}\! \int_0^1\! \frac{u_1^{X_{1i}+a_1}\exp\{(\hat{\theta}_{1n} -\hat{\theta}_{3n})u_1\}} {(\hat{\theta}_{2n}-\hat{\theta}_{3n}+\hat{\theta}_{3n}u_1 )^{X_{2i}+a_2+1}} du_1\!\right\}\notag\\[.4 cm]
&\hspace{4mm}+n\,a_2!\exp\{2( \hat{\theta}_{3n}\!-\hat{\theta}_{1n} \!-\hat{\theta}_{2n})\}\left\{\sum_{k=0}^{a_2} \frac{(-1)^k\exp\{2(\hat{\theta}_{2n}-\hat{\theta}_{3n})\}}{(a_2-k)!\ 2^{k+1}}\right.\notag\\[.35 cm]
&\hspace{9mm}\left.\times\int_0^1 \frac{u_{\!1}^{a_1}\exp(2\hat{\theta}_{1n}u_{1})\ }{(\hat{\theta}_{2n}-\hat{\theta}_{3n}+ \hat{\theta}_{3n}u_1)^{k+1}}du_{1}+\frac{(-1)^{a_2+1}} {2^{a_2+1}}\int_0^1\frac{u_{1}^{a_1}\exp\{2(\hat{\theta}_{1n}\!-\hat{\theta}_{3n}) u_1\}}{(\hat{\theta}_{2n}-\hat{\theta}_{3n}+ \hat{\theta}_{3n}u_1)^{a_2+1}}du_1\right\}.\label{Rnw-3}
\end{align}

Para las integrales de (\ref{Rnw-3}) que son de la forma $\displaystyle\int_0^1 \frac{t^k\exp(rt)}{(b+ct)^m}\,dt$, con $r,b>0$, $c\geq 0,\ k\in\mathbb{N}_0, \ m\in\mathbb{N}$, integrando por partes, se logra:
\begin{itemize}
  \item [$\bullet$] Para $\,c=0$, salvo constantes, se obtiene una integral como la dada por (\ref{Rnw-2})
  \item [$\bullet$] Para $\,c>0$, haciendo el cambio de variables $\ y=b+ct,\,$ se obtiene \begin{align} \int_0^1\frac{t^k\exp(rt)}{(b+ct)^m}\,dt& =\frac{\exp\left(-\frac{rb}{c}\right)} {c^{k+1}} \int_b^{b+c}\frac{(y-b)^k\exp\left(\frac{r}{c}y\right)}{y^m}\,dy\notag\\[.2 cm] &= \frac{\exp\left(-\frac{rb}{c}\right)}{c^{k+1}}\sum_{i=0}^k \binom{k}{i} (-b)^{k-i} \int_b^{b+c}\frac{y^i\exp\left(\frac{r}{c}y\right)}{y^m}\,dy\notag\\[.2 cm] &=\frac{\exp\!\left(-\frac{rb}{c}\right)}{c^{k+1}}\sum_{i=0}^k \!\binom{k}{i} (-b)^{k-i} \!\left\{\!\!
      \begin{array}{ll} \displaystyle\int_b^{b+c}\frac{\exp\left(\frac{r}{c}y\right)}{y^{m-i}}\,dy, &\text{si}\ \,i<m, \\[.5 cm]
        \displaystyle\int_b^{b+c}\!\!y^{i-m}\exp\!\left(\frac{r}{c}y\right)dy, & \text{si}\ i\geq m.
      \end{array}
    \right.\label{Rnw-4}
\end{align}
Para las integrales de (\ref{Rnw-4}), integrando nuevamente por partes, resulta
\begin{align}
&\int_b^{b+c}y^{i-m}\exp\!\left(\frac{r}{c}y\right)\,dy\notag\\[.2 cm]
&\quad\quad =(i-m)! \exp\!\left(\frac{rb}{c}\right)\sum_{j=0}^{i-m}\frac{(-1)^j} {(i-m-j)!}\!\left(\frac{c}{r}\right)^{j+1}\!\left\{\exp(r)(b+c)^{i-m-j}\!-b^{i-m-j} \right\}.\label{Rnw-5}
\end{align}
\begin{align}
&\int_b^{b+c}\frac{\exp(\frac{r}{c}y)}{y^{m-i}}\,dy= \frac{1}{(m-i-1)!}\left[\left(\frac{r}{c}\right)^{m-i-1} \int_{\frac{rb}{c}}^{\frac{r(b+c)}{c}} \frac{\exp(t)}{t}dt+\exp\!\left(\frac{rb}{c}\right)\right.\notag\\[.2 cm]
&\hspace{2mm}\left.\times\!
\left\{\!\!
  \begin{array}{ll}
     0, & \!\!\text{si}\ m\!-\!i=1,\\[.2 cm]
     \displaystyle\sum_{j=1}^{m-i-1}\!\frac{(m-i-j-1)!\ r^{j-1}}{c^{j-1}\, \{b(b+c)\}^{m-i-j}} \left\{(b+c)^{m-i-j}-b^{m-i-j}\exp(r)\right\}, & \!\!\text{si}\ m\!-\!i\geq 2.
  \end{array}
  \right.\!\!
\right]\!,\label{Rnw-6}
\end{align}
\end{itemize} 
As\'i, $R_{n,w}(\hat{\theta}_n)$ se puede obtener de las ecuaciones (\ref{Rnw-2}), (\ref{Rnw-4}), (\ref{Rnw-5}) y (\ref{Rnw-6}).\vskip .1 cm

Claramente, (\ref{Rnw-6}) no tiene una forma cerrada, pues contiene a la integral $\int \frac{\exp(t)}{t}dt$, la cual es muy sensible para ciertas combinaciones de los valores $r, b$ y $c$, de modo que el c\'alculo computacional de dicha integral sufre problemas de redondeo o diverge y por lo tanto no podemos obtener su valor para ciertas combinaciones de $r, b$ y $c$, lo cual nos lleva a quedar sin poder calcular $R_{n,w}(\hat{\theta}_n)$.

Por razones computacionales, la siguiente representaci\'on
\begin{equation}\label{Rnw-7}
R_{n,w}(\hat{\theta}_n) = n \sum_{i,j,k,l=0}^{\infty} \frac{\left(p_n(i,j) -p(i,j;\hat{\theta}_n)\right)\left(p_n(k,l)- p(k,l;\hat{\theta}_n)\right)}{(i+k+a_1+1)(j+l+a_2+1)}
\end{equation}
result\'o ser m\'as apropiada, donde $p(i, j;\theta)=P_{\theta}(X_1=i, X_2=j)$ y $p_n(i,j)$ es la frecuencia relativa del par $(i, j)$, dada por
\[p_n(i,j)=\frac{1}{n}\sum_{k=1}^n I{\{X_{1k}=i,X_{2k}=j\}}.\]

Adem\'as, (\ref{Rnw-7}) es menos restrictiva, pues permite que $a_1>-1$ y $a_2>-1$.

Un truncamiento de las cuatro series infinitas en $M+15$ arroj\'o valores suficientemente precisos del test estad\'istico y un buen rendimiento de la subrutina correspondiente, donde $\displaystyle M=\max \{ X_{1(n)},X_{2(n)}\}$, $X_{k(n)}=\max_{1\leq i\leq n}X_{ki} $, $k=1,2$.

\section[C\'alculo del test estad\'istico $S_{n,w}(\hat{\theta}_n)$]{C\'alculo del test estad\'istico $\boldsymbol{S_{n,w}(\hat{\theta}_n)}$ } \label{Calculo-estadistico-Snw}
Para $w$ definido como en (\ref{funcion-peso-explicita}), y para  $a_1>-1$ y $a_2>-1$, de (\ref{Estad-Snw}), obtenemos
\begin{align}
&S_{n,w}(\hat{\theta}_n)\notag\\
&=\frac{1}{n}\int_0^1\!\int_0^1  \!\left[\sum_{i=1}^n\left\{X_{1i}I{\{X_{1i}\geq 1\}}u_1^{X_{1i}-1}u_2^{X_{2i}}\!-\{\hat{\theta}_{1n}\!+  \hat{\theta}_{3n}(u_2-1)\}u_1^{X_{1i}}u_2^{X_{2i}}\right\}\right]^2 \!\! u_1^{a_1}u_2^{a_2}du\notag\\[.1 cm]
&\ \ +\frac{1}{n}\int_0^1\!\int_0^1\!\left[\sum_{i=1}^n\left\{X_{2i}I{\{X_{2i}\geq 1\}}u_1^{X_{1i}}u_2^{X_{2i}-1}\!-\!\{\hat{\theta}_{2n}\!+  \hat{\theta}_{3n}(u_1-1)\}u_1^{X_{1i}}u_2^{X_{2i}}\right\}\right]^2\! \! u_1^{a_1}u_2^{a_2}du\notag\\[.1 cm]
&=\frac{1}{n}\sum_{i=1}^n\sum_{j=1}^n \left(S_{1ij}+S_{2ij}\right),\notag
\end{align}
donde
\begin{align}
S_{1ij}&=\frac{X_{1i}\,I{\{X_{1i}\geq 1\}}\,X_{1j}\,I{\{X_{1j}\geq 1\}}}{(X_{1ij}-1) (X_{2ij}+1)}+ \frac{(\hat{\theta}_{1n}- \hat{\theta}_{3n})^2} {(X_{1ij}+1)(X_{2ij}+1)}\notag\\[.1 cm]
&\hspace{10mm}-\frac{\hat{\theta}_{3n}\Bigl(X_{1i}\, I{\{X_{1i}\geq 1\}}+X_{1j}\,I{\{X_{1j}\geq 1\}}\Bigr)} {X_{1ij}\,(X_{2ij}+2)}+\frac{2\,\hat{\theta}_{3n}(\hat{\theta}_{1n}- \hat{\theta}_{3n})} {(X_{1ij}+1)(X_{2ij}+2)}\notag\\[.1 cm]
&\hspace{10mm}- \frac{(\hat{\theta}_{1n}- \hat{\theta}_{3n})\Bigl(X_{1i}\,I{\{X_{1i}\geq 1\}}+ X_{1j}\,I{\{X_{1j}\geq 1\}}\Bigr)} {X_{1ij}\,(X_{2ij}+1)}+\frac{\hat{\theta}_{3n}^{\ 2}}{(X_{1ij}+1)(X_{2ij}+3)}\,,\notag
\end{align}

\begin{align}
S_{2ij}&=\frac{X_{2i}\,I{\{X_{2i}\geq 1\}}\,X_{2j}\,I{\{X_{2j}\geq 1\}}}{(X_{1ij}+1) (X_{2ij}-1)}+ \frac{(\hat{\theta}_{2n}- \hat{\theta}_{3n})^2} {(X_{1ij}+1)(X_{2ij}+1)}    \notag\\[.2 cm]
&\hspace{10mm}-\frac{\hat{\theta}_{3n}\Bigl(X_{2i}\, I{\{X_{2i}\geq 1\}}+X_{2j}\,I{\{X_{2j}\geq 1\}}\Bigr)} {(X_{1ij}+2)\,X_{2ij}}+\frac{2\,\hat{\theta}_{3n}(\hat{\theta}_{2n}- \hat{\theta}_{3n})} {(X_{1ij}+2)(X_{2ij}+1)}\notag\\[.2 cm]
&\hspace{10mm}- \frac{(\hat{\theta}_{2n}- \hat{\theta}_{3n})\Bigl(X_{2i}\,I{\{X_{2i}\geq 1\}}+ X_{2j}\,I{\{X_{2j}\geq 1\}}\Bigr)} {(X_{1ij}+1)\,X_{2ij}}+\frac{\hat{\theta}_{3n}^{\ 2}}{(X_{1ij}+3)(X_{2ij}+1)}\,,\notag
\end{align}
con $X_{1ij}=X_{1i}+X_{1j}+a_1\ $ y $\ X_{2ij}=X_{2i}+X_{2j}+a_2$, $1\leq i,j \leq n$.

\section[C\'alculo del test estad\'istico $W_n(\hat{\theta}_n)$]{C\'alculo del test estad\'istico $\boldsymbol{W_n(\hat{\theta}_n)}$} \label{Calculo-estadistico-Wn}
De (\ref{Estad-Wn-bivariante}) y (\ref{psi_1rs-2rs}), obtenemos
\begin{align}
W_n(\hat{\theta}_n)&=\sum_{r,s = 0}^M \left[\sum_{i=1}^n\left\{(r+1)p_n(r+1,s) - (\hat{\theta}_{1n}-\hat{\theta}_{3n})p_n(r,s) -\hat{\theta}_{3n}p_n(r,s-1)\right\}\right]^2\notag\\[.2 cm]
&\hspace{7mm} +\sum_{r,s = 0}^M \left[\sum_{i=1}^n\left\{(s+1)p_n(r,s+1) - (\hat{\theta}_{2n}-\hat{\theta}_{3n})p_n(r,s) -\hat{\theta}_{3n}p_n(r-1,s)\right\}\right]^2\!,\notag
\end{align}
donde $M=\max \{X_{1(n)}, X_{2(n)}\}$, $X_{k(n)}=\max_{1\leq i\leq n}X_{ki}$, $k=1,2$.

\newpage

\[\begin{array}{c}\\  \\ \end{array}\]

\newpage

\chapter{Extensiones} \label{Extensiones}
\section{El caso de dos variables Poisson independientes}
El caso de dos variables Poisson independientes ocurre cuando $\theta_3=0$, que fue excluido puesto que, seg\'un la definici\'on dada en la Secci\'on \ref{Notacion}, $\theta_3>0$. La raz\'on para no considerar este importante caso es que para que la aproximaci\'on bootstrap funcione, necesitamos que $\theta$ sea un punto interior del espacio param\'etrico $\Theta$.

Si estamos interesados en un test de bondad de ajuste para un modelo Poisson independiente, entonces podemos, o bien, considerar el test estad\'istico $R_{n,w}(\hat{\theta}_{0n})$, $S_{n,w}(\hat{\theta}_{0n})$ o $W_n(\hat{\theta}_{0n})$, con $\hat{\theta}_{0n}=(\hat{\theta}_{1n}, \hat{\theta}_{2n}, 0)$, es decir, el mismo test estad\'istico como antes con $\hat{\theta}_{3n}$ fijo e igual a su valor poblacional bajo la hip\'otesis nula, $\theta_3=0$, o bien, podemos usar el Teorema de Raikov y aplicar un test de bondad de ajuste para la distribuci\'on Poisson univariante a la suma de los componentes.

Los resultados de los Cap\'itulos \ref{Estadisticos-tipoCramer-von-Mises} y \ref{Estadistico-Wn} seguir\'an siendo v\'alidos para estos tests estad\'isticos.

\section[El caso general $d$-variante]{El caso general $\boldsymbol{d}$-variante}
Hasta ahora hemos asumido que los datos son bivariantes. En esta secci\'on mostramos que los tests que hemos propuesto en este texto se pueden extender de manera natural al caso $d$-variante general, $d\in \mathbb{N}$. Sea
\[X_1=Y_1+Y_{d+1}, \,\, X_2=Y_2+Y_{d+1},\,\,  \ldots,  \,\, X_d=Y_d+Y_{d+1},\]
donde $Y_1, Y_2, \ldots, Y_d, Y_{d+1}$ son v.a. Poisson (univariantes) mutuamente independientes con medias $\theta'_1=\theta_1-\theta_{d+1}>0, \ldots, \theta'_d=\theta_d-\theta_{d+1}>0$ y  $\theta_{d+1} \geq 0$, respectivamente.\vskip .2 cm

La distribuci\'on conjunta del vector $(X_1,  X_2, \ldots, X_d)$ es llamada distribuci\'on Poisson $d$-variante con par\'ametro $\theta=(\theta_1,\theta_2,\ldots, \theta_d,\theta_{d+1})$ (ver, por ejemplo, Johnson, Kotz y Balakrishnan (1997, p. 139) \cite{JoKoBa97}).

La fgp conjunta de $(X_1,  X_2, \ldots, X_d)$ est\'a dada por
\begin{equation}\label{fgp-multivariante}
g(u;\theta)=
 \exp\left\{\sum_{i=1}^d \theta_i\left(u_i-1\right)+ \theta_{d+1}
\left(\prod_{i=1}^du_i-\sum_{i=1}^du_i+d-1\right)\right\},
\end{equation}
$\forall u \in \mathbb{R}^d$.

Claramente, el test basado en el estad\'istico $R_{n,w}(\hat{\theta}_{n})$ se puede aplicar para contrastar
\[H_{0d}:(X_1,  X_2, \ldots, X_d) \text{  tiene una distribuci\'on Poisson } d\text{-variante},\]
para cualquier $d$ fijo.

El siguiente resultado muestra la extensi\'on $d$-variante de la Proposici\'on
\ref{Soluc-sistema-de-dos-EDP}, lo cual nos permitir\'a proponer una extensi\'on de los tests basados en $S_{n,w}(\hat{\theta}_{n})$ y en $W_n(\hat{\theta}_{n})$.
\begin{proposicion}\label{Soluc-sistema-d-variante}
Sea $G_d=\{g:[0,1]^d\to \mathbb{R}$, tal que $g$  es una fgp y $\frac{\partial}{\partial u_i}g (u_1,\ldots, u_d)$, existen $\forall u \in [0,1]^d$, $1\leq i\leq d\}$. Sea $g(u;\theta)$ como definida en (\ref{fgp-multivariante}). Entonces $g(u;\theta)$ es la \'unica fgp en $G_d$ que satisface el siguiente sistema
\begin{equation}\label{EDPs-fgp-d-variante}
D_{id}(u;\theta)= 0,\quad 1\leq i\leq d,\ \ \forall u \in [0,1]^d,
\end{equation}
donde
\[
  D_{id}(u;\theta)=\frac{\partial }{\partial u_i}g(u)-
  \left\{\theta_i+\theta_{d+1}\left(\prod _{j\neq i }u_j-1\right)\right\}g(u),\quad 1\leq i\leq d.
  \]
\end{proposicion}
\nt {\bf Demostraci\'on} \hspace{2pt} Consideremos el vector aleatorio $(X_1,X_2,\ldots,X_d)\in \mathbb{N}^d_0$, cuya fgp est\'a dada por $g(u_1,u_2,\ldots,u_d)=E\left(u_1^{X_1} u_2^{X_2}\cdots u_d^{X_d}\right)$. Entonces, para $i=1$ en (\ref{EDPs-fgp-d-variante}) \begin{equation}\label{EDP1-fgp-d-variante} \frac{\partial}{\partial u_1}\log g(u_1,u_2,\ldots,u_d)=\theta_1+\theta_{d+1}\left(\prod _{j\neq 1 }u_j-1\right).\end{equation}

Integrando (\ref{EDP1-fgp-d-variante}) sobre $u_1$, obtenemos
\begin{align}
&g(u_1,u_2,\ldots,u_d)= \exp\left\{\phi_1(u_2,u_3,\ldots,u_d)+\theta_1 u_1+\theta_{d+1}\left(\prod _{j=1}^d u_j-u_1\right)\right\},\notag\\[.2 cm]
&\hspace{19mm}=\exp\!\left\{\varphi_1(u_2,u_3,\ldots,u_d)+\theta_1(u_1-1) +\theta_{d+1}\!\left(\prod_{j=1}^d u_j-\!\sum_{j=1}^d u_j+d-1\right)\right\},\notag
\end{align}
donde $\,\varphi_1(u_2,u_3,\ldots,u_d)=\phi_1(u_2,u_3,\ldots,u_d)+\theta_1+ \theta_{d+1}\displaystyle\left(\sum_{j=2}^d u_j-d+1\right)$.\vskip .5 cm

Procediendo similarmente, de (\ref{EDPs-fgp-d-variante}), para $i=2,3,\ldots,d$, obtenemos
\begin{align}
&g(u_1,u_2,\ldots,u_d)= \exp\left\{\phi_2(u_1,u_3,\ldots,u_d)+\theta_2 u_2+\theta_{d+1}\left(\prod _{j=1}^d u_j-u_2\right)\right\},\notag\\[.2 cm]
&\hspace{19mm}=\exp\!\left\{\varphi_2(u_1,u_3,\ldots,u_d)+\theta_2(u_2-1) +\theta_{d+1}\!\left(\prod_{j=1}^d u_j-\sum_{j=1}^d u_j+d-1\right)\right\},\notag
\end{align}
donde $\,\varphi_2(u_1,u_3,\ldots,u_d)=\phi_2(u_1,u_3,\ldots,u_d)+\theta_2+ \theta_{d+1}\displaystyle\left(\sum_{j=1,j\neq 2}^d u_j-d+1\right)$.\vskip .6 cm

As\'i, sucesivamente hasta que
\begin{align}
&g(u_1,u_2,\ldots,u_d)= \exp\left\{\phi_d(u_1,u_2,\ldots,u_{d-1})+\theta_d u_d+\theta_{d+1}\left(\prod _{j=1}^d u_j-u_d\right)\right\},\notag\\[.2 cm]
&\hspace{18mm}=\exp\!\left\{\varphi_d(u_1,u_2,\ldots,u_{d-1})+\theta_d(u_d-1) +\theta_{d+1}\!\left(\prod_{j=1}^d u_j-\!\sum_{j=1}^d u_j+d-1\right)\right\},\notag
\end{align}
donde $\,\varphi_d(u_1,u_2,\ldots,u_{d-1})=\phi_d(u_1,u_2,\ldots,u_{d-1})+\theta_d+ \theta_{d+1}\displaystyle\left(\sum_{j=1}^{d-1} u_j-d+1\right)$.

Por lo tanto, necesariamente debe ocurrir que
\begin{align}
\varphi_1(u_2,u_3,\ldots,u_d)&=\sum_{j=2}^d \theta_j (u_j-1),\notag\\[.1 cm]
\varphi_2(u_1,u_3,\ldots,u_d)&=\sum_{\substack{j=1\\ j\neq 2}}^d \theta_j (u_j-1),\notag\\[.1 cm]
&\vdots\notag\\
\varphi_d(u_1,u_2,\ldots,u_{d-1})&=\sum_{j=1}^{d-1} \theta_j (u_j-1),\notag
\end{align}
en otras palabras, la fgp de la DP $d$-variante es la \'unica soluci\'on de (\ref{EDPs-fgp-d-variante}). $\square$\\

De la Proposici\'on \ref{Conv-FuncGenProbBiv}, $g(u)$ y sus derivadas pueden ser estimadas consistentemente por medio de la fgpe y las derivadas de la fgpe, respectivamente.

As\'i, si $H_{0d}$ fuera cierta, entonces las funciones
\[
  D_{in}(u;\hat{\theta}_n )=\frac{\partial }{\partial u_i}g_n(u_1,u_2, \ldots, u_d)-\left\{\hat{\theta}_{i,n}+\hat{\theta}_{d+1,n}\left(\prod_{j\neq i }u_j-1\right)\right\} g_n(u_1,u_2, \ldots, u_d),
\]
$1\leq i \leq d$, deber\'ian estar pr\'oximas a 0, donde $\hat{\theta}_n$ es un estimador consistente de $\theta$.

Por lo tanto, para probar $H_{0d}$ podr\'iamos considerar el siguiente test estad\'istico
\[S_{d,n,w}(\hat{\theta}_n)=n\int_{[0,1]^d} \left\{D^2_{1n}(u;\hat{\theta}_n )+\cdots+D^2_{dn}(u;\hat{\theta}_{n} )\right\}w(u)\ du,\]
donde $w(u)$ es una funci\'on de peso medible, no negativa, con integral finita en $[0,1]^d$.

Haciendo las modificaciones correspondientes, se pueden conseguir resultados similares a los establecidos en el Cap\'itulo \ref{Estadisticos-tipoCramer-von-Mises}.
\vskip .2 cm

Para el test estad\'istico basado en $W_n(\hat{\theta}_{n})$, consideremos que se verifican las hip\'otesis de la Proposici\'on \ref{Soluc-sistema-d-variante}.

Sea $(X_1,\ldots,X_d)\in \mathbb{N}^d_0\,$ un vector aleatorio y sea $\,g(u_1,\ldots,u_d)=E\left(u_1^{X_1} u_2^{X_2}\cdots u_d^{X_d}\right)$ su fgp, luego, por definici\'on
\[g(u)=\sum_{r_1,r_2,\ldots,r_d\geq 0}u_1^{r_1} u_2^{r_2}\cdots u_d^{r_d} P(r_1,r_2,\ldots,r_d),\]
donde $P(r_1,r_2,\ldots,r_d)=P(X_1=r_1,X_2=r_2,\ldots,X_d=r_d)$.\vskip .2 cm

De lo anterior y de (\ref{EDPs-fgp-d-variante}), podemos escribir
\begin{align}
D_{1d}(u;\theta)&=\sum_{r_1,r_2,\ldots,r_d\geq 0}\Bigl\{(r_1+1) P(r_1+1,r_2,r_3,\ldots,r_d)-(\theta_1-\theta_{d+1}) P(r_1,r_2,\ldots,r_d)\Bigr.\notag \\
&\hspace{46mm}\Bigl.-\theta_{d+1} P(r_1,r_2-1,r_3-1,\ldots,r_d-1) \Bigr\}u_1^{r_1} u_2^{r_2}\cdots u_d^{r_d},\notag \\[.2 cm]
D_{2d}(u;\theta)&=\sum_{r_1,r_2,\ldots,r_d\geq 0}\Bigl\{(r_2+1) P(r_1,r_2+1,r_3,\ldots,r_d)-(\theta_2-\theta_{d+1}) P(r_1,r_2,\ldots,r_d)\Bigr.\notag \\
&\hspace{46mm}\Bigl.-\theta_{d+1} P(r_1-1,r_2,r_3-1,\ldots,r_d-1) \Bigr\}u_1^{r_1} u_2^{r_2}\cdots u_d^{r_d},\notag\\
&\vdots\notag\\
D_{dd}(u;\theta)&=\sum_{r_1,r_2,\ldots,r_d\geq 0}\Bigl\{(r_d+1) P(r_1,r_2,\ldots,r_{d-1},r_d+1)-(\theta_d-\theta_{d+1}) P(r_1,r_2,\ldots,r_d)\Bigr.\notag \\
&\hspace{43mm}\Bigl.-\theta_{d+1} P(r_1-1,r_2-1,\ldots,r_{d-1}-1,r_d) \Bigr\}u_1^{r_1} u_2^{r_2}\cdots u_d^{r_d}.\notag
\end{align}

Consideremos ahora las versiones emp\'iricas $D_{1n}(u;\hat{\theta}_n), D_{2n}(u;\hat{\theta}_n), \ldots, D_{dn}(u;\hat{\theta}_n)$, de las ecuaciones anteriores.

Si $H_{0d}$ fuera cierta entonces $D_{1n}(u;\hat{\theta}_n), D_{2n}(u;\hat{\theta}_n), \ldots, D_{dn}(u;\hat{\theta}_n)$ deber\'ian ser pr\'oximas a 0, $\forall u \in [0,1]^d$. Esta proximidad a cero la podemos interpretar como ya lo hicimos al inicio del Cap\'itulo \ref{Estadistico-Wn}, para el caso bivariante. Para ello observemos que
\[D_{in}(u;\hat{\theta}_n)=\sum_{r_1,r_2,\ldots,r_d\geq 0} b_i(r_1,r_2,\ldots,r_d;\hat{\theta}_n) u_1^{r_1} u_2^{r_2}\cdots u_d^{r_d},\ \ 1\leq i\leq d\]
donde
\begin{align}
b_1(r_1,r_2,\ldots,r_d;\hat{\theta}_n)&=(r_1+1) p_n(r_1+1,r_2,r_3,\ldots,r_d)-(\hat{\theta}_{1n}-\hat{\theta}_{d+1,n}) p_n(r_1,r_2,\ldots,r_d)\notag \\[.1 cm]
&\hspace{45mm}-\hat{\theta}_{d+1,n} p_n(r_1,r_2-1,r_3-1,\ldots,r_d-1),\notag\\[.2 cm]
b_2(r_1,r_2,\ldots,r_d;\hat{\theta}_n)&=(r_2+1) p_n(r_1,r_2+1,r_3,\ldots,r_d)-(\hat{\theta}_{2n}-\hat{\theta}_{d+1,n}) p_n(r_1,r_2,\ldots,r_d)\notag \\[.1 cm]
&\hspace{45mm}-\hat{\theta}_{d+1,n} p_n(r_1-1,r_2,r_3-1,\ldots,r_d-1),\notag\\[.1 cm]
&\vdots\notag\\
b_d(r_1,r_2,\ldots,r_d;\hat{\theta}_n)&=(r_d+1) p_n(r_1,r_2,\ldots,r_{d-1},r_d+1)-\!(\hat{\theta}_{dn}\!-\hat{\theta}_{d+1,n}) p_n(r_1,r_2,\ldots,r_d)\notag \\[.1 cm]
&\hspace{44mm}-\hat{\theta}_{d+1,n} p_n(r_1-1,r_2-1,\ldots,r_{d-1}-1,r_d),\notag
\end{align}
y
\[p_n(r_1,r_2,\ldots,r_d)=\frac{1}{n}\sum_{k=1}^n I{\{X_{1k}=r_1,X_{2k}=r_2,\ldots,X_{dk}=r_d\}},\]
es la frecuencia relativa emp\'irica de la $d$-tupla $(r_1,r_2,\ldots,r_d)$.

Por lo tanto, $D_{in}(u;\hat{\theta}_n)= 0$,
$\forall u \in [0,1]^d$, $1\leq i\leq d$, s\'i y s\'olo si los coeficientes de $u_1^{r_1} u_2^{r_2}\cdots u_d^{r_d}$ en las expansiones anteriores son  nulos $\forall r_1,r_2,\ldots,r_d \geq 0$. Esto nos lleva a considerar el siguiente estad\'istico para contrastar $H_{0d}$:
\begin{align}
W_{d,n}(\hat{\theta}_n)&=\sum_{r_1,r_2,\ldots,r_d\geq 0}\left\{\sum_{i=1}^d b_i(r_1,r_2,\ldots,r_d;\hat{\theta}_n)^2\right\}\notag\\[.2 cm]
&=\sum_{r_1,r_2,\ldots,r_d= 0}^M\left\{\sum_{i=1}^d b_i(r_1,r_2,\ldots,r_d;\hat{\theta}_n)^2\right\},\notag
\end{align}
donde $M=\max \{X_{1(n)}, X_{2(n)},\ldots,X_{d(n)}\}$, $X_{k(n)}=\max_{1\leq j\leq n}X_{kj}$, $1\leq k\leq d$.

Haciendo los cambios respectivos, se pueden conseguir resultados similares a los establecidos en el Cap\'itulo \ref{Estadistico-Wn}.


\begin{thebibliography}{99}
\addcontentsline{toc}{chapter}{Bibliograf{\'\i}a}

\bibitem{AtLa06}
Athreya, K.B. y Lahiri, S.N. (2006). \emph{Measure Theory and Probability Theory}. Springer

\bibitem{BaGuHe00}
Baringhaus, L., G\"urtler, N. y Henze, N. (2000). Weighted integral test statistics and components of smooth tests of fit, \emph{Australian} \& \emph{New Zealand Journal of Statistics}, {\bf 42}, 179--192.

\bibitem{BaHe92}
Baringhaus, L. y Henze, N. (1992). A goodness of fit test for the Poisson distribution based on the empirical generating function, \emph{ Statistics} \& \emph{Probability Letters}, {\bf 13}, 269--274.

\bibitem{BePl04}
Berkhout, P. y Plug, E. (2004). A bivariate Poisson count data model using conditional probabilities, \emph{Statistica Neerlandica}, {\bf 58}, 349--364.

\bibitem{Ber09}
Berm\'udez, L. (2009). A priori ratemaking using bivariate Poisson regression models, \emph{Insurance: Mathematics and Economics}, {\bf 44}, 135--141.


\bibitem{Cro79}
Crockett, N. G., (1979). A quick test of fit of a bivariate distribution. In D. McNeil (ed.), \emph{Interactive Statistics}, 185--191. Amsterdam: North-Holland.

\bibitem{DaChLi12}
Dai, B. T., Chua, F. y Lim, E. P. (2012). Structural Analysis in Multi-Relational Social Networks, \emph{Research Collection School of Information Systems (Open Access)}.

\bibitem{Fue89}
Feuerverger, A. (1989). On the empirical saddlepoint approximation, \emph{Biometrika}, {\bf 76}, 457--464.

\bibitem{Gil74}
Gillings, D. B. (1974). Some further results for bivariate generalizations of the Neyman type A distribution, \emph{Biometrics}, {\bf 30}, 619--628.

\bibitem{Gre77}
Gregory, G. (1977). Large sample theory for $U$-statistics and tests of fit, \emph{The Annals of Statistics}, {\bf 5}, 110--123.

\bibitem{GuHe00}
G\"urtler, N. y Henze, N. (2000). Recent and classical goodness-of-fit tests for the Poisson distribution, \emph{Journal of Statistical Planning and Inference}, {\bf 90}, 207--225.

\bibitem{Hai67}
Haight, F. A. (1967). \emph{Handbook of the Poisson distribution}, New York: John Wiley \& Sons.

\bibitem{Ham72}
Hamdan, M. A. (1972). Estimation in the Truncated Bivariate Poisson Distribution, \emph{Technometrics}, {\bf 14}, 37--45.

\bibitem{HaAl69}
Hamdan, M. A. y Al-Bayyati, H. A. (1969). A note on the bivariate
Poisson distribution, \emph{The American Statistician}, {\bf 23},
No. 4, 32--33.

\bibitem{HoSi01}
Ho, L. L. y Singer, J. M. (2001). Generalizad Least Squares Methods for Bivariate Poisson Regression, \emph{Communications in Statistics - Theory and Methods}, {\bf 30}(2), 263--277.

\bibitem{Hol64}
Holgate, P. (1964). Estimation for the bivariate Poisson
distribution, \emph{Biometrika}, {\bf 51}, 241--245.

\bibitem{Hol66}
Holgate, P. (1966). Bivariate generalizations of Neyman's Type A distribution, \emph{Biometrika}, {\bf 53}, 241--245.

\bibitem{JoKo69}
Johnson, N. L. y Kotz, S. (1969). \emph{Distributions in Statistics:
Discrete Distributions}, New York: John Wiley \& Sons.

\bibitem{JoKoBa97}
Johnson, N. L., Kotz, S. y Balakrishnan, N. (1997). \emph{Discrete
Multivariate Distributions}, New York: John Wiley \& Sons.

\bibitem{JuWi93}
Jung, R. C. y Winkelmann, R. (1993). Two Aspects of Labor Mobility: A Bivariate Poisson Regression Approach, \emph{Empirical Economics}, {\bf 18}, 543--556.

\bibitem{KaNt00}
Karlis, D. y Ntzoufras, I. (2000). On modelling soccer data, \emph{Student}, {\bf 3}, 229--244.

\bibitem{KaNt03a}
Karlis, D. y Ntzoufras, I. (2003a). Analysis of sports data by using bivariate Poisson models, \emph{The Statistician}, {\bf 52}, Part 3, 381–-393.

\bibitem{KaNt03b}
Karlis, D. y Ntzoufras, I. (2003b). Bayesian and Non-Bayesian Analysis of Soccer Data using Bivariate Poisson Regression Models, \emph{http://www.docstoc.com/docs/39143794/Bayesianand-Non-Ba-yesian-Analysis-of-Soccer-Data-using-Bivariate.}

\bibitem{KaNt05}
Karlis, D. y Ntzoufras, I. (2005). Bivariate Poisson and Diagonal Inflated Bivariate Poisson Regression Models in R, \emph{Journal of Statistical Software}, {\bf 14} (10), 1--36.

\bibitem{KaTs08}
Karlis, D. y Tsiamyrtzis, P. (2008). Exact Bayesian modeling for bivariate Poisson data and extensions, \emph{Statistics and Computing}, {\bf 18}, 27--40.

\bibitem{KoKo92}
Kocherlakota, S. y Kocherlakota, K. (1992). \emph{Bivariate Discrete Distributions}, New York: Marcel Dekker.

\bibitem{KuMaMu00}
Kundu, S., Majumdar, S. y Mukherjee, K. (2000). Central limits theorems revisited, \emph{Statistics} \& \emph{Probability Letters}, {\bf 47}, 265--275.

\bibitem{LaPaRa99}
Lakshminarayana, J., Pandit, S. N. N. y Rao, S. (1999). On a bivariate Poisson distribution, \emph{Communications in Statistics - Theory and Methods}, {\bf 28}(2), 267--276.

\bibitem{LehRom05}
Lehmann, E.L. y Romano, J.P. (2005). \emph{Testing Statistical Hypotheses}, Springer.

\bibitem{LoKe86}
Loukas, S. y Kemp, C. D. (1986). The Index of Dispersion Test for the Bivariate Poisson Distribution, \emph{Biometrics}, {\bf 42}, 941--948.

\bibitem{LoKePa86}
Loukas, S., Kemp, C. D. y Papageorgiou, H. (1986). Even point
estimation for the bivariate Poisson distribution,
\emph{Biometrika}, {\bf 73}, 222-223.

\bibitem{Mah82}
Maher, M. J. (1982). Modelling association football scores, \emph{Statistica Neerlandica}, {\bf 36}, 109--118.

\bibitem{NaPe93}
Nakamura, M. y P\'erez-Abreu, V. (1993). Use of an Empirical Probability Generating Function for Testing a Poisson Model, \emph{Canadian Journal of Statistics}, {\bf 21}, 149--156.

\bibitem{PaLo88}
Papageorgiou, H. y Loukas, S. (1988). Conditional even point estimation for bivariate discrete distributions, \emph{Communications in Statistics--Theory and Methods}, {\bf 17}, 3403--3412.

\bibitem{PaHo89}
Paul, S. R. y Ho, N. I. (1989). Estimation in the bivariate Poisson distribution and hypothesis
testing concerning independence, \emph{Communications in Statistics--Theory and Methods}, {\bf 18}, 1123--1133.

\bibitem{R}
R Development Core Team, R: A language and environment for statistical computing. R Foundation for Statistical Computing, Vienna, Austria. ISBN 3-900051-07-0, 2009, URL http://www.R-project.org.

\bibitem{RaBe95}
Rayner, J. C. W y Best, D. J. (1995). Smooth Tests for the Bivariate Poisson Distribution, \emph{Australian $\&$ New Zealand Journal of Statistics}, {\bf 37}, 233--245.

\bibitem{RaThBe09}
Rayner, J. C. W, Thas, O. y Best, D. J. (2009). \emph{Smooth Tests of Goodness of Fit}, John Wiley \& Sons.

\bibitem{RuSa00}
Rue, H. y Salvesen, $\varnothing$. (2000). Prediction and retrospective analysis of soccer matches in a league, \emph{Statistician}, {\bf 49}, 399--418.

\bibitem{RuPeOR91}
Rueda, R., P\'erez-Abreu, V., O'Reilly, F. (1991). Goodness of fit for the Poisson distribution based on the probability generating function, \emph{Communications in Statistics--Theory and Methods}, {\bf A20}, 3093--3110.

\bibitem{SaKh93}
Sahai, H. y Khurshid, A. (1993). Confidence Intervals for the Mean of a Poisson Distribution: A Review, \emph{Biometrical Journal}, {\bf 35}, 857--867.

\bibitem{Ser80}
Serfling, R. J. (1980). \emph{Approximation Theorems of Mathematical Statistics}, Wiley, New York.

\bibitem{StOr87}
Stuart, A. y Ord, J. K. (1987). \emph{Kendall's advanced theory of statistics}, Londres: Charles Griffin $\&$ C\'ia. Ltda.

\bibitem{vanWel96}
van der Vaart, A.W. y Wellner, J.A. (1996).  \emph{Weak Convergence and Empirical Processes}, New York: Springer-Verlag.

\end{thebibliography}
\end{document}